\documentclass[amssymb,aps,prd,floatfix,onecolumn]{revtex4}
\usepackage[intlimits]{amsmath}
\usepackage{amssymb}
\usepackage{graphicx}
\usepackage{tensor}
\usepackage[dvipsnames]{xcolor}
\usepackage{hyperref}
\usepackage[utf8]{inputenc}
\usepackage{natbib}
\usepackage{enumitem}

\begin{document}

\newcommand{\cc}[1]{\overline{#1}}

\title{Kerr Geodesics in Terms of Weierstrass Elliptic Functions}
\author{Adam Cie\'{s}lik}
\email{adam.cieslik@doctoral.uj.edu.pl}
\affiliation{Szkoła Doktorska Nauk Ścisłych i Przyrodniczych, Uniwersytet Jagiello\'{n}ski}
\affiliation{Instytut Fizyki Teoretycznej, Uniwersytet Jagiello\'{n}ski, \L ojasiewicza 11, 30-348 Krak\'ow, Poland}

\author{Eva Hackmann}
\email{eva.hackmann@zarm.uni-bremen.de}
\affiliation{Center of Applied Space Technology and Microgravity (ZARM), University of Bremen, Am Fallturm, 28359 Bremen, Germany}

\author{Patryk Mach}
\email{patryk.mach@uj.edu.pl}
\affiliation{Instytut Fizyki Teoretycznej, Uniwersytet Jagiello\'{n}ski, \L ojasiewicza 11,  30-348 Krak\'ow, Poland}

\begin{abstract}
    We derive novel analytical solutions describing timelike and null geodesics in the Kerr spacetime. The solutions are parameterized explicitly by constants of motion---the energy, the angular momentum, and the Carter constant---and initial coordinates. A single set of formulas is valid for all null and timelike geodesics, irrespectively of their radial and polar type. This uniformity has been achieved by applying a little-known result due to Biermann and Weierstrass, regarding solutions of a certain class of ordinary differential equations. Different from other expressions in terms of Weierstrass functions, our solution is explicitly real for all types of geodesics. In particular, for the first time the so-called transit orbits are now expressed by explicitly real Weierstrass functions.
\end{abstract}

\maketitle

\section{Introduction}
Few results in General Relativity have more astrophysical implications than the theory of timelike and null Kerr geodesics. Properties of geodesic motion in the Kerr spacetime are key in modeling accretion disks \cite{novikov_thorne_1973,page_thorne_1974} or the propagation of light in the Kerr geometry \cite{gralla_lupsasca_2020b,gralla_lupsasca_2020c}. Geodesic motion is used as a zero-order approximation in the self-force approach \cite{mino_2003} and serves as a basis for our understanding of the Penrose process \cite{piran_shaham_katz_1975}.

The theory of Kerr geodesics developed quickly after a fundamental discovery made by Carter that the Hamilton-Jacobi equation is completely separable in the Kerr spacetime \cite{Carter_1968, carter_1968b}. The geometrical nature of this fact was understood by Walker and Penrose in terms of Killing tensors \cite{walker_penrose_1970}, and equatorial orbits were soon analyzed, see e.g. \cite{defelice_1968, bardeen_1970, isco}. An analysis of non-equatorial geodesics is, of course, much more involved. Vortical orbits were first discussed  by De Felice and Calvani in \cite{defelice_calvani_1972} and later in \cite{calvani_defelice_1977}; in \cite{wilkins1972} Wilkins described bound orbits. Another account of timelike and null geodesics can be found in Bardeen's lectures \cite{bardeen_les_houches}. An early review of the theory of geodesics in black hole spacetimes was given by Sharp in \cite{sharp_1979}, and an extensive text-book discussion of various generic types of Kerr geodesics was provided in Chandrasekhar's book \cite{chandrasekhar_1983}.

Much has been achieved in subsequent years. Another detailed text-book introduction to the theory of Kerr geodesics was published by O'Neill in \cite{Neill}. Negative energy geodesics within the ergosphere were studied in \cite{contopoulos_1984} and later in \cite{grib_pavlov_vertogradov_2014, vertogradov_2015}. Spherical orbits were investigated in \cite{teo_2003, teo_2021, tavlayan_tekin_2020}. A particularly convenient geodesic parametrization, allowing for partial decoupling of geodesic equations, was introduced by Mino in \cite{mino_2003}. In \cite{schmidt_2002} Schmidt derived action-angle variables for Kerr geodesics and computed corresponding fundamental frequencies. A combination of Schmidt's results and the Mino parametrization was given in \cite{drasco_hughes_2004}. Analytic solutions for bound timelike orbits (and the associated characteristic frequencies) were obtained in \cite{fujita_hikida_2009, rana_mangalam_2019}, using Jacobian elliptic functions and Legendre's integrals. Various solutions expressed in terms of Jacobian elliptic functions were also given in \cite{slezakova_2006}. General methods of obtaining analytical solutions in terms of Weierstrass elliptic functions were described by Hackmann and L\"{a}mmerzahl in \cite{hackmann_2010, hackmann_lammerzahl_2015, lammerzahl_hackmann_2016}.

The phase-space picture of Kerr geodesics was investigated by Levin, Perez-Giz, Stein, and Warburton in \cite{levin_perez_giz_2009, perez_giz_levin_2009, stein_warburton_2020}. In particular, Ref.\ \cite{levin_perez_giz_2009} discusses equatorial homoclinic orbits and the separatrix between stable orbits and orbits plunging into the black hole. In \cite{stein_warburton_2020} Stein and Warburton considered non-equatorial orbits as well. Special classes of orbits characterized by constants of motion that also describe circular orbits were recently analysed by Mummery and Balbus in \cite{mummery_balbus_2022, mummery_balbus_2023}. The separatrix problem has also been discussed earlier in the context of gravitational radiation by Glampedakis, Kennefick and O'Shaughnessy \cite{glampedakis_kennefick_2002,oshaughnessy_2003}, and it appears naturally in the study of accretion of the Vlasov gas onto the Kerr black hole \cite{cieslik_mach_odrzywolek_2022}. Periodic orbits in the Kerr spacetime were studied in \cite{levin_perez_giz_2008}.

In \cite{gralla_lupsasca_2020} Gralla and Lupsasca revisited null geodesics in the Kerr exterior, providing a convenient classification and analytic solutions. A very detailed classification of radial motion for timelike and null geodesics has recently been provided in \cite{compere_liu_long_2022}.

In addition, many authors investigated geodesic motion in the near-horizon approximation for high-spin black holes (see, e.g., \cite{hod_2013, hadar_porfyriadis_strominger_2014, gralla_porfyriadis_warburton_2015, kapec_lupsasca_2020, compere_druart_2020}). References \cite{kapec_lupsasca_2020, compere_druart_2020} contain solutions for the generic polar motion in the Kerr spacetime as well. Last but not least, an important progress was recently made in the analysis of intricate properties of null Kerr geodesics, related to strictly observational problems---gravitational lensing, observable properties of photon rings, etc.\ \cite{rauch_blandford_1994, gralla_lupsasca_2020b, gralla_lupsasca_2020c, gralla_lupsasca_marrone_2020, gates_shadar_lupsasca_2020, himwich_et_al_2020}.

In this paper we give exact solutions of the geodesic equations in the Kerr spacetime. Using an old, but relatively little known result by Biermann and Weierstrass \cite{biermann_1865} (see also \cite{whittaker_1927} and \cite{Greenhill_1892, Reynolds_1989, CM2022} for the proof), we were able to find a single set of formulas, valid for all generic timelike and null geodesics, that yield an explicitly real solution for arbitrary admissible initial data. Our analysis is a sequel to Ref.\ \cite{CM2022}, which is also based on the Biermann-Weierstrass theorem and provides solutions for generic timelike and null  geodesics in the Schwarzschild spacetime. Benefits of this new approach are at least twofold. All solutions can be effectively written in terms of Weierstrass elliptic functions $\wp$, $\sigma$, and $\zeta$, and a single set of formulas remains valid for all generic geodesics. Solutions are explicitly specified by prescribing constants of motion---the energy, the angular momentum, the Carter constant---and the initial position. This means, in practice, that no \textit{a priori} knowledge about the type of the orbit is required in order to select an appropriate formula. Secondly, radial motions of the so-called transit type \cite{Neill}, for which the radial potential has no real zeros, can be described in a straightforward way, by explicitly real formulas.

%This aspect represents an improvement with respect to the existing approach using Weierstrass functions, described by one of us in \cite{hackmann_2010}.

Many geodesic types, which are usually treated as special, non-generic cases in other approaches, are incorporated naturally in our formalism. Interesting examples of such types include ``whirling'' near-separtartix orbits. 

All our formulas have been coded in \textit{Wolfram Mathematica} \cite{Mma}, and a collection of sample solutions is plotted in Sec.\ \ref{sec:examples} of this paper. A \textit{Wolfram Mathematica} package implementing our formulas is now avilable at the GitHub \cite{git}.

The remainder of this paper is organized as follows. Section \ref{sec:preliminaries} contains preliminary material: we introduce our notation conventions, discuss a simple classification of orbital types, and recall the Biermann-Weierstrass theorem. Section \ref{sec:solutions} provides general solutions to the geodesic equations. We discuss the general method of deriving our solutions, but many technical details are given in Appendix \ref{mainappendix}. Sections \ref{sec:implementation} and \ref{sec:examples} give a short description of our implementation in \textit{Wolfram Mathematica} and a collection of examples, respectively. Section \ref{sec:discussion} provides a short discussion of our results. Appendix \ref{appendix:limit} is devoted to Schwarzschild and extreme-Kerr limits of our solutions. In Apppendix \ref{appendix:Integrals} we provide a table o integrals to which we refer in our derivation.

\section{Preliminaries}
\label{sec:preliminaries}

\subsection{Metric conventions}

We use standard geometric units with $c = G = 1$, where $c$ denotes the speed of light, and $G$ is the gravitational constant. The signature of the metric is $(-,+,+,+)$.

We will work in Boyer-Lindquist coordinates $x^\mu = (t,r,\theta,\varphi)$. The Kerr metric can be written as
\begin{equation}
\label{kerr4d}
    g = g_{\mu \nu} dx^\mu dx^\nu = g_{t t} dt^2 + 2 g_{t \varphi} dt d\varphi + g_{r r} dr^2 + g_{\theta \theta} d\theta^2 + g_{\varphi \varphi} d\varphi^2,
\end{equation}
where
\begin{subequations}
\begin{eqnarray}
g_{t t} & = & -1 + \frac{2Mr}{\rho^2}, \\
g_{t \varphi} & = & - \frac{2 M a r \sin^2 \theta}{\rho^2}, \\
g_{r r} & = & \frac{\rho^2}{\Delta}, \\
g_{\theta \theta} & = & \rho^2, \\
g_{\varphi \varphi} & = & \left( r^2 + a^2 + \frac{2 M a^2 r \sin^2 \theta}{\rho^2} \right) \sin^2 \theta,
\end{eqnarray}
\end{subequations}
and we denote
\begin{subequations}
\begin{eqnarray}
\Delta & = & r^2 - 2 M r + a^2, \\
\rho^2 & = & r^2 + a^2 \cos^2 \theta.
\end{eqnarray}
\end{subequations}
The Kerr spacetime is characterized by the mass $M$ and the angular momentum $J = M a$. The two solutions of the equation $\Delta = 0$, $r_H^\pm = M \pm \sqrt{M^2 - a^2}$, correspond to the inner and outer Kerr horizons.

\subsection{Equations of motion}

Timelike and null geodesic equations can be expressed in the Hamiltonian form
\begin{subequations}
\begin{eqnarray}
\frac{d x^\mu}{d \tilde{\tau}} & = & \frac{\partial H}{\partial p_\mu}, \\
\frac{d p_\nu}{d \tilde{\tau}} & = & - \frac{\partial H}{\partial x^\nu},
\end{eqnarray}
\end{subequations}
where $p^\mu = d x^\mu / d \tilde{\tau}$, $H(x^\alpha,p_\beta) = \frac{1}{2} g^{\mu\nu}(x^\alpha) p_\mu p_\nu = - \frac{1}{2} m^2$, and $m$ denotes the particle rest mass. The above normalization of the four-momentum $p^\mu$ is a useful convention. The four-velocity $u^\mu$ is normalized as $g_{\mu \nu} u^\mu u^\nu = -\delta_1$, where
\begin{equation}
    \delta_1 = \begin{cases} 1 & \text{for timelike geodesics,} \\ 0 & \text{for null geodesics.} \end{cases}
\end{equation}
Thus, for timelike geodesics, $p^\mu = m u^\mu$, and the proper time can be expressed as $\tau = m \tilde \tau$.

By standard arguments, $H$, $p_t \equiv - E$ and $p_\varphi \equiv l_z$ are constants of motion. The fourth constant---the so-called Carter constant $\mathcal{K}$---follows from the separation of variables in the Hamilton-Jacobi equation \cite{Carter_1968, chandrasekhar_1983}. In explicit terms, geodesic equations can be written as
\begin{equation*}
        \begin{aligned}
            \rho^2  \frac{dr}{d\tilde{\tau}}  & = \epsilon_r \sqrt{R(r)}, \\
            \rho^2  \frac{d\theta}{d\tilde{\tau}}  & = \epsilon_{\theta} \sqrt{\Theta(\theta)},\\
            \rho^2  \frac{d\varphi}{d\tilde{\tau}} & =   \frac{ a \left[ \left(r^2 + a^2 \right) E - a l_z \right] }{\Delta} + \frac{1}{\sin^2 \theta} \left( l_z - a E \sin^2 \theta  \right),\\
            \rho^2  \frac{dt}{d\tilde{\tau}} & =  \frac{ \left(r^2  + a^2 \right) \left[ \left(r^2 + a^2 \right) E - a l_z \right] }{\Delta} +  a \left( l_z - a E \sin^2 \theta  \right).
        \end{aligned} 
\end{equation*}
Here
\begin{equation}
    \label{W_r1}
    R(r) \equiv \left[ (r^2 + a^2) E - a l_z \right]^2 - \Delta (m^2 r^2 + \mathcal{K})
\end{equation}
and
\begin{equation}
    \label{W_theta1}
    \Theta(\theta) \equiv \mathcal{K} - m^2 a^2 \cos^2 \theta - \left( \frac{l_z}{\sin \theta} - a \sin \theta E \right)^2.
\end{equation}
With a slight abuse of terminology, we will refer to $R(r)$ and $\Theta(\theta)$ as radial and polar potentials, respectively. The signs $\epsilon_r$ and $\epsilon_\theta$ correspond to directions of motion in $r$ and $\theta$, respectively. More precisely,
\begin{equation}
\label{W_r2}
   p_r = \epsilon_r \frac{\sqrt{R(r)}}{ \Delta}, 
\end{equation}
and
\begin{equation}
\label{W_theta2}
    p_\theta = \epsilon_{\theta} \sqrt{\Theta(\theta)}.
\end{equation}

There is a convenient way to partially decouple the above equations, introduced by Mino in \cite{mino_2003}. It consists in reparametrizing geodesics so that
\begin{equation}\label{coord_time_1}
    \rho^2 \frac{dx^\mu}{d \tilde{\tau} } = \frac{dx^\mu}{d\tilde s}
\end{equation}
or
\begin{equation}\label{coord_time_2}
    \tilde{\tau} = \int_0^{\tilde s} \rho^2 ds.
\end{equation}
This yields
\begin{subequations}\label{eq_mot}
    \begin{eqnarray}
        \frac{dr}{d\tilde s}  &=& \epsilon_r \sqrt{R(r)},\\
        \frac{d\theta}{d\tilde s}  &=& \epsilon_{\theta} \sqrt{\Theta(\theta)},\\
        \frac{d\varphi}{d\tilde s} &=&   \frac{ a \left[ \left(r^2 + a^2 \right) E - a l_z \right] }{\Delta} + \frac{1}{\sin^2 \theta} \left( l_z - a E \sin^2 \theta  \right),\\
        \frac{dt}{d\tilde s} &=&   \frac{ \left(r^2  + a^2 \right) \left[ \left(r^2 + a^2 \right) E - a l_z \right] }{\Delta} +  a \left( l_z - a E \sin^2 \theta  \right).
    \end{eqnarray}
\end{subequations}

It is convenient to work in dimensionless rescaled variables. For timelike geodesics, we define dimensionless variables as in \cite{Olivier1,Olivier3}, i.e., by
\begin{equation}
\label{rescaling}
    a = M \alpha, \quad t = M T, \quad r = M \xi,  \quad E  =  m \varepsilon, \quad \mathcal{K} =  M^2 m^2 \kappa, \quad l_z = M m \lambda_z.
\end{equation}
In addition, the rescaled Mino time $s$ is defined by
\begin{equation}
    \tilde{s} = \frac{1}{Mm} s.
\end{equation}
For null geodesics $m = 0$, and the parameter $m$ in Eqs.\ (\ref{rescaling}) should be replaced with any positive mass parameter $\tilde{m}>0$.

In terms of dimensionless variables, geodesic equations \eqref{eq_mot} can be written as
\begin{subequations}\label{eq_mot_dimless}
    \begin{eqnarray}
        \frac{d\xi}{ds}  &=& \epsilon_r \sqrt{\tilde R}, \label{eq_mot_dimless1}\\
        \frac{d\theta}{ds}  &=& \epsilon_{\theta} \sqrt{\tilde \Theta}, \label{eq_mot_dimless2}\\
        \frac{d\varphi}{ds} &=& \frac{ \alpha \left[ \left(\xi^2 + \alpha^2 \right) \varepsilon - \alpha \lambda_z \right] }{\xi^2 - 2\xi + \alpha^2} + \frac{1}{\sin^2 \theta} \left( \lambda_z - \alpha \varepsilon \sin^2 \theta  \right), \label{eq_mot_dimless3}\\
        \frac{dT}{ds} &=&  \frac{ \left(\xi^2  + \alpha^2 \right) \left[ \left(\xi^2 + \alpha^2 \right) \varepsilon - \alpha \lambda_z \right] }{\xi^2 - 2 \xi + \alpha^2} + \alpha \left( \lambda_z - \alpha \varepsilon \sin^2 \theta  \right),\label{eq_mot_dimless4}
    \end{eqnarray}
\end{subequations}
where
\begin{eqnarray}
    \tilde R & \equiv & \left[ (\xi^2 + \alpha^2) \varepsilon - \alpha \lambda_z \right]^2 - (\xi^2 - 2\xi + \alpha^2)(\delta_1 \xi^2 + \kappa), \\
    \tilde \Theta & \equiv & \kappa - \delta_1 \alpha^2 \cos^2 \theta - \frac{1}{\sin^2 \theta} \left( \lambda_z - \alpha  \varepsilon \sin^2 \theta \right)^2. 
\end{eqnarray}
In Section \ref{sec:solutions} we give explicit solutions of Eqs.\ (\ref{eq_mot_dimless}) for the functions $\xi(s)$, $\theta(s)$, $\varphi(s)$, and $T(s)$. They are specified by prescribing $\delta_1$ and the constants of motion $\varepsilon$, $\lambda_z$, and $\kappa$, which in turn can be computed from initial velocities. Defining, for timelike geodesics, components of the veolcity $v^i = u^i/u^t = p^i/p^t$, $i = r, \theta, \varphi$, we get
\begin{equation}
    \left( \frac{p^t}{m} \right)^2 = - \frac{1}{g_{tt} + 2 g_{t \varphi} v^\varphi + g_{rr} (v^r)^2 + g_{\theta \theta} (v^\theta)^2 + g_{\varphi \varphi} (v^\varphi)^2}.
\end{equation}
Constants $\varepsilon$, $\lambda_z$, and $\kappa$ can be expressed as
\begin{eqnarray}
    \varepsilon & = & - \frac{p^t}{m} \left( g_{tt} + g_{t \varphi} v^\varphi \right), \\
    \lambda_z & = & \frac{p^t}{M m} \left( g_{t\varphi} + g_{\varphi \varphi} v^\varphi \right), \\
    \kappa & = & \frac{(p^t)^2}{M^2 m^2} \left\{ \rho^4 (v^\theta)^2 + \left[  \frac{g_{\varphi \varphi} v^\varphi + g_{t \varphi}}{\sin \theta} + a \sin \theta \left( g_{tt} + g_{t\varphi} v^\varphi \right)\right]^2 \right\} +  \alpha^2 \cos^2 \theta.
\end{eqnarray}

It is well known \cite{chandrasekhar_1983, gralla_lupsasca_2020} that for null geodesics solutions of Eqs.\ (\ref{eq_mot}) depend on $E$, $l_z$, and $\mathcal K$ via the ratios $l_z/E$ and $\mathcal{K}/E^2$. This corresponds to the choice $\tilde m = E$ in Eqs.\ (\ref{rescaling}) and (\ref{eq_mot_dimless}), which, assuming $E > 0$, yields $\varepsilon = 1$. Constants $E$, $\lambda_z$ and $\kappa$ can be then expressed as
\begin{eqnarray}
E & = & - g_{tt} p^t - g_{t \varphi} p^\varphi, \\
\lambda_z & = & \frac{1}{M E} \left( g_{t\varphi} p^t + g_{\varphi \varphi} p^\varphi \right), \\
\kappa & = & \frac{\rho^4}{M^2 E^2} \left( p^\theta \right)^2 + \frac{1}{\sin^2 \theta} \left( \lambda_z - \alpha \sin^2 \theta \right)^2,
\end{eqnarray}
where $p^\mu$ is a null vector.

\subsection{Types of orbits}

Although the main idea of this paper is to provide a single set of formulas describing solutions of all generic timelike and null geodesic trajectories, it is useful to introduce a general classification of different types of possible orbits and some terminology, which can facilitate the discussion. A detailed discussion of various orbital types can be found, e.g., in \cite{Neill}.

Generic orbits can be classified according to the type of radial motion. Allowing, in the discussion, for negative values of $\xi$, we adopt the following terminology from \cite{Neill} (p. 209, Def.\ 4.6.3): A geodesic is said to be of transit type, if $\xi(s)$ goes from $\pm \infty$ to $\mp \infty$, as the parameter $s$ goes from $-\infty$ to $+\infty$. A geodesics, for which $\xi(s)$ goes from $\pm \infty$ to $\pm \infty$ is called a flyby. Both types represent unbound orbits. A geodesic is said to be interval-bound, if $\xi_1 \le \xi(s) \le x_2$, for some $-\infty < r_1 < r_2 < + \infty$. All other geodesics are exceptional in some sense, and this includes orbits terminating in the $\rho^2 = 0$ singularity or geodesics related with an occurrence of multiple zeros of $\tilde R$. The character of generic orbits is mainly governed by the number of real zeros of the radial potential $\tilde R$. The following cases are generically possible:
\begin{enumerate}[label=\Roman*.]
    \item The potential $\tilde R$ has no real zeros, and $\tilde R > 0$ for all $\xi$. This allows for transit orbits.
    \item The potential $\tilde R$ has two real zeros $\xi_1 < \xi_2$, and $\tilde R > 0$ for $\xi < \xi_1$ and $\xi_2 < \xi$. This case allows for flyby orbits.
    \item The potential $\tilde R$ has four real zeros $\xi_1 < \xi_2 < \xi_3 < \xi_4$, and $\tilde R > 0$ for $\xi_1 < \xi < \xi_2$ and $\xi_3 < \xi < \xi_4$. Interval-bound orbits are possible.
    \item The potential $\tilde R$ has four real zeros $\xi_1 < \xi_2 < \xi_3 < \xi_4$, and $\tilde R > 0$ for $\xi < \xi_1$, $\xi_2 < \xi < \xi_3$, and $\xi_4 < \xi$. Both flyby and interval-bound orbits are possible.
    \item The potential $\tilde R$ has two real zeros $\xi_1 < \xi_2$, and $\tilde R > 0$ for $\xi_1 < \xi < \xi_2$. Only interval-bound orbits are possible.
\end{enumerate}

Since Boyer-Lindquist coordinates used in this paper are singular at Kerr horizons ($\xi = \xi_H^\pm$), solutions of the whole system of equations (\ref{eq_mot_dimless}) cannot be continued through $\xi = \xi_H^\pm$. On the other hand, the radial and polar equations (\ref{eq_mot_dimless1}) and (\ref{eq_mot_dimless2}) remain unaffected by horizon singularities of Boyer-Lindquist coordinates, and this fact allows for the general classification given above, without invoking explicitly regular coordinate systems.

A very detailed classification of the radial geodesic motion has recently been given in \cite{compere_liu_long_2022} for timelike trajectories. In contrast to the terminology summarized above, the authors of \cite{compere_liu_long_2022} concentrate on possible configurations of roots of $\tilde R$ in the region outside the outer Kerr horizon, which is of course of main physical relevance. A similar classification for null geodesics was given in \cite{gralla_lupsasca_2020} and subsequently slightly expanded in \cite{compere_liu_long_2022}.

A classification with respect to the range of the $\theta$ coordinate is simpler \cite{chandrasekhar_1983, Neill}. Defining $\mu = \cos \theta$, one can express $\tilde \Theta$ as $\tilde \Theta = g(\mu)/\sin^2 \theta$, where $g(\mu)$ is a biquadratic polynomial defined in Eq.\ (\ref{gdef}) below. It allows for one or two zeros in the range $0 \le \mu^2 \le 1$. Moreover, $g(\mu = 0) = \kappa - (\alpha \varepsilon - \lambda_z)^2$ and $g(\mu = 1) = - \lambda_z^2 \le 0$. The condition $\tilde \Theta > 0$, or equivalently $g(\mu) > 0$, leads to two generic types of motion. For $\kappa - (\alpha \varepsilon - \lambda_z)^2 > 0$, the trajectory oscillates around the equator. In this case $\mu^2 < \mu^2_+ \le 1$, where $\mu_+$ is a zero of the polynomial $g(\mu)$. For $\kappa - (\alpha \varepsilon - \lambda_z)^2 < 0$, if the motion is at all possible, it is restricted to one hemisphere, and the allowed range of $\mu^2$ is $0 < \mu_-^2 \le \mu^2 \le \mu_+^2 \le 1$, where again $\mu_\pm$ denote zeros of $g(\mu)$.

\subsection{Biermann--Weierstrass theorem}

Solutions derived in this work are largely based on the following result due to Biermann and Weierstrass. The original formulation of this theorem was published in \cite{biermann_1865}. More detailed proofs can be found in \cite{Greenhill_1892, Reynolds_1989, CM2022}.

Let 
\begin{equation}
    f(x) = a_0 x^4 + 4 a_1 x^3 +6 a_2 x^2 + 4a_3 x + a_4
\end{equation}
be a quartic polynomial, and let $g_2$ and $g_3$ denote Weierstrass invariants of $f$, i.e.,
\begin{subequations}
\label{invariants_theorem}
\begin{eqnarray}
        g_2 & = &  a_0 a_4 - 4a_1 a_3 + 3 a_2^2, \\
        g_3 & = & a_0 a_2 a_4 + 2a_1 a_2 a_3 -a_2^3 -a_0 a_3^2 - a_1^2 a_4.
\end{eqnarray}
\end{subequations}
Let
\begin{equation}
\label{zthm}
     z(x) = \int^x_{x_0} \frac{dx^\prime}{\sqrt{f(x^\prime)}},
\end{equation}
where $x_0$ is any constant, not necessarily a zero of $f(x)$. Then,
\begin{equation}
\label{glowne}
     x = x_0 + \frac{- \sqrt{f(x_0)} \wp'(z) + \frac{1}{2} f'(x_0) \left[ \wp(z) - \frac{1}{24}f''(x_0) \right] + \frac{1}{24} f(x_0) f'''(x_0)  }{2 \left[ \wp(z) - \frac{1}{24} f''(x_0) \right]^2 - \frac{1}{48} f(x_0) f^{(4)}(x_0) },
\end{equation}
where $\wp(z) =\wp(z;g_2,g_3)$ is the Weierstrass function corresponding to invariants (\ref{invariants_theorem}), i.e., the function $z(x)$ can be inverted. In addition
\begin{subequations}
\label{BW_wp}
    \begin{equation}
        \wp(z)  =  \frac{\sqrt{f(x) f(x_0)} + f(x_0)}{2(x-x_0)^2} + \frac{f'(x_0)}{4(x-x_0)} + \frac{f''(x_0)}{24},\\
    \end{equation}
    \begin{equation}
        \wp'(z)  =  \textstyle  - \left[ \frac{f(x)}{(x-x_0)^3} -\frac{f'(x)}{4(x-x_0)^2} \right] \sqrt{f(x_0)} - \left[ \frac{f(x_0)}{(x-x_0)^3} + \frac{f'(x_0)}{4(x-x_0)^2} \right]\sqrt{f(x)}.
    \end{equation}    
\end{subequations}

%=======================================================
%=======================================================
\section{Solutions of the geodesic equations}
\label{sec:solutions}

\subsection{Radial motion}

A solution for $\xi(s)$ can be obtained by a direct application of the Biermann-Weierstrass theorem to Eq.\ (\ref{eq_mot_dimless1}). The polynomial $\tilde R(\xi)$ can be written as
\begin{equation}
\label{f_general}
    \tilde R(\xi) = a_0 \xi^4 + 4a_1 \xi^3 +6a_2\xi^2 + 4a_3\xi + a_4,
\end{equation}
where
\begin{subequations}\label{fcoeffs}
    \begin{eqnarray}
        a_0 &=& \varepsilon^2 - \delta_1,\\
        a_1 &=& \frac{1}{2} \delta_1,\\
        a_2 &=& - \frac{1}{6} (\delta_1 \alpha^2 + \kappa - 2\alpha^2 \varepsilon^2 + 2\alpha\varepsilon\lambda_z),\\
        a_3 &=&  \frac{1}{2}\kappa,\\
        a_4 &=& \alpha^4 \varepsilon^2 - \alpha^2 \kappa - 2\alpha^3 \varepsilon \lambda_z + \alpha^2 \lambda_z^2 = - \alpha^2 \left[ \kappa - (\alpha \varepsilon - \lambda_z)^2 \right].
    \end{eqnarray}
\end{subequations}

For a segment of a trajectory for which $\epsilon_r$ is constant, we get
\begin{equation}
\label{psi_integral}
s = \epsilon_r \int_{\xi_0}^\xi \frac{d \bar \xi}{\sqrt{\tilde R( \bar \xi)}},
\end{equation}
where $\xi_0$ is an arbitrarily chosen (initial) radius corresponding $s = 0$. Weierstrass invariants of the polynomial $\tilde R$ read
\begin{subequations}
 \label{invariants_phys}
\begin{eqnarray}   
        g_{\tilde R,2} & = & a_0 a_4 -4a_1 a_3 + 3a_2^2, \\
        g_{\tilde R,3} & = & a_0 a_2 a_4 + 2a_1 a_2 a_3 - a_2 ^3 - a_0a_3^2 -a_1^2a_4.
\end{eqnarray}
\end{subequations}
Using the Biermann-Weierstrass theorem,
we can write the formula for $\xi = \xi(s)$ as 
\begin{equation}
\label{xi_s}
    \xi(s) =  \xi_0 + \frac{- \epsilon_r \sqrt{\tilde R(\xi_0)} \wp_{\tilde R}'(s) + \frac{1}{2} \tilde R'(\xi_0 ) \left[ \wp_{\tilde R}(s) - \frac{1}{24} \tilde R''(\xi_0 )\right] + \frac{1}{24} \tilde R(\xi_0 ) \tilde R'''(\xi_0 )  }{2 \left[ \wp_{\tilde R}(s) - \frac{1}{24} \tilde R''(\xi_0 ) \right]^2 - \frac{1}{48} \tilde R(\xi_0 ) \tilde R^{(4)}(\xi_0 ) },
\end{equation}
where $\wp_{\tilde R}(s) = \wp(s;g_{\tilde R,2},g_{\tilde R,3})$, and the polynomial $\tilde R$ is defined in Eqs.\ (\ref{f_general}) and (\ref{fcoeffs}).

It is important to stress that the above solution remains valid also when the sign $\epsilon_r$ changes along a trajecotry. In this case $\epsilon_r$ in Eq.\ (\ref{xi_s}) should be understood as an initial sign value, corresponding to $s = 0$. A longer discussion of this fact in the context of Schwarzschild geodesics can be found in \cite{CM2022}.

Note that if $\xi_0$ is chosen as a root of the polynomial $\tilde R$, i.e., $\tilde R(\xi_0) = 0$, expression (\ref{xi_s}) can be reduced to
\begin{equation}
    \xi(s) = \xi_0 + \frac{\tilde R^\prime (\xi_0)}{4 \wp_{\tilde R}(s) - \frac{1}{6} \tilde R''(\xi_0) }.
\end{equation}
This can be a useful parametrization for trajectories with radial turning points. On the other hand, if no real zeros of $\tilde R$ exist, Eq.\ (\ref{xi_s}) still provides a valid, explicitly real solution. This is the case of transit orbits of radial type I.

\subsection{Polar motion}
\label{subsection:theta}

Equation (\ref{eq_mot_dimless2}) can be transformed to the Biermann-Weierstrass form by a substitution $\nu = \cos^2 \theta$ or $\mu = \cos \theta$. The first option yields a convenient form of a third order polynomial under the square root in Eq.\ (\ref{eq_mot_dimless2}), but it does not constitute a one to one function in the region $0 \le \theta \le \pi$. In this work we choose the second possibility and define $\mu = \cos \theta$, which yields a one to one mapping for $0 \le \theta \le \pi$. We get, in this case,
\begin{equation}
    \frac{d \mu}{d s} = - \epsilon_\theta \sin \theta \sqrt{\tilde \Theta} = - \epsilon_\theta \sqrt{\sin^2 \theta \tilde \Theta} = - \epsilon_\theta \sqrt{g(\mu)},
\end{equation}
where
\begin{equation}
\label{gdef}
    g(\mu) = b_0 \mu^4 + 6 b_2 \mu^2 + b_4
\end{equation}
and
\begin{subequations}
\begin{eqnarray}
    b_0 & = & - \alpha^2 (\varepsilon^2 - \delta_1) = - \alpha^2 a_0, \\
    b_2 & = & \frac{1}{6} \left( -\alpha^2 \delta_1 + 2 \alpha^2 \varepsilon^2 - \kappa - 2 \alpha \varepsilon \lambda_z \right) = a_2, \\
    b_4 & = & -\alpha^2 \varepsilon^2 + \kappa + 2 \alpha \varepsilon \lambda_z - \lambda_z^2 = \kappa - (\alpha \varepsilon - \lambda_z)^2 = - \frac{a_4}{\alpha^2}.
\end{eqnarray}
\end{subequations}
The fact that the coefficients $b_0$, $b_2$, and $b_4$ are directly related to the coefficients $a_0$, $a_2$, and $a_4$ is somewhat surprising, since Eqs.\ (\ref{eq_mot_dimless1}) and (\ref{eq_mot_dimless2}) are decoupled. The appropriate Weierstrass invariants read
\begin{subequations}\label{invariants_g}
\begin{eqnarray}
    g_{g,2} & = &  b_0 b_4 + 3 b_2^2, \\
    g_{g,3} & = & b_0 b_2 b_4 - b_2^3,
\end{eqnarray}
\end{subequations}
and the solution for $\mu = \mu(s)$ can be written as
\begin{equation}
\label{mu_s}
    \mu(s) =  \mu_0 + \frac{\epsilon_\theta \sqrt{g(\mu_0)} \wp_g'(s) + \frac{1}{2} g'(\mu_0 ) \left[ \wp_g(s) - \frac{1}{24}g''(\mu_0 )\right] + \frac{1}{24} g(\mu_0 ) g'''(\mu_0 )  }{2 \left[ \wp_g(s) - \frac{1}{24} g''(\mu_0 ) \right]^2 - \frac{1}{48} g(\mu_0 ) g^{(4)}(\mu_0 ) },
\end{equation}
where $\wp_g(s) = \wp (s; g_{g,2},g_{g,3})$ and $\mu_0 = \mu(0) = \cos \theta_0 = \cos \theta(0)$ is the initial value corresponding to $s = 0$. Here, as in the radial case, the sign $\epsilon_\theta$ is to be understood as referring to the polar momentum component at the initial location $\theta_0$.

Note that due to the relation between $b_0$, $b_2$, $b_4$ and $a_0$, $a_2$, $a_4$, we have
\begin{subequations}
\begin{eqnarray}
    g_{\tilde R,2} & = & g_{g,2} - \delta_1 \kappa, \\
    g_{\tilde R,3} & = & g_{g,3} + 2 a_1 a_2 a_3 - a_0 a_3^2 - a_1^2 a_4.
\end{eqnarray}
\end{subequations}

\subsection{Azimuthal motion}
\label{sec:varphi}

In order to find a solutions for $\varphi(s)$, we split the right-hand side of Eq.\ \eqref{eq_mot_dimless3} into two parts depending on $\xi$ and $\theta$, respectively. Integrating the result with respect to the Mino time $s$, we get
\begin{eqnarray}
    \varphi(s) - \varphi(0) & = & \int_0^s \frac{ \alpha \left[ \left(\xi(\bar s)^2 + \alpha^2 \right) \varepsilon - \alpha \lambda_z \right] }{\xi(\bar s)^2 - 2\xi(\bar s) + \alpha^2} d \bar s +  \int_0^s \frac{1}{\sin^2\theta(\bar s)} \left[ \lambda_z - \alpha \varepsilon \sin^2 \theta(\bar s)  \right] d \bar s \nonumber \\
    & = & I_\xi(s) + I_\theta(s),
\label{varphi_int}
\end{eqnarray}
where
\begin{equation}
\label{Ixi1}
I_\xi(s) = \alpha \int_0^s \frac{2\xi(\bar s)\varepsilon - \alpha \lambda_z}{\xi(\bar s)^2 - 2 \xi(\bar s) + \alpha^2} d \bar s,  
\end{equation}
and
\begin{equation}
\label{itheta}
    I_\theta(s) = \lambda_z \int_0^s \frac{d \bar s}{\sin^2 \theta(\bar s)}.
\end{equation}

The first step in computing the integral $I_\xi(s)$ consists in applying the partial fraction decomposition with respect to $\xi(s)$. Note that $\xi^2 - 2 \xi + \alpha^2 = (\xi - \xi^+_H)(\xi - \xi^-_H)$, where $\xi^{\pm}_H = 1 \pm \sqrt{1-\alpha^2}$ denote the dimensionless radii of the inner and outer Kerr horizons. This yields, assuming $-1 < \alpha < 1$, the following partial fraction decomposition
\begin{eqnarray}
    I_\xi (s) & = & \frac{\alpha \varepsilon}{\xi_H^+ - \xi_H^-} \int_0^s \left\{ \left(\xi_H^{+2} + \alpha^2 - \frac{\alpha\lambda_z}{\varepsilon}\right) \frac{1}{\xi(\bar s) -\xi_H^+} - \left(\xi_H^{-2} + \alpha^2 - \frac{\alpha\lambda_z}{\varepsilon}\right) \frac{1}{\xi(\bar s) -\xi_H^-} \right\}  d \bar s \nonumber \\
     & = &  \frac{\alpha \varepsilon}{\xi_H^+ - \xi_H^-} \left[ \left(\xi_H^{+2} + \alpha^2 - \frac{\alpha\lambda_z}{\varepsilon}\right) N_H^+(s) - \left(\xi_H^{-2} + \alpha^2 - \frac{\alpha\lambda_z}{\varepsilon}\right) N_H^-(s) \right],
    \label{Ixi2}
\end{eqnarray}
where
\begin{equation}
\label{N_H}
    N_H^{\pm}(s) := \int_0^s \frac{1}{\xi(\bar s) - \xi_H^{\pm}} d \bar s.
\end{equation}

One way of computing integrals $N_H^\pm(s)$ would be to express the integrand in terms of derivatives of the Weierstrass $\zeta$ function, and to adhere to a general integration scheme described in \cite{hackmann_2010}. Here, we decided to follow a more straightforward approach, which once again makes use of the Biermann-Weierstrass result. We start with the following substitution
\begin{equation}
\label{sub1}
    u_{\pm}(s) = \frac{1}{\xi (s) -  \xi_H^{\pm}},
\end{equation}
so that
\begin{equation}
    N_H^{\pm} = \int_0^s u_{\pm}(\bar s) d \bar s.
\end{equation}
Differentiating $u_\pm$ with respect to $s$, we get, using Eq.\ (\ref{eq_mot_dimless1}),
\begin{equation}
\label{u_mot}
    \frac{du_{\pm}}{ds} = -\epsilon_r u_{\pm}^2 \sqrt{\tilde R\left( \frac{1}{u_{\pm}} +  \xi_H^{\pm} \right)} =  -\epsilon_r \sqrt{h\left( u_{\pm} \right)},
\end{equation}
where 
\begin{equation}
\label{g_general}
    h(u_{\pm}) \equiv c_0 u_{\pm}^4 + 4 c_1 u_{\pm}^3 +6 c_2 u_{\pm}^2 + 4 c_3 u_{\pm} + c_4,
\end{equation}
and 
\begin{subequations}\label{gcoeffs}
    \begin{eqnarray}
        c_0 & = & (2\xi_H^\pm \varepsilon - \alpha \lambda_z)^2, \\
        c_1 & = & \frac{1}{2} \left\{ \xi_H^\pm  \left[ - \alpha^2 \delta_1 + 2 \alpha^2 \varepsilon^2  - 2 \alpha \lambda_z \varepsilon + \xi_H^\pm \left( -2 \delta_1 \xi_H^\pm +3 \delta_1 + 2 \xi_H^\pm \varepsilon^2 \right) \right] - \kappa (\xi_H^\pm - 1) \right\}, \\
        c_2 & = & \frac{1}{6} \left[ - \alpha^2 \left( \delta_1 - 2 \varepsilon^2 \right) - \kappa - 2\alpha\varepsilon\lambda_z + 6\left( \delta_1 - \delta_1 \xi_H^\pm + \varepsilon^2 \xi_H^\pm \right) \xi_H^{\pm}   \right], \\
        c_3 & = & \delta_1 \left( \frac{1}{2} - \xi_H^\pm \right) + \varepsilon^2 \xi_H^\pm, \\
        c_4 & = & \varepsilon^2 - \delta_1.
    \end{eqnarray}
\end{subequations}
A simpler expression for $c_1$ can be obtained by replacing the terms $\xi_H^\pm$ with $\xi_H^\pm = 1 \pm \sqrt{1 - \alpha^2}$. This yields
\begin{eqnarray}
    c_1 & = &  -2 \alpha ^2 \varepsilon ^2 + \frac{1}{2} \alpha^2 \delta_1 \left( \pm \sqrt{1 - \alpha ^2} + 2 \right) - \delta_1
   \left( \pm \sqrt{1 - \alpha ^2} + 1 \right) \nonumber \\
   & & \pm \frac{1}{2} \sqrt{1 - \alpha^2} \left(8
   \varepsilon ^2-\kappa \right) - \alpha  \lambda_z \varepsilon 
   \left( \pm \sqrt{1 - \alpha^2} + 1 \right) + 4 \varepsilon ^2.
\end{eqnarray}

One can check that  invariants of the polynomial $h$  are the same as for the polynomial $\tilde R$, i.e., they are given by Eqs.\ (\ref{invariants_phys}). This is not surprising. It can be easily proved \cite{Greenhill_1892} that a substitution (we keep the original notation of \cite{Greenhill_1892} at this point)
\begin{equation}
\label{linear_subs}
    x = \frac{l x^\prime + m}{l^\prime x^\prime + m^\prime} 
\end{equation} 
transforms a differential form $dx/\sqrt{X(x)}$, where $X(x)$ denotes a 4th order polynomial in $x$, into
\begin{equation}
\frac{(l m^\prime - l^\prime m) dx^\prime}{\sqrt{X^\prime(x^\prime)}},
\end{equation}
where $X^\prime(x^\prime)$ is a 4th order polynomial of $x^\prime$. The Weierstrass invariants $g_2$ and $g_3$ of the polynomial $X(x)$ and the invariants $g_2^\prime$, $g_3^\prime$ of $X^\prime(x^\prime)$ are related:
\begin{subequations}
\begin{eqnarray}
    g_2^\prime & = & (l m^\prime - l^\prime m)^4 g_2, \\
    g_3^\prime & = & (l m^\prime - l^\prime m)^6 g_3.
\end{eqnarray}
\end{subequations}
In our case, $\xi(s) = \left[ \xi_H^\pm u_\pm(s) + 1 \right]/u_\pm(s)$, i.e., $m = 1$, $l = \xi_H^\pm$, $l^\prime = 1$, and $m^\prime = 0$ in substitution (\ref{linear_subs}). Consequently $l^\prime m - l m^\prime = 1$, and $g_2^\prime = g_2$, $g_3^\prime = g_3$.

Before proceeding further, let us note that Eqs.\ (\ref{sub1}) and (\ref{u_mot}) give 
\begin{equation}
    \frac{1}{u_{\pm}^2} = \frac{\sqrt{\tilde R(\xi)}}{\sqrt{h(u_{\pm})}}, \quad \xi = \frac{1}{u_\pm} + \xi_H^\pm,
\end{equation}
and thus
\begin{equation}
   h(u_\pm) = u_\pm^4 \tilde R(\xi).
\end{equation}
It follows that $h(u_\pm)$ is positive, as long as $\tilde R(\xi)$ remains positive. Note, however, that $\xi \to \xi_H^\pm$, corresponds to $u_\pm \to \infty$, i.e., the change of variables given by Eq.\ (\ref{sub1}) is singular at one of horizons. 

Keeping the above fact in mind, we can now apply the Biermann-Weierstrass theorem to Eq.\ \eqref{u_mot} and express $u_{\pm}$ as a function of $s$: 
\begin{equation}
\label{u_s}
    u_{\pm}(s) =  u_{0\pm} + \frac{ \epsilon_r \sqrt{h(u_{0\pm})} \wp_{\tilde R}'(s) + \frac{1}{2} h'(u_{0\pm} ) \left[ \wp_{\tilde R}(s) - \frac{1}{24}h''(u_{0\pm} )\right] + \frac{1}{24} h(u_{0\pm} ) h'''(u_{0\pm})}{2 \left[ \wp_{\tilde R}(s) - \frac{1}{24} h''(u_{0\pm} ) \right]^2 - \frac{1}{48} h(u_{0\pm} ) h^{(4)}(u_{0\pm} ) },
\end{equation}
where $u_{0 \pm} = u_\pm(0) = 1/(\xi_0 - \xi_H^\pm)$. Formula (\ref{u_s}) is written in terms of the Weierstrass function $\wp_{\tilde R} = \wp(s; g_{\tilde R,2},g_{\tilde R,3})$, since both polynomials $\tilde R$ and $h$ are characterized by the same Weierstrass invariants.

Expression (\ref{u_s}) can be integrated. To simplify subsequent calculations, we introduce the following new symbols:
\begin{subequations}
    \begin{eqnarray}
        \mathcal{A}_{1\pm} &=& \frac{1}{48}h(u_{0\pm}) h'''(u_{0\pm}) - \frac{1}{96} h'(u_{0\pm})h''(u_{0\pm}), \\    
        \mathcal{A}_{2\pm} &=& \frac{1}{4}h'(u_{0\pm}), \\
        \mathcal{A}_{3\pm} &=& \frac{1}{2}\epsilon_r \sqrt{h(u_{0\pm})},\\
        \mathcal{A}_{4\pm} &=& \frac{1}{24}h''(u_{0\pm}), \\
        \mathcal{A}_{5\pm} &=&  \frac{1}{48}h(u_{0\pm})h^{(4)}(u_{0\pm}),\\
        \wp_{1\pm} &=&   \mathcal{A}_{4\pm} + \sqrt{\frac{ \mathcal{A}_{5\pm}}{2}},\\
        \wp_{2\pm} &=&   \mathcal{A}_{4\pm} - \sqrt{\frac{ \mathcal{A}_{5\pm}}{2}}.
    \end{eqnarray}
\end{subequations}
Hence Eq.\ \eqref{u_s} takes the form
\begin{equation}
\label{u_s_2}
    u_{\pm}(s) = u_{0\pm} + \frac{2\mathcal{A}_{3\pm} \wp_{\tilde R}'(s) + 2\mathcal{A}_{2\pm}\wp_{\tilde R}(s) + 2\mathcal{A}_{1\pm}   }{2 \left[ \wp_{\tilde R}(s) - \wp_{1 \pm} \right] \left[ \wp_{\tilde R}(s) - \wp_{2 \pm} \right]},
\end{equation}
and integral \eqref{N_H} can be written as
\begin{eqnarray}
    N_H^{\pm} & = & \int_0^s \left\{ u_{0\pm} + \frac{ \mathcal{A}_{1\pm} + \mathcal{A}_{2\pm}\wp_{\tilde R}(\bar s) +  \mathcal{A}_{3\pm} \wp_{\tilde R}'(\bar s)}{\left[ \wp_{\tilde R}(\bar s) - \wp_{1 \pm} \right] \left[ \wp_{\tilde R}(\bar s) - \wp_{2 \pm} \right]} \right\} d \bar s \nonumber \\
    &=&  u_{0\pm} s +  \mathcal{A}_{1\pm} K_{1}(s; p_{1\pm}, p_{2 \pm}) + \mathcal{A}_{2\pm}  K_{2}(s; p_{1\pm}, p_{2 \pm})  + \mathcal{A}_{3\pm}  K_{3}(s;p_{1\pm}, p_{2 \pm}),
    \label{nhfinal}
\end{eqnarray}
where
\begin{subequations}\label{K_int}
    \begin{eqnarray}
        K_{1}(s; p_{1\pm}, p_{2 \pm}) &:=& \int_0^s \frac{ 1 }{ \left[ \wp(\bar s) - \wp_{1\pm} \right] \left[ \wp(\bar s) - \wp_{2 \pm} \right]} d \bar s,\\
        K_{2}(s; p_{1\pm}, p_{2 \pm}) &:=& \int_0^s \frac{\wp(\bar s) }{ \left[ \wp(\bar s) - \wp_{1\pm} \right] \left[ \wp(\bar s) - \wp_{2 \pm} \right]} d \bar s,\\
        K_{3}(s; p_{1\pm}, p_{2 \pm}) &:=&  \int_0^s \frac{\wp'(\bar s) }{ \left[ \wp(\bar s) - \wp_{1\pm} \right] \left[ \wp(\bar s) - \wp_{2 \pm} \right]} d \bar s,
    \end{eqnarray}
\end{subequations}
and $p_{1 \pm}$ and $p_{2 \pm}$ satisfy $\wp\left( p_{1 \pm}; g_{\tilde R,2}, g_{\tilde R,3} \right) = \mathcal{A}_{4\pm} + \sqrt{\frac{ \mathcal{A}_{5\pm}}{2}}$ and $\wp \left( p_{2 \pm}; g_{\tilde R,2}, g_{\tilde R,3} \right) = \mathcal{A}_{4\pm} - \sqrt{\frac{ \mathcal{A}_{5\pm}}{2}}$. Weierstrass functions appearing in the expressions for $K_1$, $K_2$, and $K_3$ in Eq.\ (\ref{nhfinal}) should be computed assuming invariants $g_{\tilde R,2}$ and $g_{\tilde R,3}$. Integrals \eqref{K_int} are calculated in Appendix \ref{appendix:Integrals} [Eqs.\ (\ref{intC1}), (\ref{intC2}), (\ref{intC7}), and (\ref{intK1}--\ref{intK3})]. 

Note that $h^{(4)}(u_{0\pm}) = 24 c_0 = 24 (2 \xi_H^\pm \varepsilon - \alpha \lambda_z)^2$ is non-negative, and hence $\mathcal{A}_{5 \pm}$ must also be non-negative. Thus both $\wp_{1\pm}$ and $\wp_{2\pm}$ are real. Nevertheless $p_{1\pm}$ or $p_{2\pm}$ may be complex, even though both Weierstrass invariants $g_{\tilde R,2}$, $g_{\tilde R,3}$ are real. This happens, for instance, when $\wp(p; g_{\tilde R,2}, g_{\tilde R,3})$ is strictly positive on the real line, and $\wp_{1\pm}$ or $\wp_{2\pm}$ becomes negative. However, even in this case, the integral $K_1$ given by Eqs.\ (\ref{intC1}) and (\ref{intK1}) is explicitly real. A potentially complex term in the expression for $K_1$ [Eq.\ (\ref{intK1})] has the form
\[ I_1(s;p) = \frac{1}{\wp'(p)} \left[ 2 \zeta(p) s + \ln \frac{\sigma(s - p)}{\sigma(s + p)} \right], \]
where $s$ is real. In specific examples $I_1(s;p)$ can indeed become complex, but the imaginary part of $I_1(s;p)$ is independent of $s$, and hence it cancels out in the definite integral $K_1$. This fact follows immediately from Eq.\ (\ref{intC1}), which we write as
\[ \frac{d}{ds} I_1(s;p) = \frac{1}{\wp(s) - \wp(p)}. \]
Since both $\wp(p)$ and $\wp(s)$ are real, one gets
\[ \frac{d}{ds} \, \mathrm{Im} \, I_1(s;p)  = 0. \]
In a similar way, one can demonstrate that $K_2$ must be real. The integral $K_3$ is given by an explicitly real formula.

The integral $I_\theta$ defined in Eq.\ (\ref{itheta}) can be computed using the substitution $\mu = \cos \theta$. We have
\begin{equation}
    \frac{1}{\sin^2 \theta(s)} = \frac{1}{1 - \mu^2(s)} \equiv q(s).
\end{equation}
The derivative of $q(s)$ reads
\begin{equation}
    \frac{dq}{ds} = 2 q^2 \mu \frac{d\mu}{ds} = - 2 \epsilon_\theta q^2 \mu \sqrt{g(\mu)} = - \epsilon_\theta \epsilon_\mu \sqrt{4 q^4 \mu^2 g(\mu)},
\end{equation}
where $\epsilon_\mu = \mathrm{sgn} \, \mu$. On the other hand
\begin{equation}
    4 q^4 \mu^2 g(\mu) = 4 q (q-1) \left[ b_0 (q-1)^2 + 6 b_2 (q^2 - q) + b_4 q^2 \right] \equiv w(q).
\end{equation}
Thus $q$ satisfies the equation
\begin{equation}
    \frac{dq(s)}{ds} = - \epsilon_\theta \epsilon_\mu \sqrt{w(q)},
\end{equation}
where
\begin{equation}
    w(q) \equiv d_0 q^4 + 4 d_1 q^3 + 6 d_2 q^2 + d_3 q,
\end{equation}
and
\begin{subequations}
\label{coeffsd}
\begin{eqnarray}
    d_0 & = & - 4 \lambda_z^2, \\
    d_1 & = & - \alpha^2 \delta_1 + \kappa + 2 \alpha \varepsilon \lambda_z + \lambda_z^2, \\
    d_2 & = & \frac{2}{3} \left[ \alpha^2 (2 \delta_1 - \varepsilon^2) - \kappa - 2 \alpha \varepsilon \lambda_z \right], \\
    d_3 & = & \alpha^2 (-\delta_1 + \varepsilon^2).
\end{eqnarray}
\end{subequations}
Weierstrass invariants of the polynomial $w(q)$ read
\begin{subequations}
\begin{eqnarray}
    g_{w,2} & = & - 4 d_1 d_3 + 3 d_2^2, \\
    g_{w,3} & = & 2 d_1 d_2 d_3 - d_2 ^3 - d_0 d_3^2.
\end{eqnarray}
\end{subequations}
Thus,
\begin{equation}
\label{qs}
    q(s) = q_0 + \frac{ \epsilon_\theta \epsilon_\mu \, \sqrt{w(q_0)} \wp_w'(s) + \frac{1}{2} w'(q_0 ) \left[ \wp_w(s) - \frac{1}{24}w''(q_0 )\right] + \frac{1}{24} w(q_0 ) w'''(q_0 )  }{2 \left[ \wp_w(s) - \frac{1}{24} w''(q_0 ) \right]^2 - \frac{1}{48} w(q_0) w^{(4)}(q_0)},
\end{equation}
where $\wp_w(s) = \wp(s;g_{w,2},g_{w,3})$. Since $w(q) = 4 q^2 \mu^2 g(\mu)$, $w(q)$ is non-negative, whenever $g(\mu)$ remains non-negative. Note also that $q \to +\infty$ at the axes, while $\mu = 0$ at the equatorial plane. We define
\begin{subequations}
    \begin{eqnarray}
        \mathcal{B}_{1} &=& \frac{1}{48}w(q_{0}) w'''(q_{0}) - \frac{1}{96} w'(q_{0})w''(q_{0}), \\
        \mathcal{B}_{2} &=& \frac{1}{4}w'(q_{0}), \\
        \mathcal{B}_{3} &=& \frac{1}{2}\epsilon_\theta \epsilon_\mu \sqrt{w(q_{0})},\\
        \mathcal{B}_{4} &=& \frac{1}{24}w''(q_{0}), \\
        \mathcal{B}_{5} &=&  \frac{1}{48}w(q_{0})w^{(4)}(q_{0}),\\
        \wp_{w,1} &=&   \mathcal{B}_{4} + \sqrt{\frac{ \mathcal{B}_{5}}{2}},\\
        \wp_{w,2} &=&   \mathcal{B}_{4} - \sqrt{\frac{ \mathcal{B}_{5}}{2}}.
    \end{eqnarray}
\end{subequations}
Hence
\begin{eqnarray}
    I_\theta(s) & = & \lambda_z \int_0^s q(\bar s) d \bar s = \lambda_z \int_0^s \left\{ q_0 + \frac{ \mathcal{B}_1 + \mathcal{B}_2 \wp_w(\bar s) +\mathcal{B}_3 \wp_w'(\bar s) }{ \left[ \wp_w(\bar s) - \wp_{w,1} \right] \left[ \wp_w(\bar s) - \wp_{w,2} \right]} \right\} d \bar s \nonumber \\
    & = & \lambda_z \left[ q_{0} s +  \mathcal{B}_{1} K_{1}(s; p_{w,1}, p_{w,2}) + \mathcal{B}_{2}  K_{2}(s; p_{w,1}, p_{w,2})  + \mathcal{B}_{3}  K_{3}(s;p_{w,1}, p_{w,2}) \right],
    \label{ithetafinal}
\end{eqnarray}
where $p_{w,1}$ and $p_{w,2}$ satisfy $\wp\left( p_{w,1}; g_{w,2}, g_{w,3} \right) = \mathcal{B}_{4} + \sqrt{\frac{ \mathcal{B}_{5}}{2}}$ and $\wp \left( p_{w,2}; g_{w,2}, g_{w,3} \right) = \mathcal{B}_{4} - \sqrt{\frac{ \mathcal{B}_{5}}{2}}$. Moreover, Weierstrass functions appearing in the expressions for $K_1$, $K_2$, $K_3$ in the above formula should be computed with the invariants $g_{w,2}$ and $g_{w,3}$.

Integrals $K_1$, $K_2$, and $K_3$ appearing in Eq.\ (\ref{ithetafinal}) are again real, although the constants  $p_{w,1}$, $p_{w,2}$ are in general complex. The situation encountered here is, however, different than the one described for integrals $N_H^\pm(s)$. We have $\mathcal B_5 \le 0$, and thus $\wp_{w,1}$ and $\wp_{w,2}$ are, in general, complex. The fact that the integral $K_1$ is real can be demonstrated as follows. Observe that
\begin{align}
\wp(p_{1}) = \mathcal{B}_{4} + \sqrt{\frac{ \mathcal{B}_{5}}{2}} = \cc{\wp(p_{2})} = \wp(\cc{p}_{2}),
\end{align}
where the bar denotes complex conjugation, and where, for simplicity, we omit the Weierstrass invariants $g_{w,2}$ and $g_{w,3}$, as well as the subscript $w$ in the symbols $p_{w,1}$ and $p_{w,2}$. Therefore, $p_{1}$ and $\cc{p}_{2}$ are the two solutions, in the period parallelogram, of the equation $\wp(x)=\mathcal{B}_{4} + \sqrt{\frac{\mathcal{A}_{5}}{2}}$, and hence they are related by
\begin{align}
\cc{p}_{2} = z_{NM} - p_{1},
\end{align}
where $z_{NM}=2N\omega + 2M\omega'$ is a period of $\wp$, and $N$, $M$ are integers. Keeping in mind that $\cc{z}_{NM} = z_{nm}$ is a period of $\wp$ as well (this happens for real Weierstrass invariants), we find the following relations:
\begin{align}
\wp'(p_{2}) & = \wp'(z_{nm} - \cc{p}_{1}) = -\wp'(\cc{p}_{1}) = - \cc{\wp'(p_{1})}, \\
\zeta(p_{2}) & = \zeta(z_{nm} - \cc{p}_{1}) = -\zeta(\cc{p}_{1}) + 2n\eta + 2m\eta' = - \cc{\zeta(p_{1})} + 2n\eta + 2m\eta', \\
\sigma(x+p_{2}) & = \sigma(x-\cc{p}_{1} + z_{nm}) \nonumber \\
& =  (-1)^{n+m+nm} \sigma(x-\cc{p}_{1})  \exp\big[2(n\eta+m\eta')(n\omega+m\omega'+x-\cc{p}_{1}) \big], \\
\sigma(x-p_{2}) & = \sigma(x+\cc{p}_{1}-z_{nm}) \nonumber\\
& = (-1)^{-n-m+nm} \sigma (x+\cc{p}_{1}) \exp\big[2(-n\eta-m\eta')(-n\omega-m\omega'+x+\cc{p}_{1}) \big], \\
\frac{\sigma(x-p_{2})}{\sigma(x+p_{2})} & = \frac{\sigma (x+\cc{p}_{1})}{\sigma(x-\cc{p}_{1})} \exp\big[ -4(n\eta+m\eta')x \big],
\end{align}
where $\eta$ and $\eta'$ are the periods of the second kind. Inserting the above formulas in the expressions for integrals $K_1$ and $K_2$ [see Eqs.\ (\ref{intC1}), (\ref{intC7}), (\ref{intK1}), and (\ref{intK2})], we find, for instance,
\begin{align}
K_1 & = \frac{1}{\wp(p_{1})-\wp(\cc{p}_{1})} \Bigg[ 2s \left( \frac{\zeta(p_{1})}{\wp'(p_{1})} - \frac{\zeta(\cc{p}_{1})}{\wp'(\cc{p}_{1})} \right) + \frac{4s(n\eta+m\eta')}{\wp'(\cc{p}_{1})} + \frac{1}{\wp'(p_{1})} \ln \frac{\sigma(s-p_{1})}{\sigma(s+p_{1})} \nonumber \\
& + \frac{1}{\wp'(\cc{p}_{1})} \ln \left(\frac{\sigma(s+\cc{p}_{1})}{\sigma(s-\cc{p}_{1})} e^{-4s(n\eta+m\eta')} \right) \Bigg]\nonumber\\
& = \frac{1}{\wp(p_{1})-\wp(\cc{p}_{1})} \Bigg[ 2s \left( \frac{\zeta(p_{1})}{\wp'(p_{1})} - \frac{\zeta(\cc{p}_{1})}{\wp'(\cc{p}_{1})} \right) + \frac{1}{\wp'(p_{1})} \ln \frac{\sigma(s-p_{1})}{\sigma(s+p_{1})} - \frac{1}{\wp'(\cc{p}_{1})} \ln \frac{\sigma(s-\cc{p}_{1})}{\sigma(s+\cc{p}_{1})}\Bigg]. \label{K1real}
\end{align}
We see that pairs of the type $x-\cc{x}$ appear in Eq.\ (\ref{K1real}) both in the nominator and the denominator and, therefore, $K_1$ is real. Similarly, the integral $K_2$ is basically of the same form. Integral $K_3$ does not involve any sigma functions and reads
\begin{align}
K_3 & = \frac{\ln [\wp(s)-\wp(p_{1})] - \ln  [\wp(s)-\wp(\cc{p}_{1})]}{\wp(p_{1}) - \wp(\cc{p}_{1})}.
\end{align}

Summarizing all results from this section, ultimately, we can write Eq.\ \eqref{varphi_int} as
\begin{equation}
    \varphi(s) - \varphi(0)  =  \frac{\alpha \varepsilon}{\xi_H^+ - \xi_H^-} \left[ \left(\xi_H^{+2} + \alpha^2 - \frac{\alpha\lambda_z}{\varepsilon}\right) N_H^+ (s) - \left(\xi_H^{-2} + \alpha^2 - \frac{\alpha\lambda_z}{\varepsilon}\right) N_H^- (s)\right] + I_\theta(s).
    \label{varphi_int2}
\end{equation}
with integrals $N_H^\pm$ given in eq. \eqref{nhfinal} and integral $I_\theta$ given in eq. \eqref{ithetafinal}.

The case with $\alpha = \pm 1$ (the extreme Kerr limit) requires a separate treatment, as the partial fraction decomposition given by Eq.\ (\ref{Ixi2}) has a different form. Appropriate formulas are given in Appendix \ref{appendix:limit}.

\subsection{Time coordinate}

The solution for $T(s)$ can be obtained in a similar way. A direct integration of Eq.\ (\ref{eq_mot_dimless4}) yields
\begin{eqnarray} \label{tau_int}
    T(s) - T(0) & = & \int_0^s \frac{ \left[\xi^2(\bar s)  + \alpha^2 \right] \left\{ \left[\xi^2(\bar s) + \alpha^2 \right] \varepsilon - \alpha \lambda_z \right\} }{\xi^2(\bar s) - 2 \xi (\bar s) + \alpha^2} d \bar s + \int_0^s \alpha \left[ \lambda_z - \alpha \varepsilon \sin^2 \theta (\bar s)  \right] d \bar s =J_\xi(s) - \alpha^2 \varepsilon J_\theta(s)
\end{eqnarray}
where

\begin{equation}
 J_\xi(s) = \int_0^s \frac{\left[\xi^2(\bar s) + \alpha^2 \right]^2 \varepsilon - 2\alpha \lambda_z \xi(\bar s)}{\xi(\bar s)^2-2\xi(\bar s)+\alpha^2} d \bar s   
\end{equation}

\begin{equation}\label{J_theta}
    J_\theta(s) = \int_0^s \sin^2 \theta(\bar s) d \bar s.
\end{equation}

For $-1 < \alpha < 1$, the expression

\begin{equation}
    \frac{ (\xi^2+\alpha^2)^2 \varepsilon - 2\alpha \lambda_z \xi }{\xi^2 - 2 \xi + \alpha^2} = \frac{ (\xi^2+\alpha^2)^2 \varepsilon - 2\alpha \lambda_z \xi }{(\xi - \xi_H^+)(\xi - \xi_H^-)}
\end{equation}

has the following partial fraction decomposition

\begin{eqnarray}
    \frac{ (\xi^2+\alpha^2)^2 \varepsilon - 2\alpha \lambda_z \xi }{(\xi - \xi_H^+)(\xi - \xi_H^-)} = (4 + \alpha^2) \varepsilon +  2 \varepsilon \xi + \varepsilon \xi^2 + \frac{A_-}{\xi - \xi_H^-} + \frac{A_+}{\xi - \xi_H^+},
\end{eqnarray}

where
\begin{equation}
    A_\pm = \frac{ 2 \left( \mp \alpha^2 + 2 \sqrt{1-\alpha^2} \pm 2 \right) \varepsilon -\alpha \left( \sqrt{1 - \alpha^2} \pm 1 \right) \lambda_z}{\sqrt{1 - \alpha^2}} = \frac{2\varepsilon (\xi_H^\pm)^2 - \alpha \lambda_z \xi_H^\pm}{\xi_H^\pm-1}.
\end{equation}

Consequently,

\begin{equation}
    J_\xi(s) = (4 + \alpha^2) \varepsilon s + 2 \varepsilon N_1(s) + \varepsilon N_2(s) + A_- N_H^-(s) + A_+ N_H^+(s),
\end{equation}

where, apart from the integrals $N_H^\pm$ computed in the previous section, we have denoted
\begin{equation}
\label{N_1}
    N_1(s) = \int_0^s \xi(\bar s) d \bar s
\end{equation}
and
\begin{equation}
\label{N_2}
    N_2(s) = \int_0^s \xi(\bar s)^2 d \bar s.
\end{equation}
Detailed expressions for $N_1(s)$, $N_2(s)$, and $J_\theta(s)$ are given in Appendix \ref{mainappendix}.

Finally, we can write Eq.\ \eqref{tau_int} as
\begin{eqnarray}
    T(s) - T(0) & = & (4 + \alpha^2) \varepsilon s + 2 \varepsilon N_1(s) + \varepsilon N_2(s) + A_- N_H^-(s) + A_+ N_H^+(s) - \alpha^2 \varepsilon  J_\theta(s).
    \label{tau_s}
\end{eqnarray}
As before, the extreme Kerr case with $\alpha = \pm 1$ has to be considered separately. The appropriate partial fraction expansion of the radial part of Eq.\ (\ref{tau_int}) can be found in Appendix \ref{appendix:limit}.

\subsection{Proper time}

The relation between the proper time $\tilde \tau$ and the rescaled Mino time $s$ can also be integrated. Equations \eqref{coord_time_1} or \eqref{coord_time_2} give
\begin{equation*}
     \frac{d\tilde{\tau}}{ds} = \frac{M}{m} \left( \xi^2 + \alpha^2 \cos^2 \theta \right)
\end{equation*}
and, upon integration,
\begin{equation*}
\tilde{\tau}(s) - \tilde{\tau}(0) = \frac{M}{m} \int^s_0 \left[ \xi(\bar s)^2  + \alpha^2 \cos^2 \theta(\bar s) \right] \bar s,
\end{equation*}
which can be expressed in terms of $N_2(s)$ and $J_\theta(s)$ as
\begin{equation}
    \tilde{\tau}(s) -\tilde{\tau}(0) = \frac{M}{m} \left\{ N_2(s) +  \alpha^2 \left[ s -J_{\theta}(s) \right] \right\}.
\end{equation}

\section{Implementation in Wolfram Mathematica}
\label{sec:implementation}

All formulas expressing the solutions obtained in Sec.\ \ref{sec:solutions} have been encoded in a \textit{Wolfram Mathematica} \cite{Mma} package, which is now available at the GitHub \cite{git}. The implementation of our formulas is essentially straightforward, but it is tedious, due to the number of different integrals appearing in our calculation. Perhaps the only place which requires special attention is related with computing the terms of the form
\begin{equation}
\label{sigma_log}
    \ln \frac{\sigma(s-y;g_2,g_3)}{\sigma(s+y;g_2,g_3)},
\end{equation}
where $\sigma$ denotes the Weierstrass function sigma, $s$ is the Mino time, and $y$ is, in general, a complex parameter. Expressions of this type appear in integrals (\ref{intC1}), (\ref{intC3}), and (\ref{intC7}). A proper implementation of our formulas, yielding continuous solutions for $\varphi(s)$ and $T(s)$, requires careful selection of the complex logarithm branch in Eq.\ (\ref{sigma_log}). A ``brute force'' approach to this problem is to start at $s = 0$ and to follow the appropriate logarithm branch up to a given non-zero value of $s$, assuring that both real and imaginary parts of Eq.\ (\ref{sigma_log}) remain continuous. The logarithm in Eq.\ (\ref{sigma_log}) can be thus computed as
\begin{equation}
\label{sigma_log2}
    \ln \frac{\sigma(s-y;g_2,g_3)}{\sigma(s+y;g_2,g_3)} = \mathrm{Log} \, \frac{\sigma(s-y;g_2,g_3)}{\sigma(s+y;g_2,g_3)} + 2 \pi k i,
\end{equation}
where $\mathrm{Log}$ denotes the principal value of the complex logarithm, and $k \in \mathbb Z$, but one has to select an appropriate integer $k$, which can change as $s$ increases. Another straightforward (but numerically expensive) approach is to express (\ref{sigma_log}) as
\begin{equation}
\label{sigma_log_int}
    \ln \frac{\sigma(s-y;g_2,g_3)}{\sigma(s+y;g_2,g_3)} = \int_0^s \left[ \zeta (\bar s - y; g_2, g_3) - \zeta(\bar s + y; g_2, g_3) \right] d \bar s,
\end{equation}
and evaluate the above integral numerically.

There is a partial workaround to the above problem, based on the properties of the Weierstrass function sigma. Let $\omega_1$ and $\omega_3$ denote half-periods of $\wp(z)$, and let $\omega_1$ be real. The function sigma can be expressed as
\begin{eqnarray}
    \sigma(z) & = & \frac{2 \omega_1}{\pi} \exp\left(\frac{\eta_1 z^2}{2\omega_1}\right) \sin\left(\frac{\pi z}{2\omega_1}\right) \prod_{n=1}^\infty \frac{1 - 2 q^{2n} \cos \left( \frac{\pi z}{\omega_1} \right) + q^{4n}}{\left( 1 - q^{2n} \right)^2} \nonumber \\
    & = & \frac{2 \omega_1}{\pi} \exp\left(\frac{\eta_1 z^2}{2\omega_1}\right) \frac{\theta_1 \left( \pi z/(2 \omega_1), q \right)}{\theta_1^\prime (0,q)},
\end{eqnarray}
where $q = e^{i \pi \omega_3/\omega_1}$, $\eta_1 = \zeta(\omega_1)$, $\theta_1(z,q)$ denotes the Jacobi theta function, and where we omit temporarily the Weierstrass invariatns $g_2$ and $g_3$ in the argument of $\sigma(z)$ \cite{DLMF}. For arguments $z$ of the form $z = s \pm y$, where $y$ is a (possibly) complex parameter, and $s$ is real, we get a product of an exponential factor $\exp \left[ \eta_1 z^2/(2 \omega_1) \right)]$ and the remaining part, periodic in $s$. To disentangle these two different behaviors, we define
\begin{equation}  \tilde \sigma(z) = \sigma(z) \exp \left[ - \eta_1 z^2/(2 \omega_1) \right]. \end{equation}
Thus
\begin{equation}
\ln \frac{\sigma(s-y)}{\sigma(s+y)} = \ln \frac{\tilde \sigma(s-y)}{\tilde \sigma(s+y)}  - \frac{2 \eta_1 s y}{\omega_1}.
\end{equation}
Here the key task is to control the phase of $\tilde \sigma (s - y) / \tilde \sigma (s + y)$. The principal value of the argument of this ratio, $\mathrm{Arg} \, \left[ \tilde \sigma (s - y) / \tilde \sigma (s + y) \right]$, is discontinuous at $s = 0$ and $s = 2 \omega_1$. This can be seen by expanding $\tilde \sigma (s - y) / \tilde \sigma (s + y)$ with respect to $s$ around $s = 0$, which gives, up to terms linear in $s$,
\begin{equation}
\frac{\tilde \sigma(s-y)}{\tilde \sigma(s+y)} = -1 + 2 \left[ \zeta(y) - \frac{\eta_1 y}{\omega_1} \right] s + O (s^2).
\end{equation}
Hence, $\mathrm{Arg} \, \left[ \tilde \sigma (s - y) / \tilde \sigma (s + y) \right]$ changes at $s = 0$ from $-\pi$ to $\pi$, if $\mathrm{Im} \, \left[ \zeta(y) - \eta_1 y/\omega_1 \right] > 0$, and from $\pi$ to $-\pi$, if  $\mathrm{Im} \, \left[ \zeta(y) - \eta_1 y/\omega_1 \right] < 0$. By periodicity, the same happens also at $s = 2 \omega_1$. In the simplest case $\mathrm{Arg} \, \left[ \tilde \sigma (s - y) / \tilde \sigma (s + y) \right]$ would exhibit no additional discontinuities within the range $0 < s < 2 \omega_1$. If, however, the argument $\mathrm{Arg} \, \left[ \tilde \sigma (s - y) / \tilde \sigma (s + y) \right]$ changes sufficiently fast, additional jumps may also occur. To correct for this general behavior, we write
\begin{equation}
\ln \frac{\tilde \sigma(s-y)}{\tilde \sigma(s+y)} = \ln \left\{ \frac{\tilde \sigma(s-y)}{\tilde \sigma(s+y)} \exp \left[  k \pi i \left( \frac{s}{\omega_1} - 1 \right) \right] \right\} - k \pi i \left( \frac{s}{\omega_1} - 1 \right),
\end{equation}
where $k$ is a fixed integer.

A final prescription for a continuous choice of the logarithm branch can be written as
\begin{equation}
\label{logfinal}
\ln \frac{\sigma(s-y)}{\sigma(s+y)} = \mathrm{Log} \, \left\{ \frac{\tilde \sigma(s -y)}{\tilde \sigma(s + y)} \exp \left[ k \pi i \left( \frac{s}{\omega_1} - 1 \right) \right] \right\} - k \pi i \left( \frac{s}{\omega_1} - 1 \right) - \frac{2 \eta_1 s y}{\omega_1}.
\end{equation}
In practical examples, the above prescription works well, saving time with respect to the ``brute force'' algorithm described above. Controlling the changes of the phase of $\sigma(s - y)/\sigma(s + y)$ within one period $2 \omega_1$ suffices to compute the suitable value of $k$. Let
\begin{equation}
    c = \mathrm{Im} \, \left[ \ln \frac{\sigma(s + 2 \omega_1 - y)}{\sigma(s + 2 \omega_1 + y)} \right] - \mathrm{Im} \, \left[ \ln \frac{\sigma (s - y)}{\sigma (s + y)} \right] = \mathrm{Arg} \, \frac{\sigma(s + 2 \omega_1 - y)}{\sigma(s + 2 \omega_1 + y)} - \mathrm{Arg} \, \frac{\sigma (s - y)}{\sigma (s + y)},
\end{equation}
which can be computed, e.g., from Eq.\ (\ref{sigma_log_int}). Assuming Eq.\ (\ref{logfinal}) and the equality
\begin{equation}
    \mathrm{Arg} \, \left\{ \frac{\tilde \sigma(s + 2 \omega_1 -y)}{\tilde \sigma(s + 2 \omega_1 + y)} \exp \left[ k \pi i \left( \frac{s + 2 \omega_1}{\omega_1} - 1 \right) \right] \right\} = \mathrm{Arg} \, \left\{ \frac{\tilde \sigma(s -y)}{\tilde \sigma(s + y)} \exp \left[ k \pi i \left( \frac{s}{\omega_1} - 1 \right) \right] \right\},
\end{equation}
one gets immediately
\begin{equation}
    c = - 2 k \pi - \mathrm{Im} \, (4 \eta_1 y),
\end{equation}
and thus
\begin{eqnarray}
    k =  - \frac{1}{2 \pi} \left[ \mathrm{Im} \, \left( 4 \eta_1 y \right) + c \right].
\end{eqnarray}

\section{Examples}
\label{sec:examples}

\begin{table*}
\caption{\label{tab1} Parameters of geodesics shown in Figs.\ \ref{figA} to \ref{figNullD}. Black hole horizons are located at $\xi_H^- = 1 - \sqrt{1 - \alpha^2} = 0.4$ and $\xi_H^+ = 1 + \sqrt{1 - \alpha^2} = 1.6$.}
\begin{ruledtabular}
    \begin{tabular}{ccccccccc}
        Fig.\ No.\ & $\delta_1$ & $\varepsilon^2$ & $\kappa$ & $\lambda_z$ & $\alpha$ & Radial type & Real zeros of $\tilde R(\xi)$ & $\xi_0$ \\
        \hline
        \ref{figA} & 1 & 1.1 & 12 & $-1$ & 0.8 & II, flyby & $-24.3351$, 0.254136 & 8\\
        \ref{figB} & 1 & 0.95 & 12 & 3 & 0.8 & III, interval-bound & 0.22019, 1.63896, 8.44487, 29.696 & 10 \\
        $\ref{figC}$ & 1 & 0.95 & 12 & 3 & 0.8 & III, interval-bound & 0.22019, 1.63896, 8.44487, 29.696 & 1.55 \\
        \ref{figD} & 1 & 1.1 & 12 & 3 & 0.8 & IV, flyby & $-26.4861$, 0.230431, 1.67987, 4.57578 & 10\\
        \ref{figE} & 1 & 0.5 &  12 & $-1$ & 0.8 & V, interval-bound & 0.291099, 2.3974 & 2.3 \\
        \ref{figF} & 1 & 30 & 12 & $-0.05$ & 0.8 & I, transit & --- & 10 \\
        \ref{figNullA} & 0 & 1 & 60 & 4.47214 & 0.8 & IV, flyby & $-8.927$, 0.296172, 1.60191, 7.02891 & 10 \\
        \ref{figNullB} & 0 & 1 & 60 & 4.47214 & 0.8 & IV, interval-bound & $-8.927$, 0.296172, 1.60191, 7.02891 & 1.5 \\
        \ref{figNullC} & 0 & 1 & 0.6 & $-0.111803$ & 0.8 & II, flyby & $-0.721257$,  $-0.137167$ & 10 \\
        \ref{figNullD} & 0 & 1 & 0.4 & $-0.00912871$ & 0.8 & I, transit & --- & 10 \\
    \end{tabular}
    \end{ruledtabular}
\end{table*}

\begin{table*}
\caption{\label{tab2} Parameters of special equatorial geodesics shown in Fig.\ \ref{figZW}.}
\begin{ruledtabular}
    \begin{tabular}{ccccccc}
    Fig.\ No.\ & $\delta_1$ & $\varepsilon^2$ & $\kappa$  & $\lambda_z$ & $\alpha$ & $\xi_0$ \\
    \hline
    \ref{figZW} & 1 & 0.77064 & 2.81619 & 2.38044 & 0.8 & 2.89664 \\
    \ref{figZW} & 0 & 1 & 5.94042 & 3.2373 & 0.8 & 1.80109
    \end{tabular}
    \end{ruledtabular}
\end{table*}

\begin{figure}[t]
\centering
\includegraphics[width=0.4\linewidth]{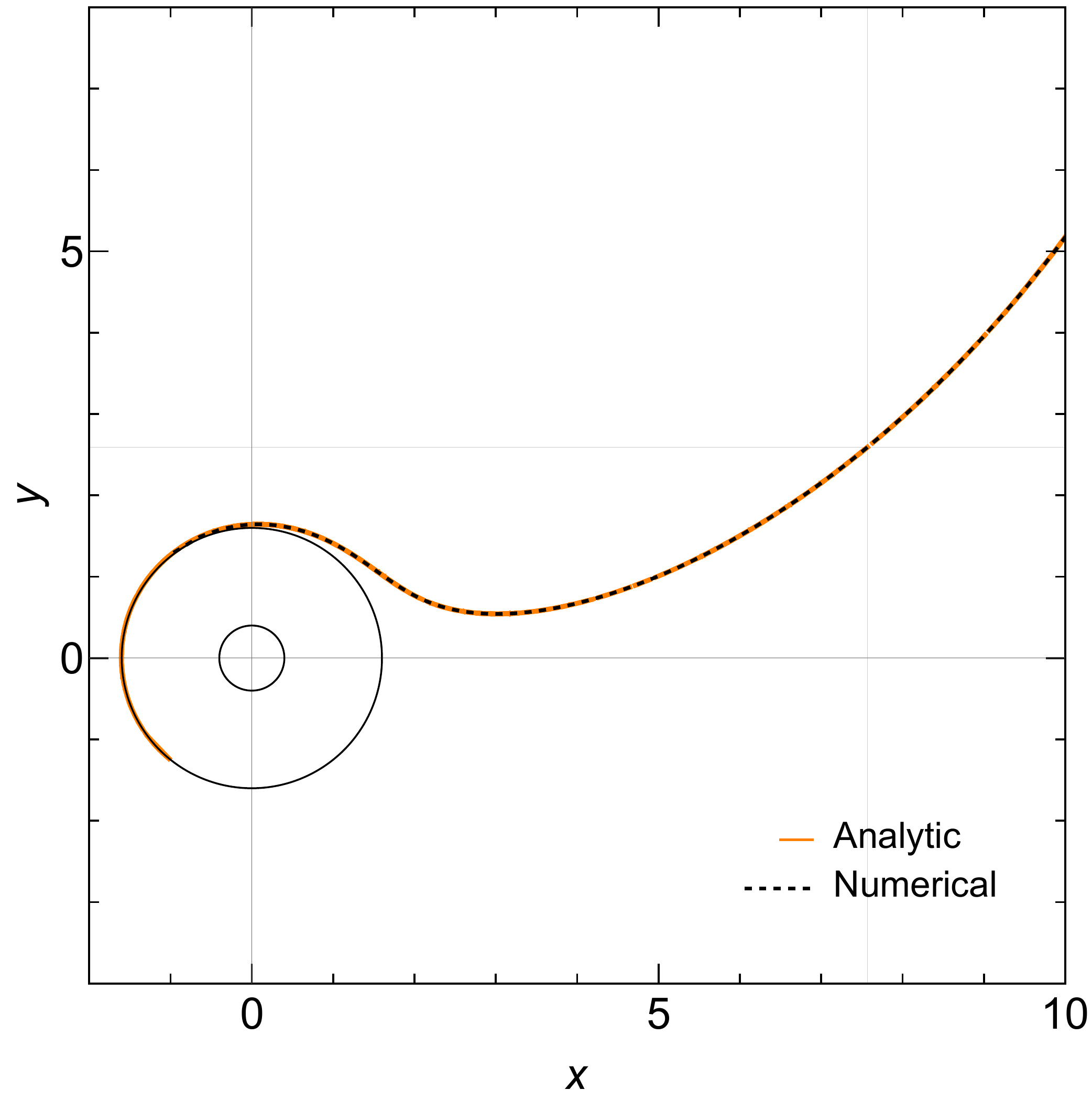}
\includegraphics[width=0.4\linewidth]{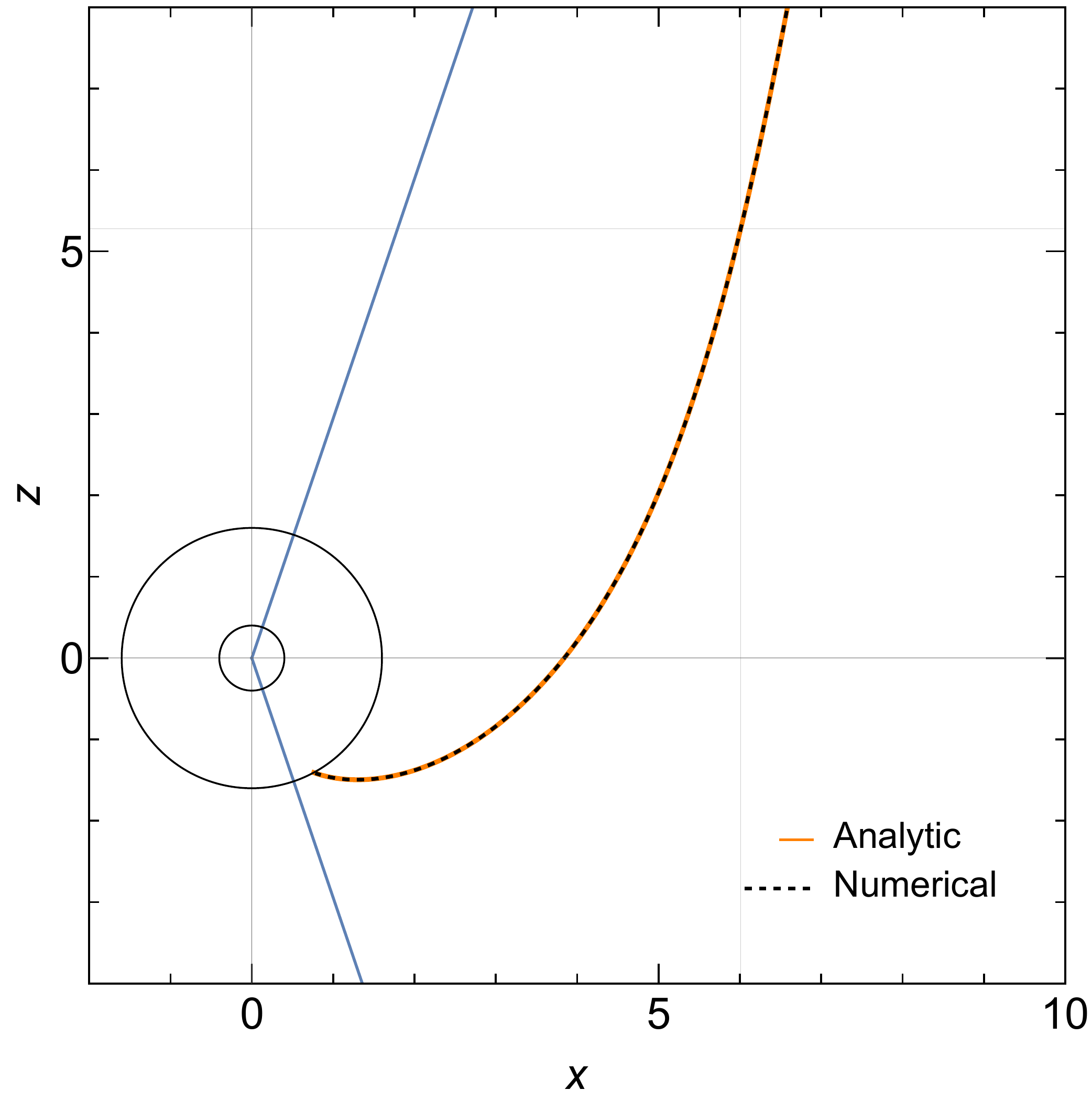}
\includegraphics[width=0.4\linewidth]{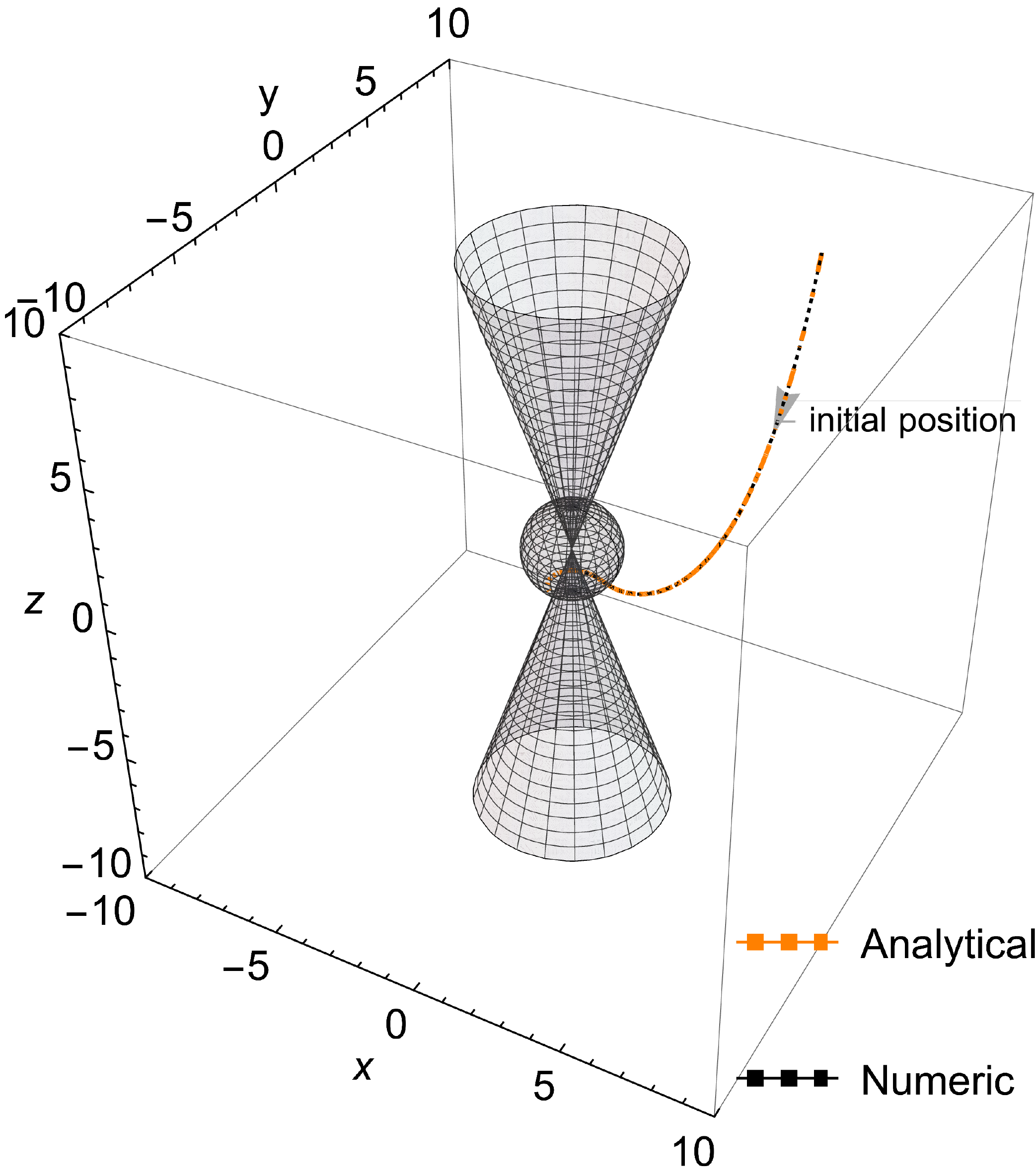}
\includegraphics[width=0.4\linewidth]{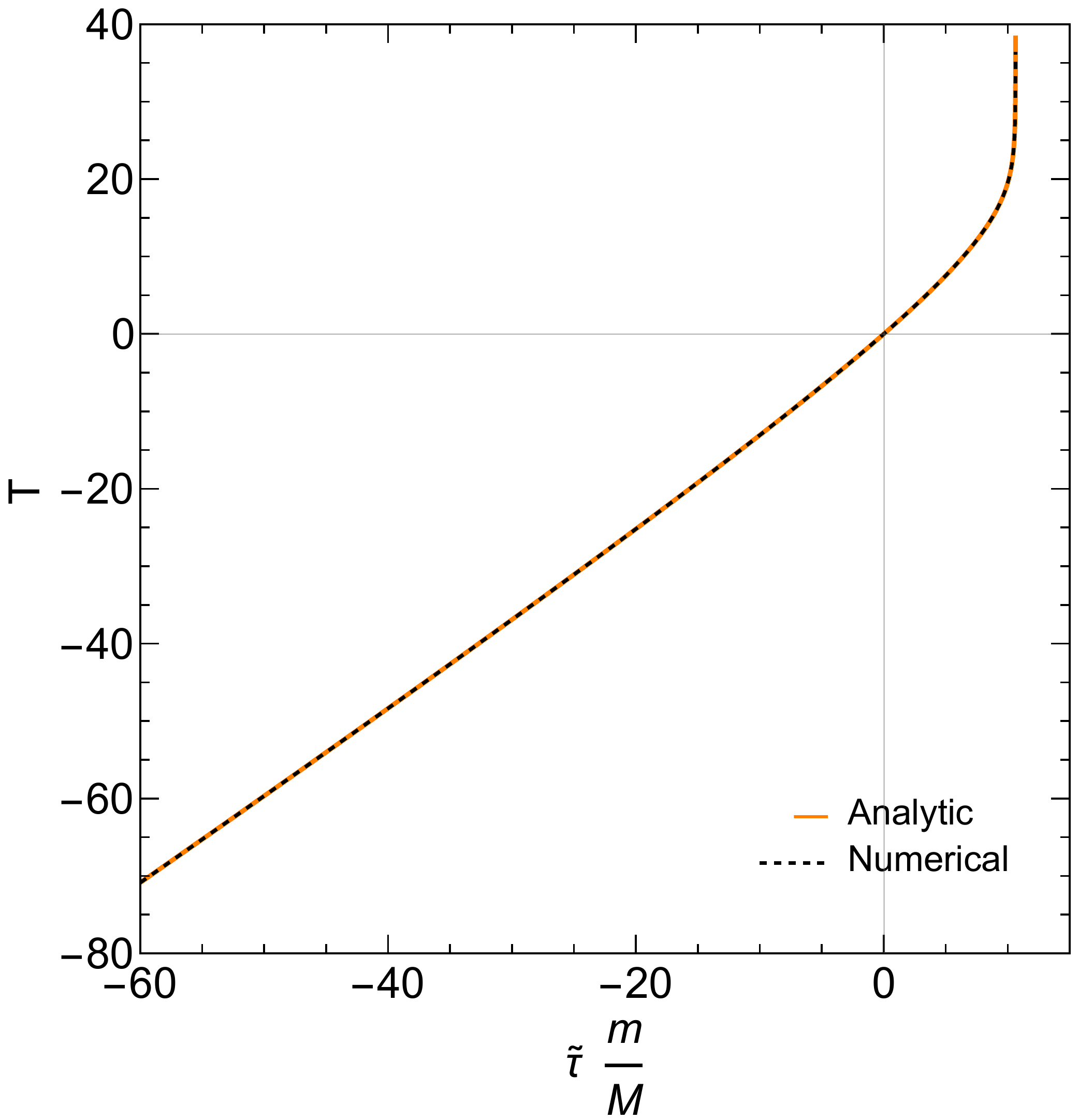}
\caption{\label{figA} An example of a timelike  unbound absorbed orbit with $\varepsilon^2 = 1.1$, $\lambda_z = -1$, $\alpha=0.8$, $\kappa=12$ and initial position at $\xi_0 = 8$, $\theta_0 = 0.85$, $\varphi_0 = 0.33$, $\epsilon_r = -1$, $\epsilon_{\theta} = 1$. Solid color lines correspond to solutions obtained with Eqs.\ \eqref{xi_s}, \eqref{mu_s}, \eqref{varphi_int2}. Dotted lines depict corresponding numerical solutions. The motion is depicted in two planes: $xy$ on the right and  $xz$ on the left. The intersection of thin black lines marks the initial position. Blue lines in the $xz$ plane and cones in the three-dimensional plot correspond to extremal values of $\theta$. Gray spheres and dark circles depict black hole horizons. The figure in the lower right corner shows to the coordinate time $T$ versus the proper time $\tilde{\tau}$. }
\end{figure}

\begin{figure}[t]
\centering
\includegraphics[width=0.4\linewidth]{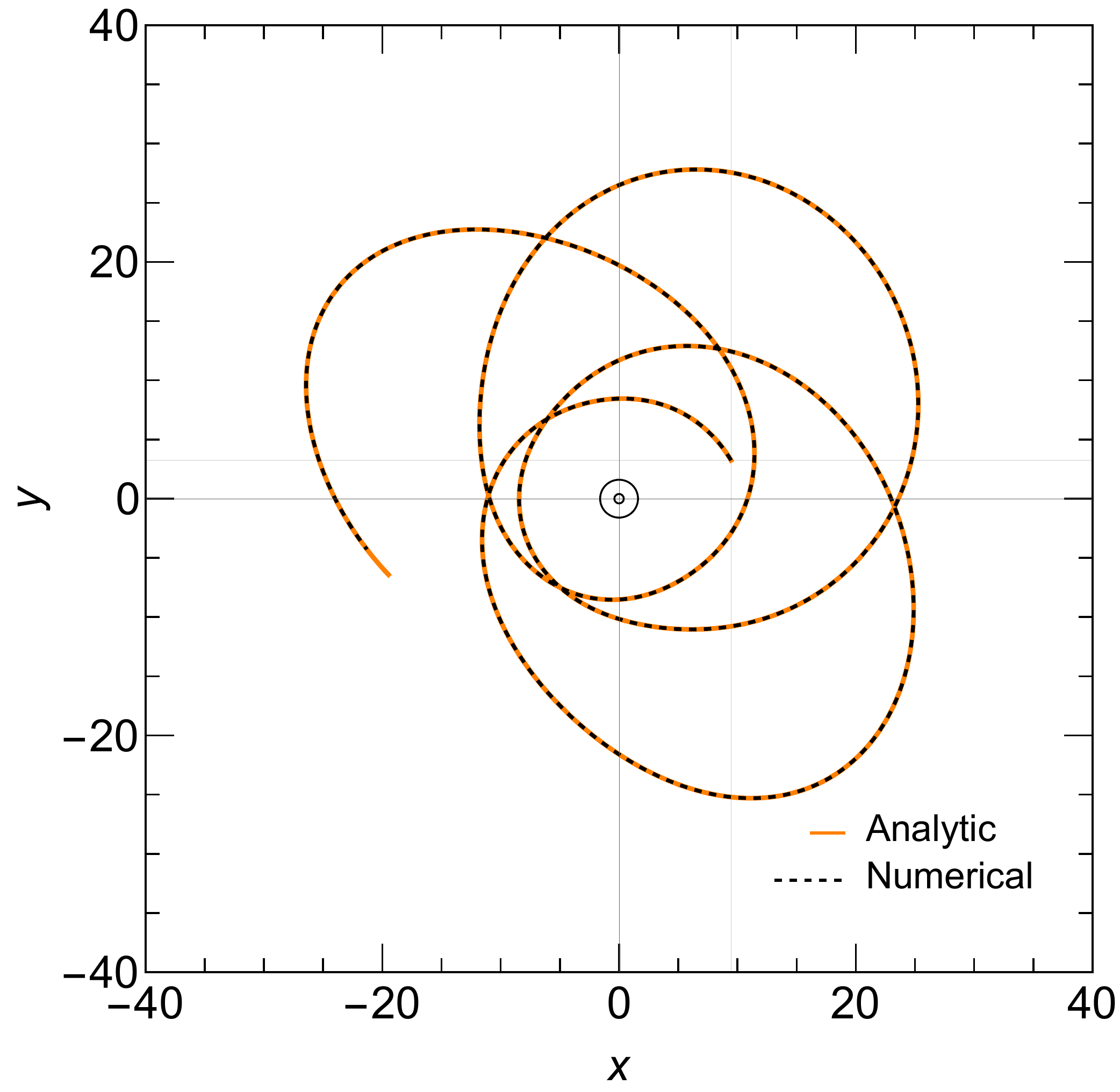}
\includegraphics[width=0.4\linewidth]{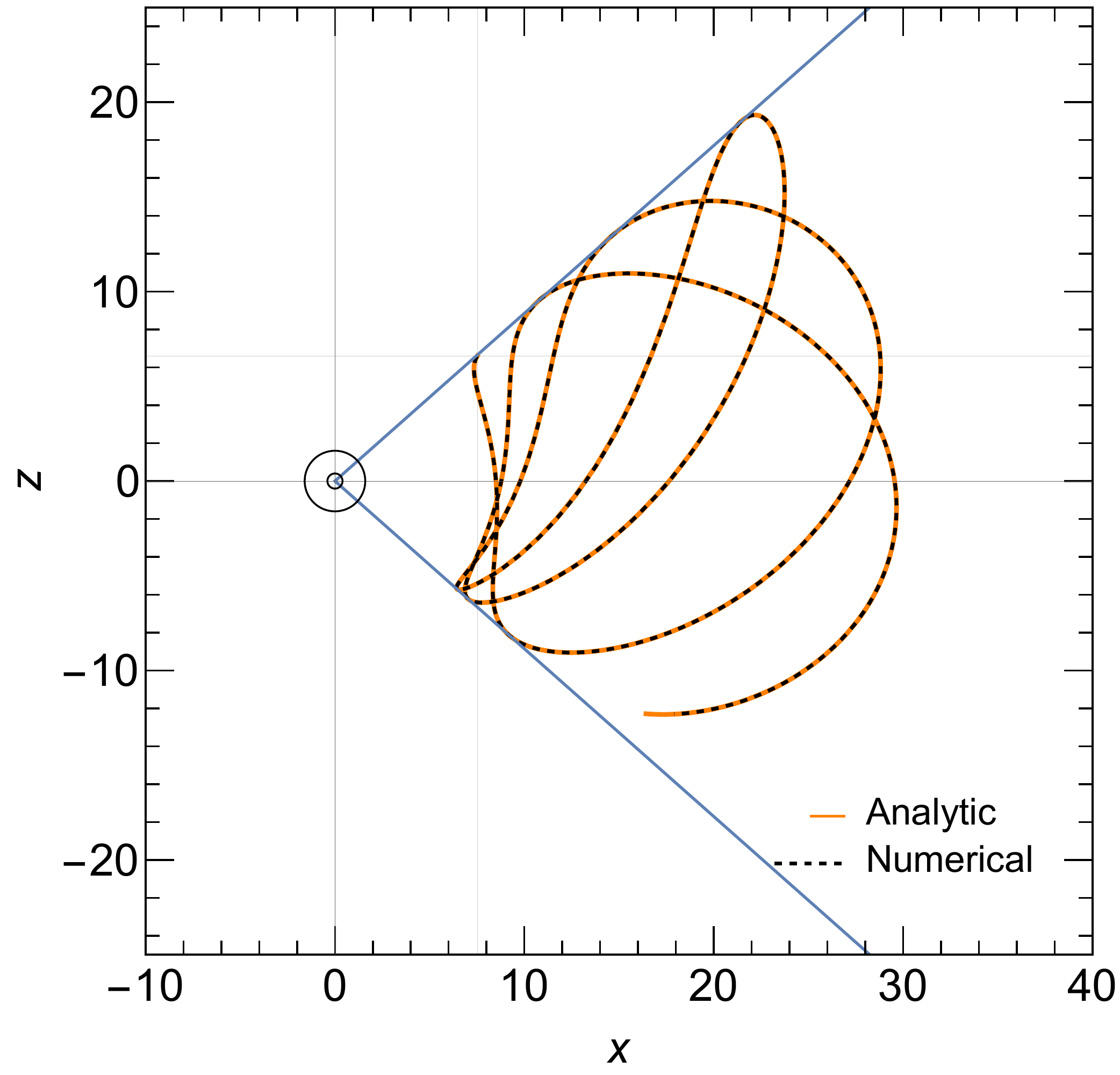}
\includegraphics[width=0.4\linewidth]{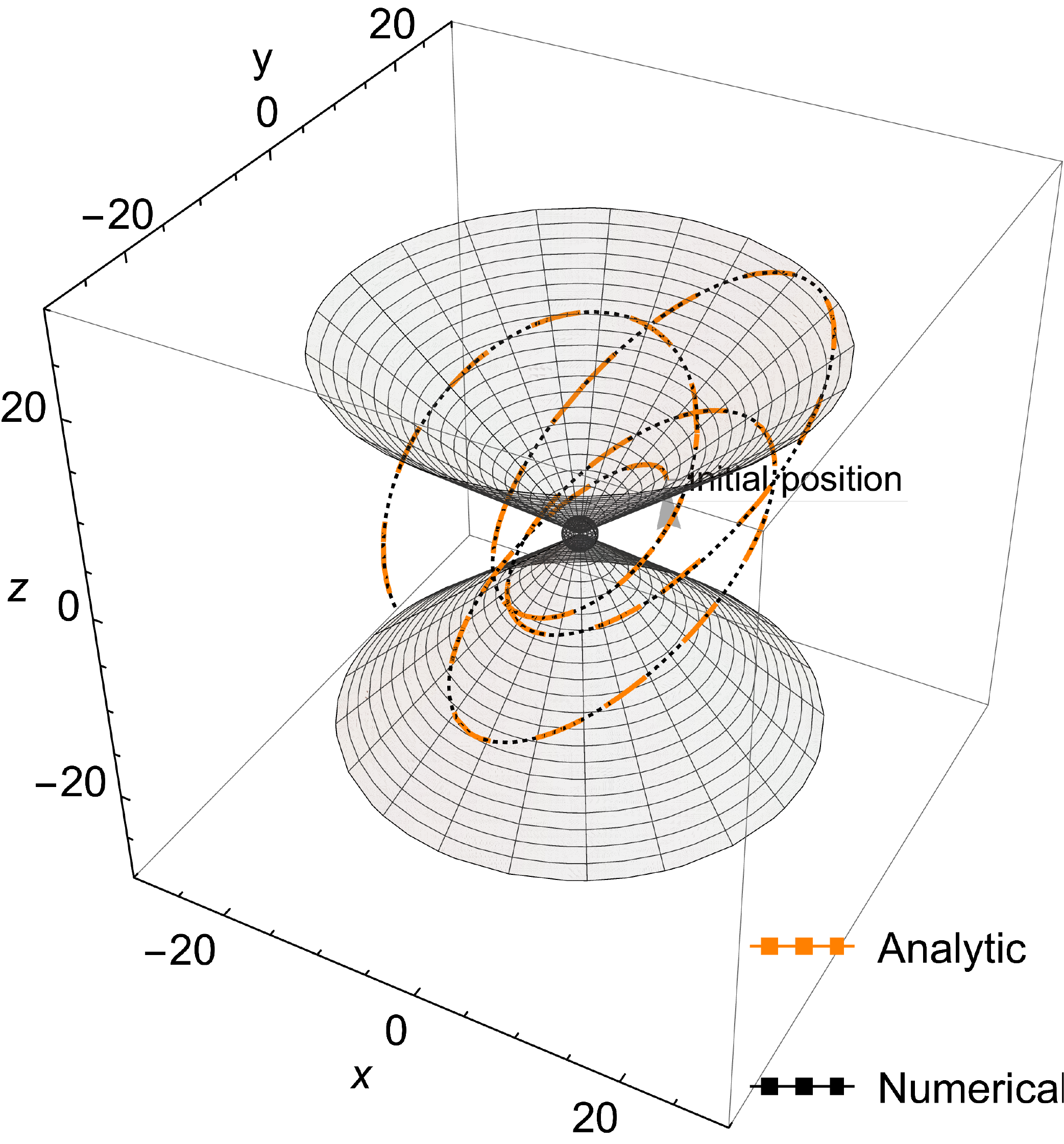}
\includegraphics[width=0.4\linewidth]{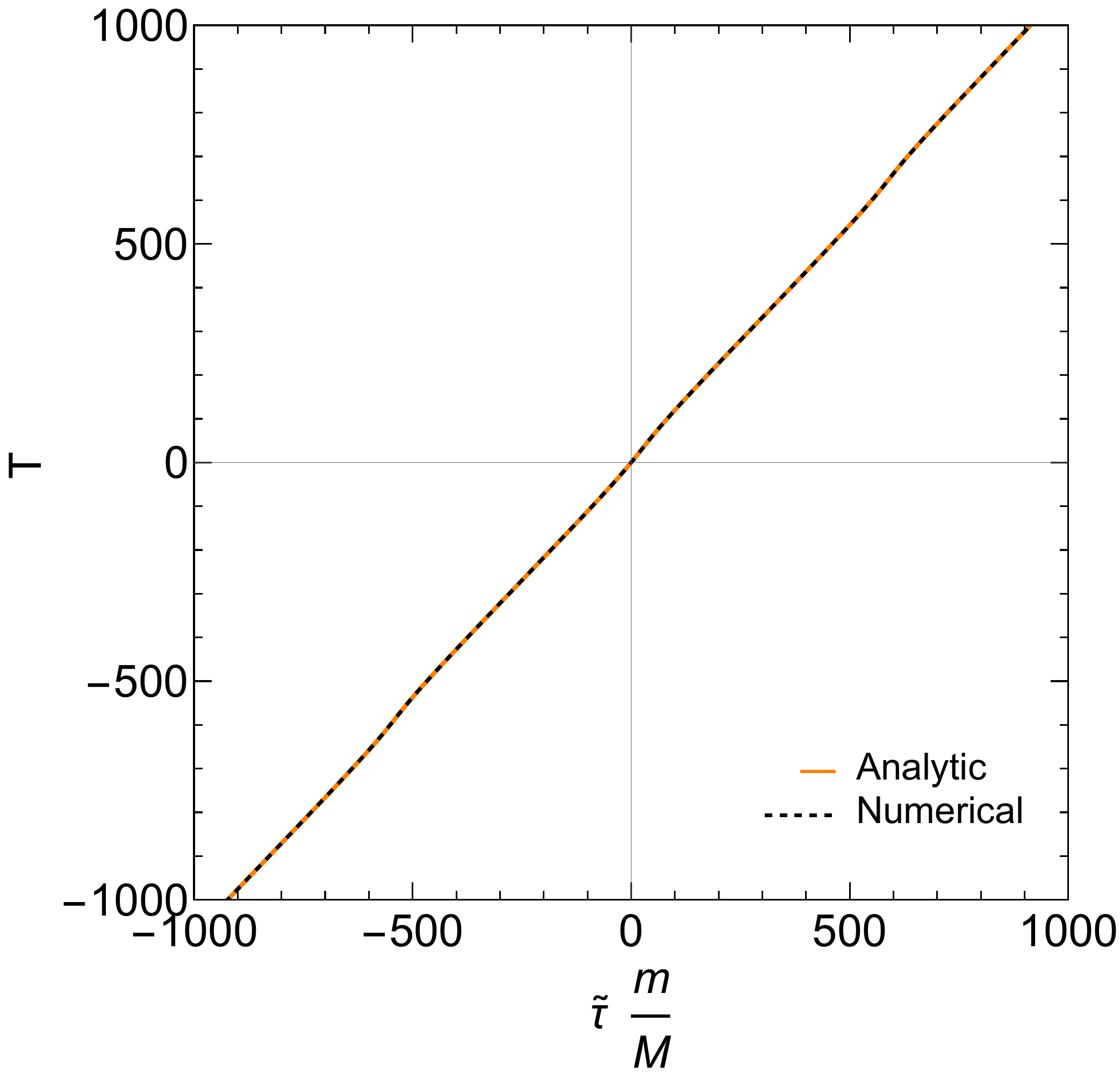}
\caption{\label{figB} Same as in Fig.\ \ref{figA}, but for a timelike bound orbit with $\varepsilon^2 = 0.95$, $\lambda_z = 3$, $\alpha=0.8$, $\kappa=12$, and an initial position at $\xi_0 = 10$, $\theta_0 = 0.85$, $\varphi_0 = 0.33$, $\epsilon_r = -1$, $\epsilon_{\theta} = 1$.}
\end{figure}

\begin{figure}[t]
\centering
\includegraphics[width=0.4\linewidth]{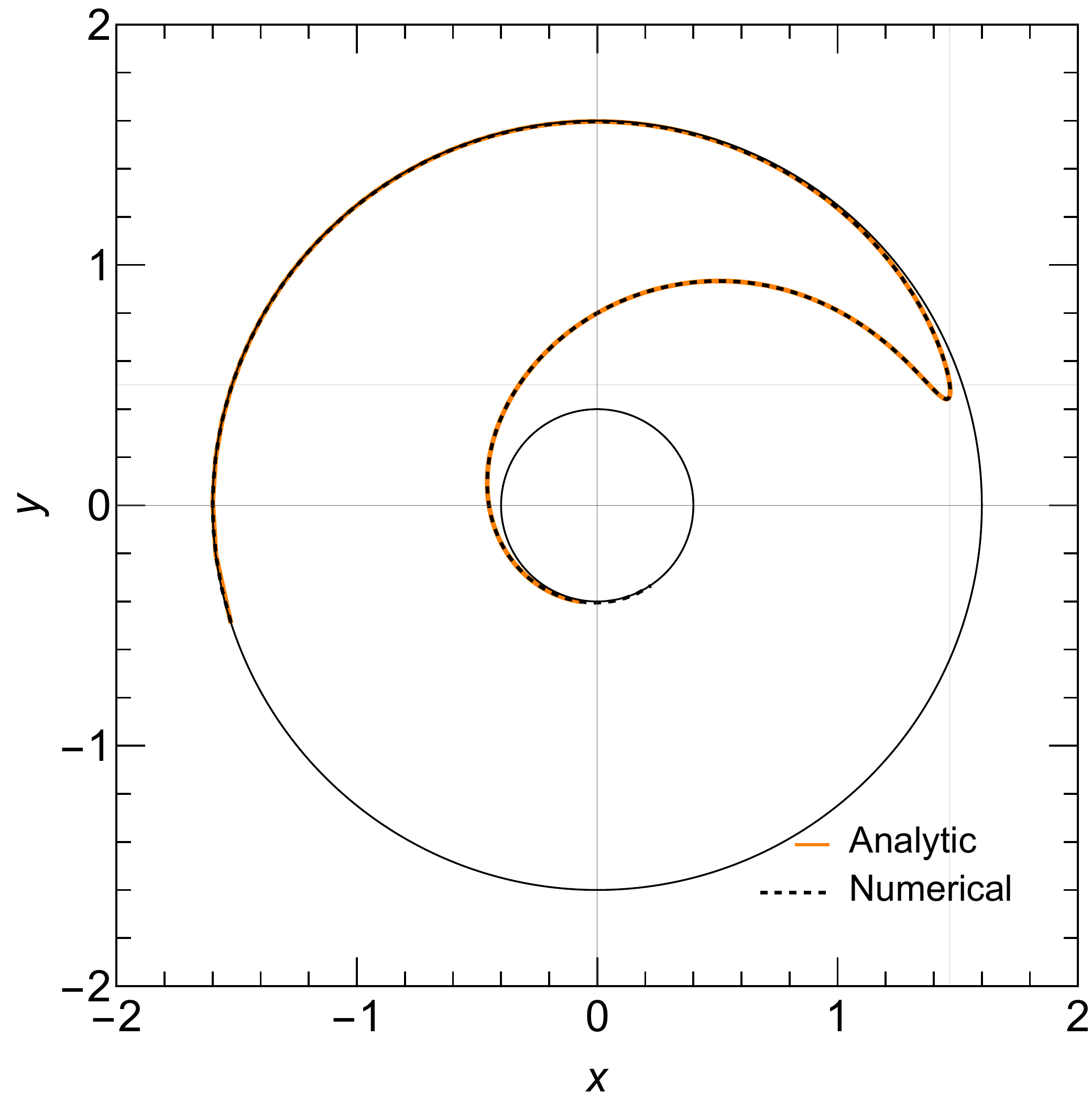}
\includegraphics[width=0.4\linewidth]{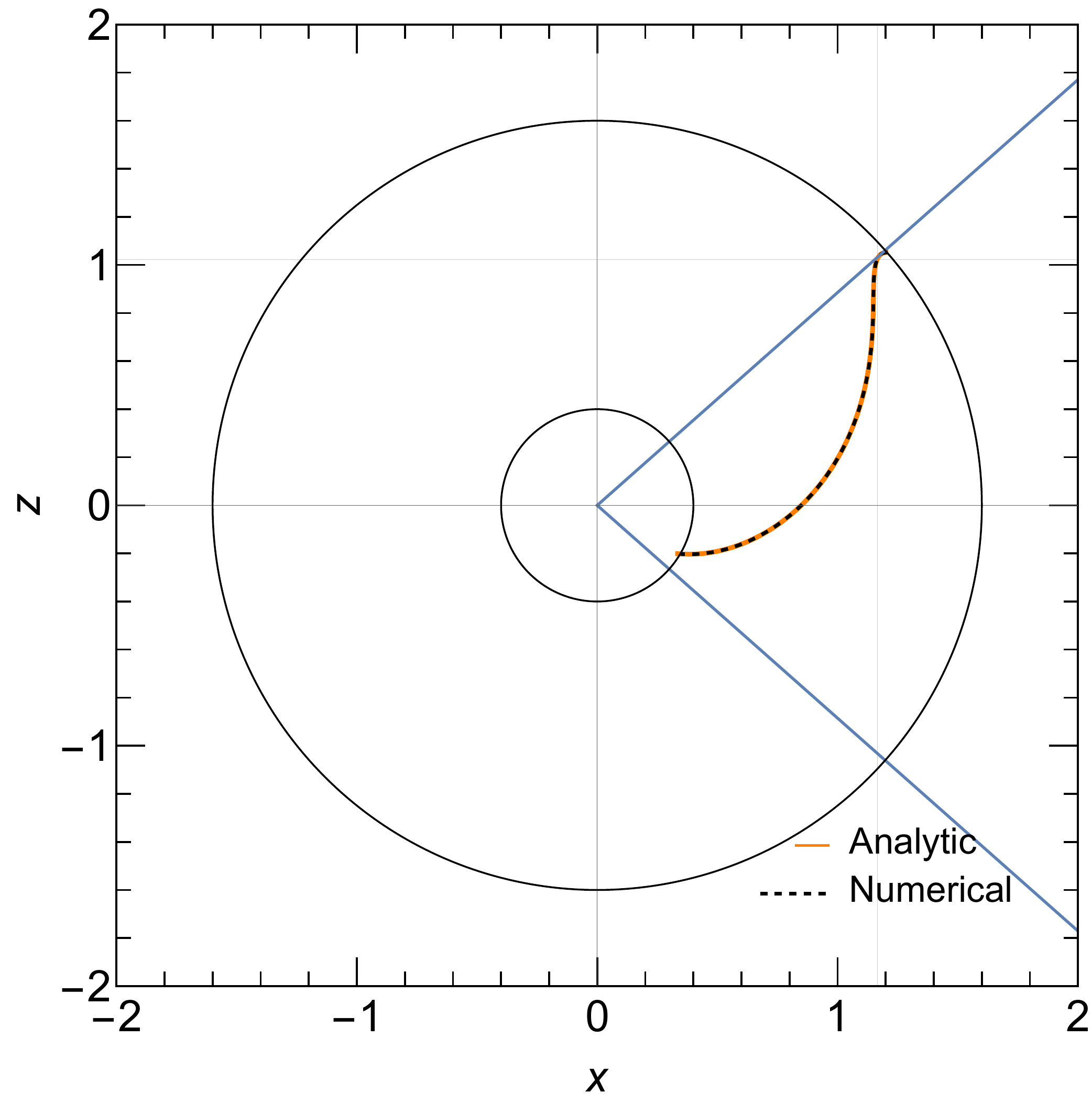}
\includegraphics[width=0.4\linewidth]{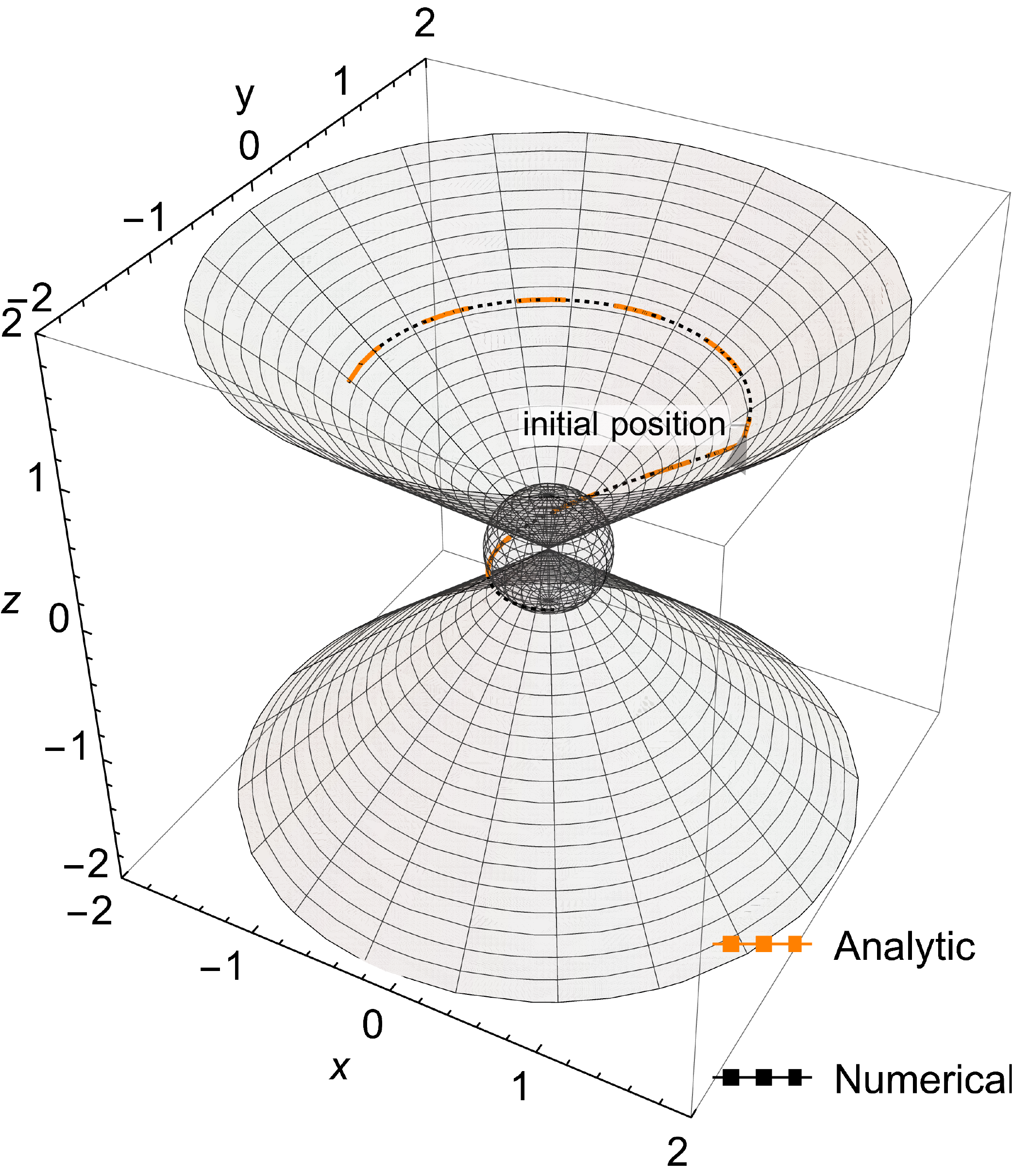}
\includegraphics[width=0.4\linewidth]{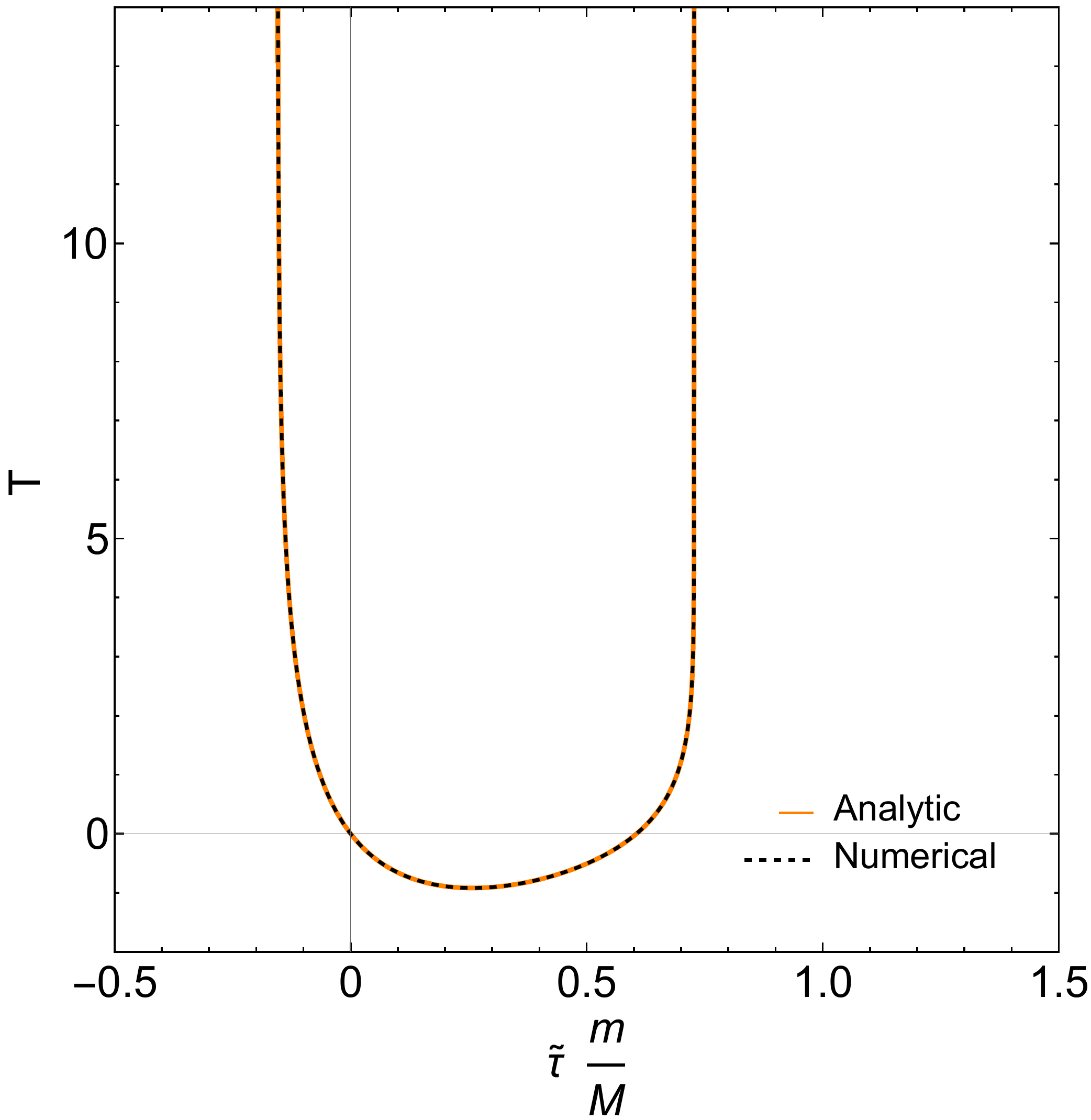}
\caption{\label{figC} Same as in Fig.\ \ref{figA}, but for an inner bound orbit with $\varepsilon^2 = 0.95$, $\lambda_z = 3$, $\alpha=0.8$, $\kappa=12$, and an initial position at $\xi_0 = 1.55$, $\theta_0 = 0.85$, $\varphi_0 = 0.33$, $\epsilon_r = -1$, $\epsilon_{\theta} = 1$.}
\end{figure}

\begin{figure}[t]
\centering
\includegraphics[width=0.4\linewidth]{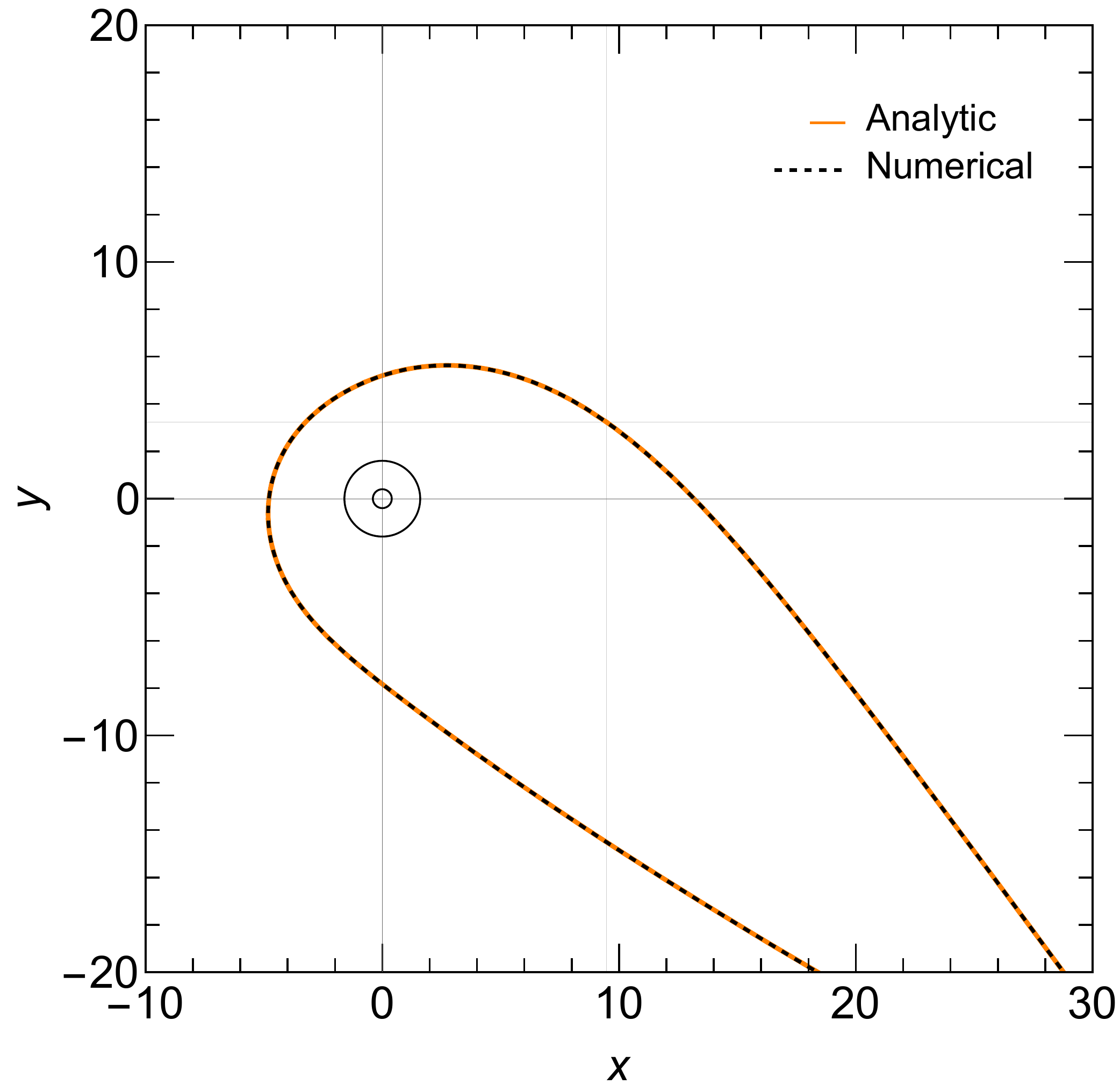}
\includegraphics[width=0.4\linewidth]{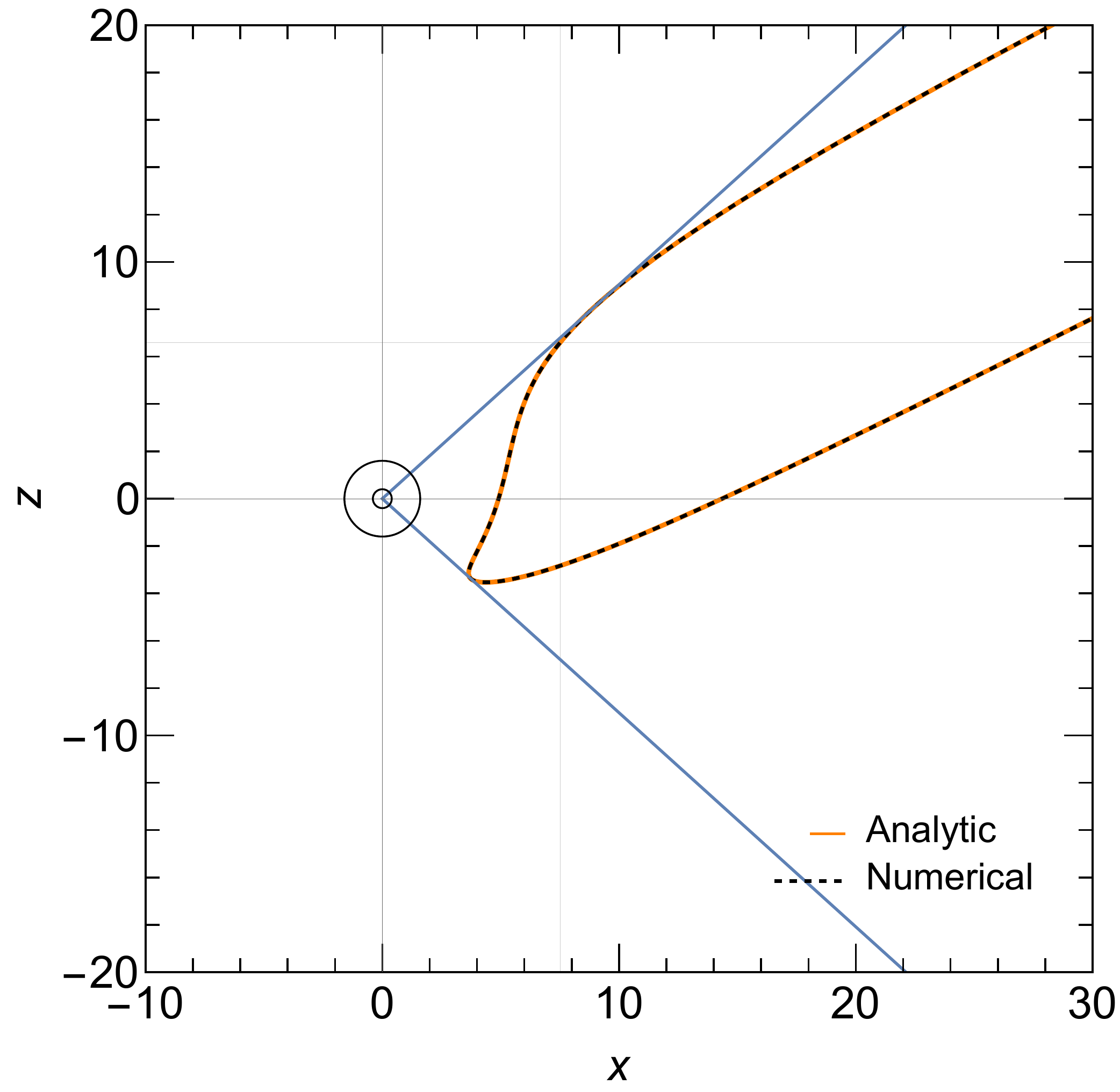}
\includegraphics[width=0.4\linewidth]{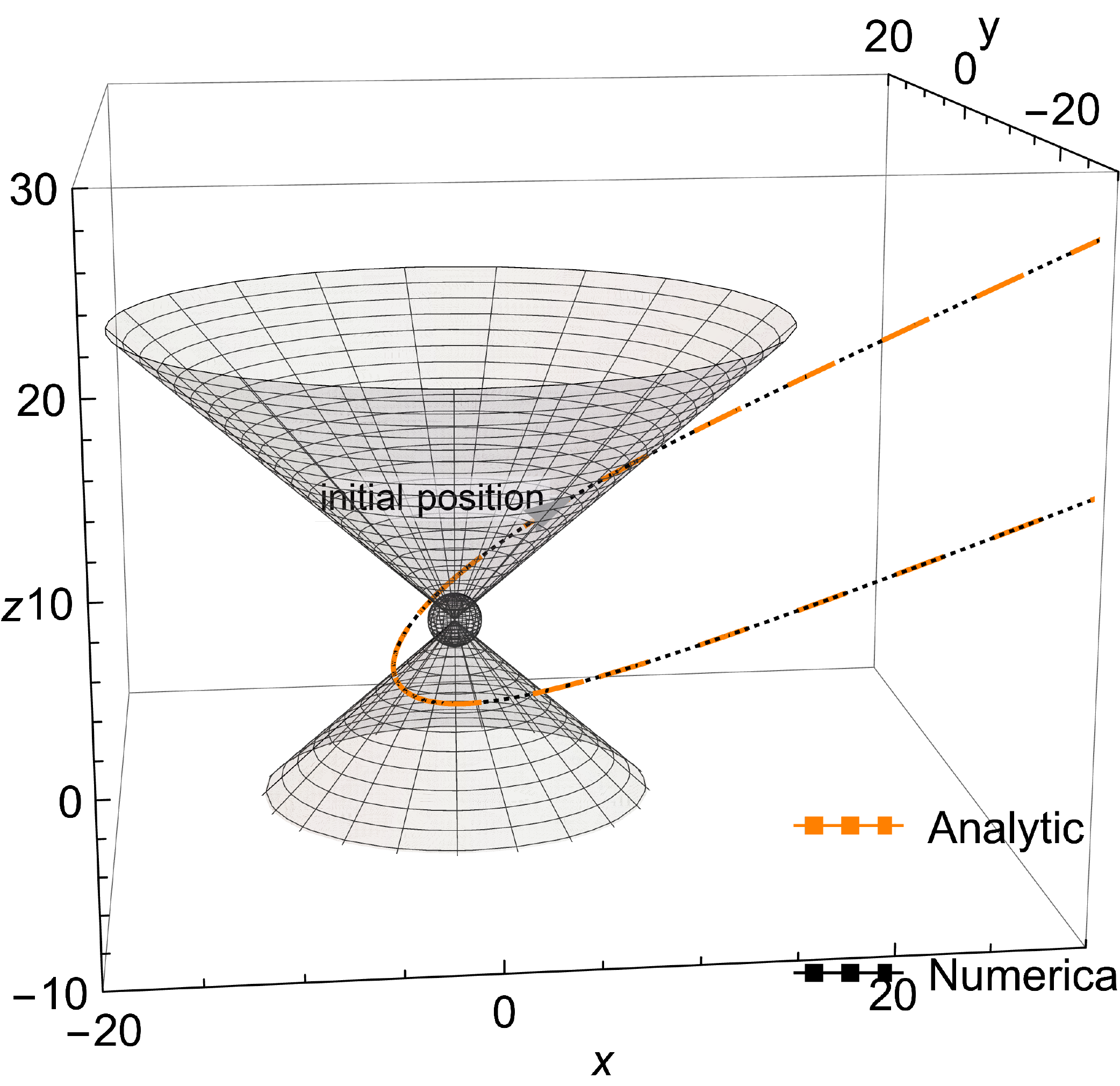}
\includegraphics[width=0.4\linewidth]{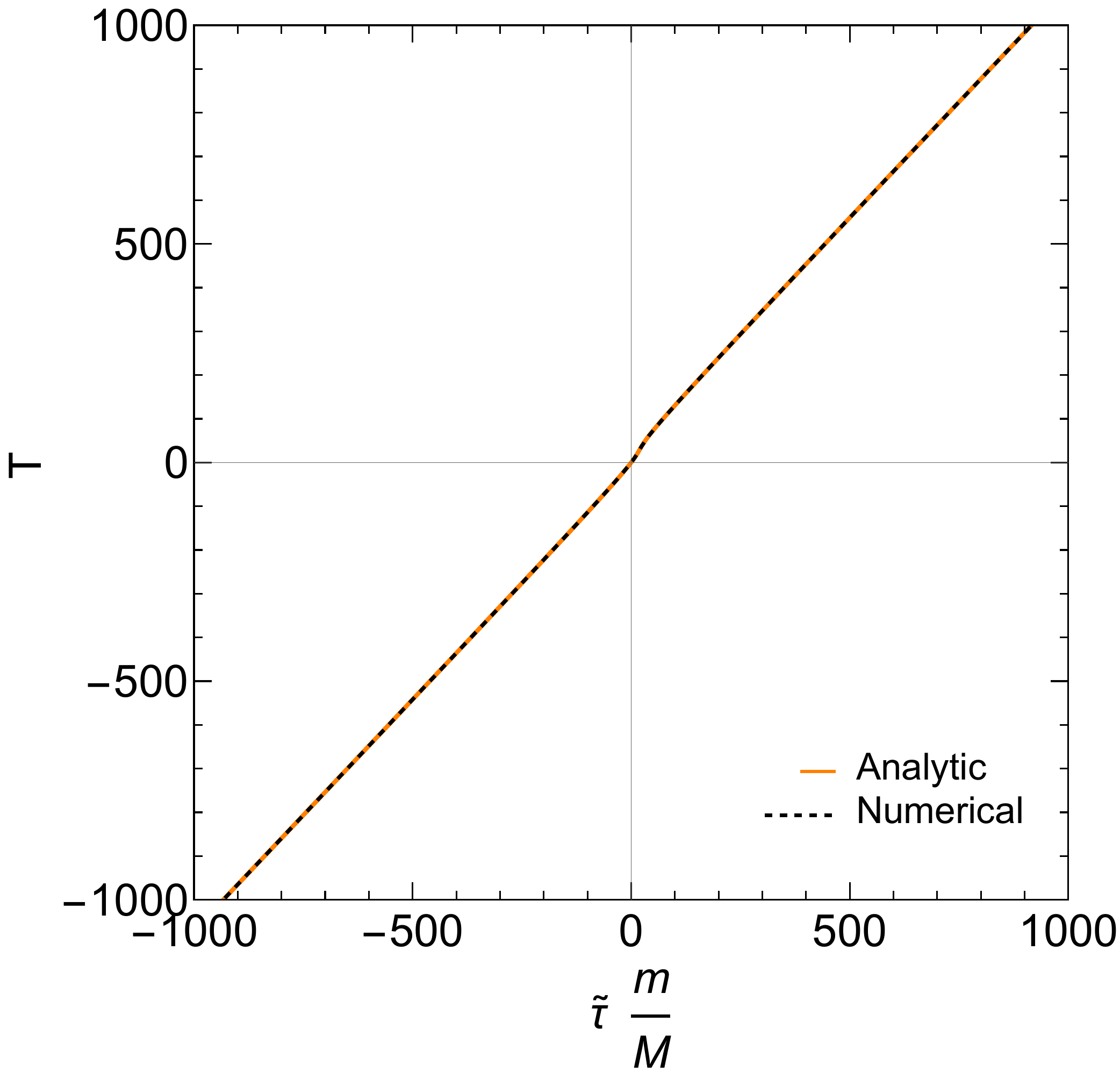}
\caption{\label{figD} Same as in Fig.\ \ref{figA}, but for a timelike unbound scattered orbit with $\varepsilon^2 = 1.1$, $\lambda_z = 3$, $\alpha=0.8$, $\kappa=12$, and an initial position at $\xi_0 = 10$, $\theta_0 = 0.85$, $\varphi_0 = 0.33$, $\epsilon_r = -1$, $\epsilon_{\theta} = 1$.}
\end{figure}

\begin{figure}[t]
\centering
\includegraphics[width=0.4\linewidth]{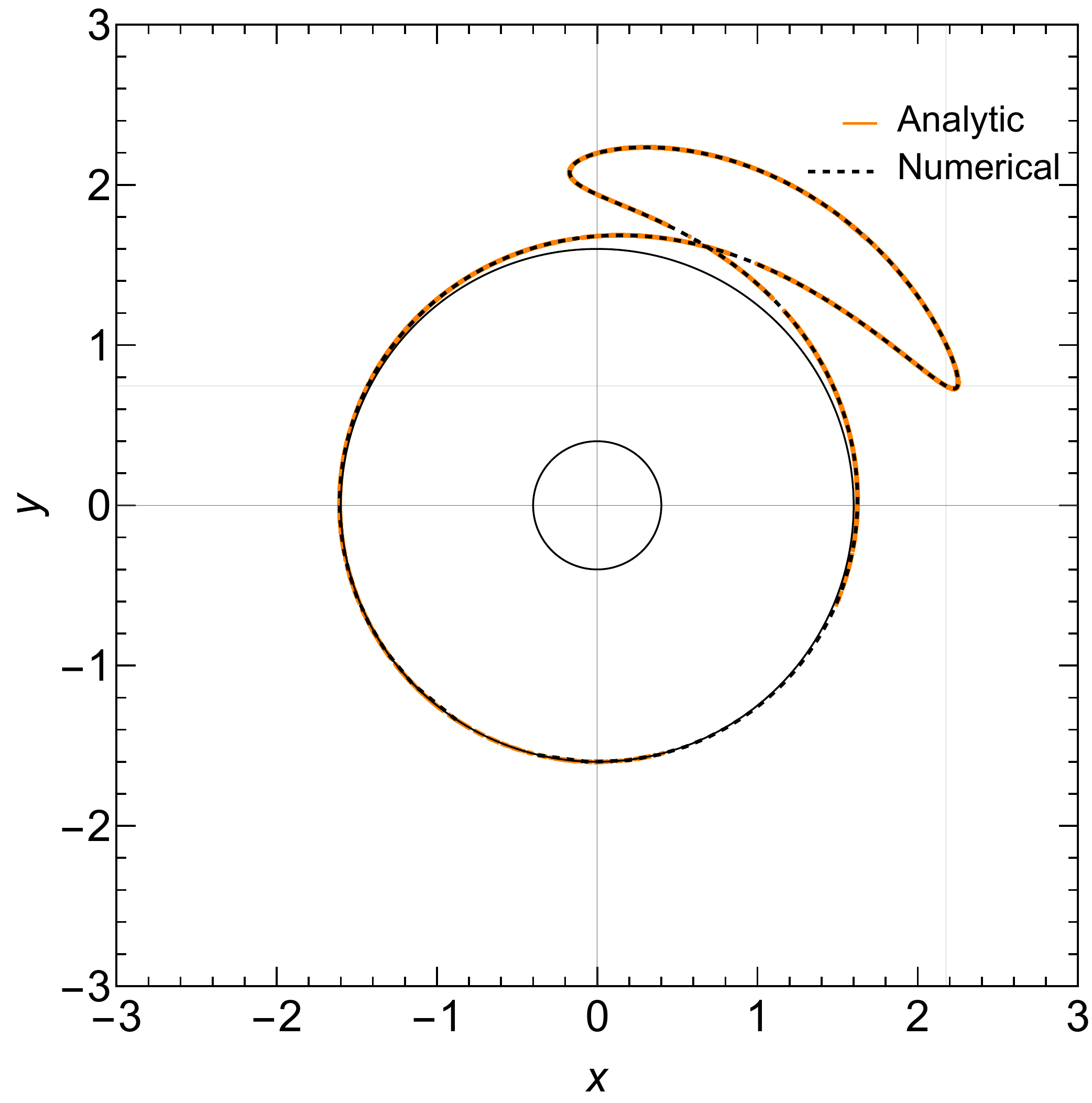}
\includegraphics[width=0.4\linewidth]{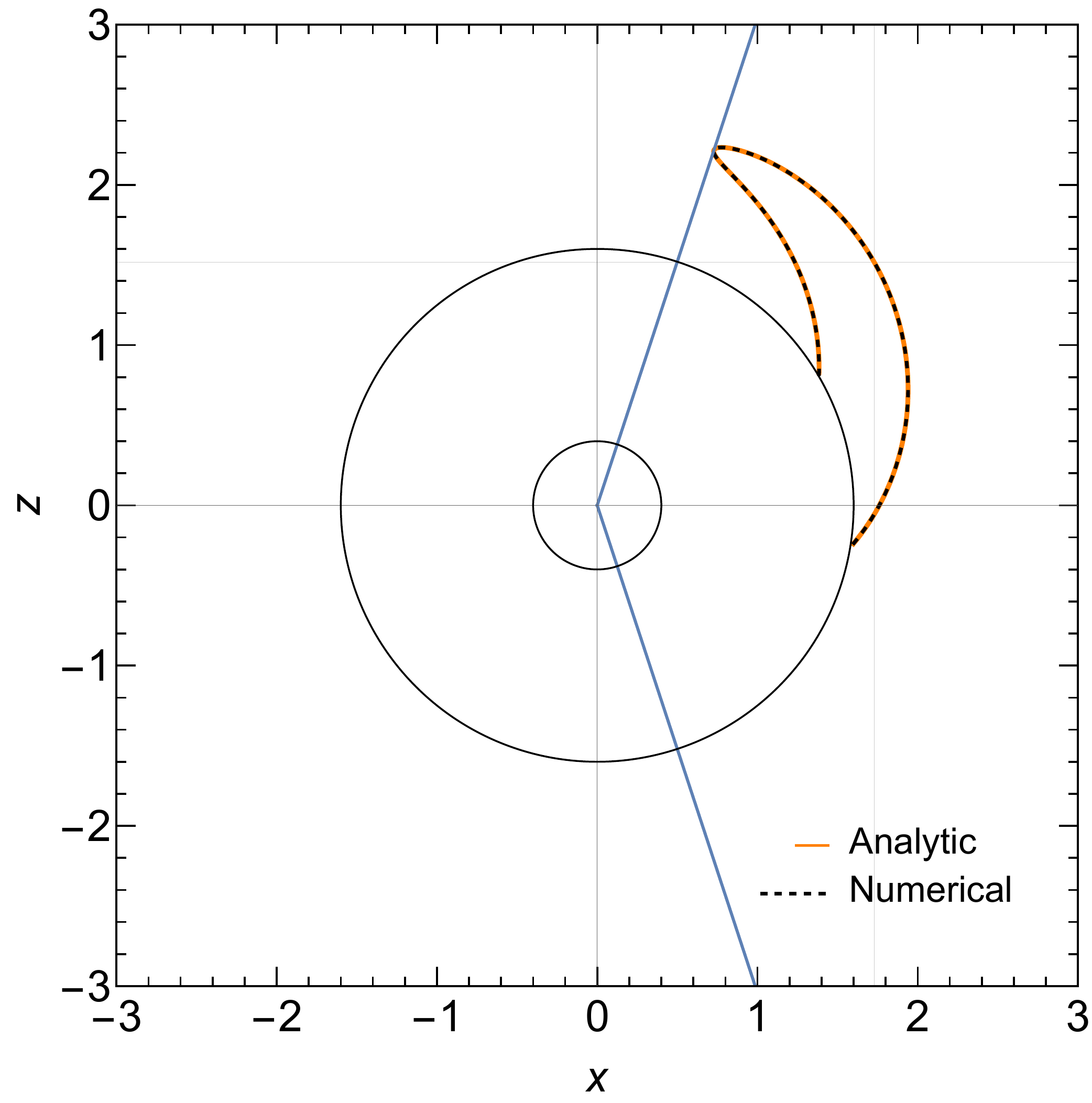}
\includegraphics[width=0.4\linewidth]{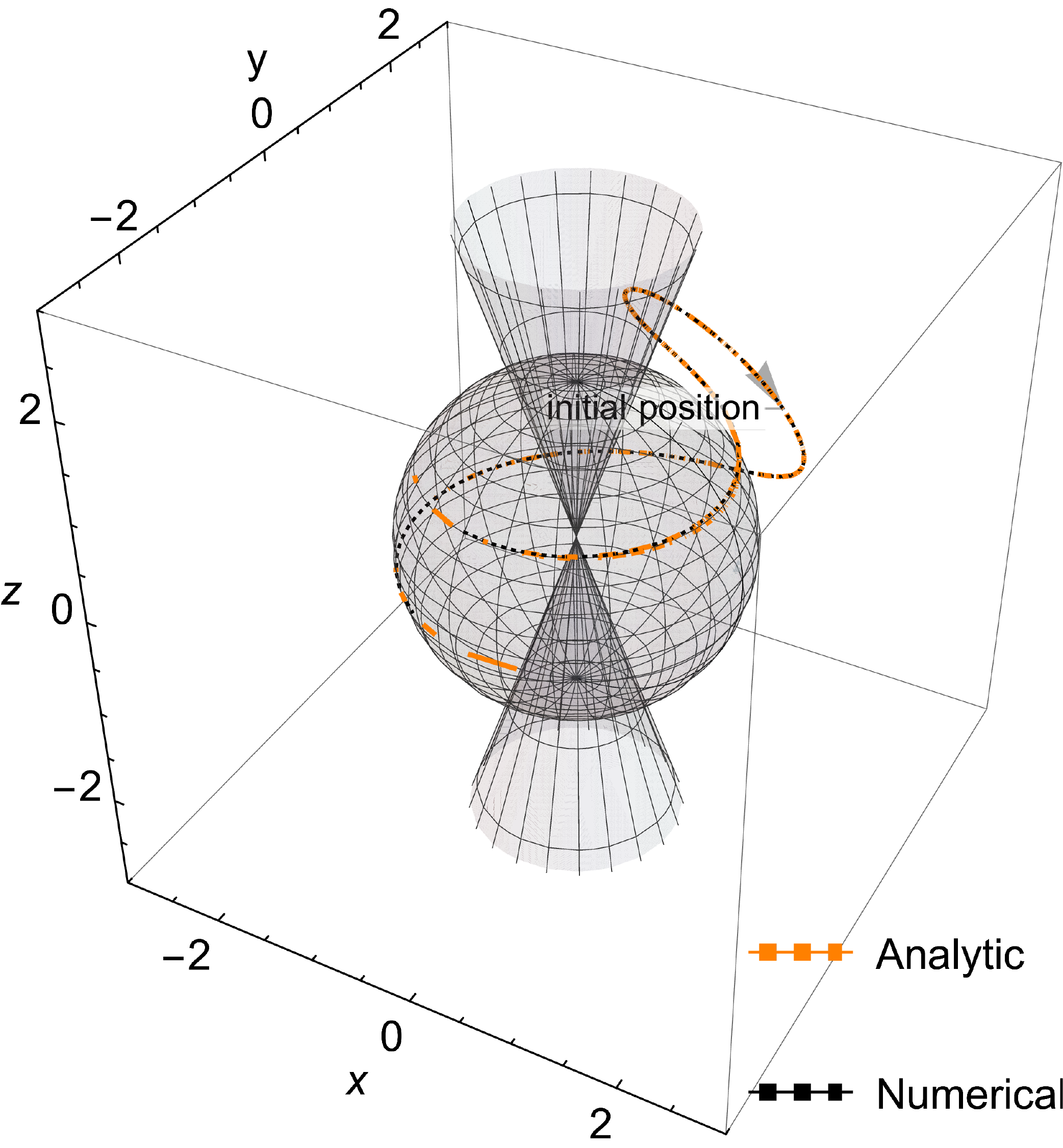}
\includegraphics[width=0.4\linewidth]{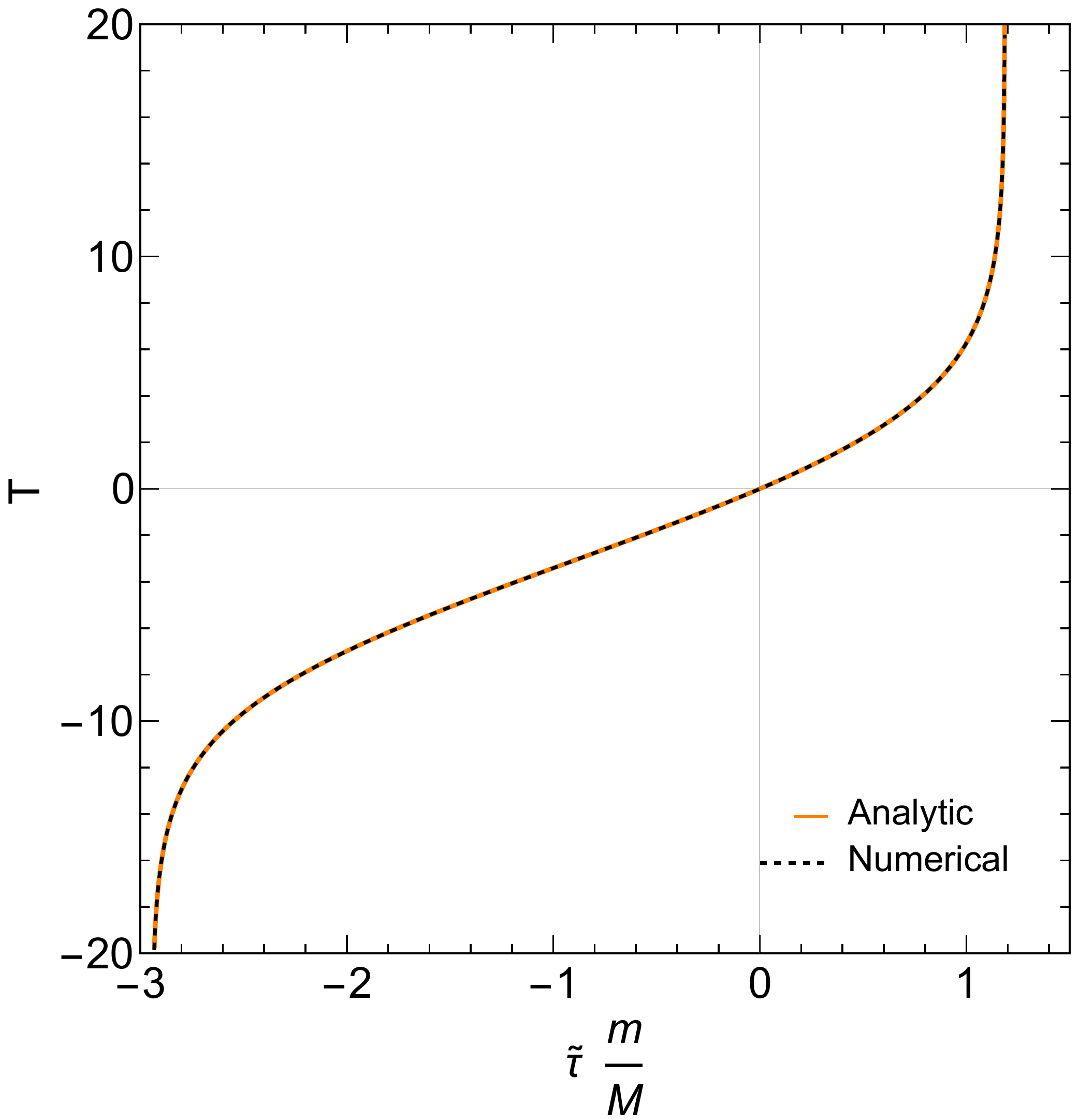}
\caption{\label{figE} Same as in Fig.\ \ref{figA}, but for a timelike bound orbits with $\varepsilon^2 = 0.5$, $\lambda_z = -1$, $\alpha=0.8$, $\kappa=12$, and an initial position at $\xi_0 = 2.3$, $\theta_0 = 0.85$, $\varphi_0 = 0.33$, $\epsilon_r = -1$, $\epsilon_{\theta} = 1$.}
\end{figure}

\begin{figure}[t]
\centering
\includegraphics[width=0.4\linewidth]{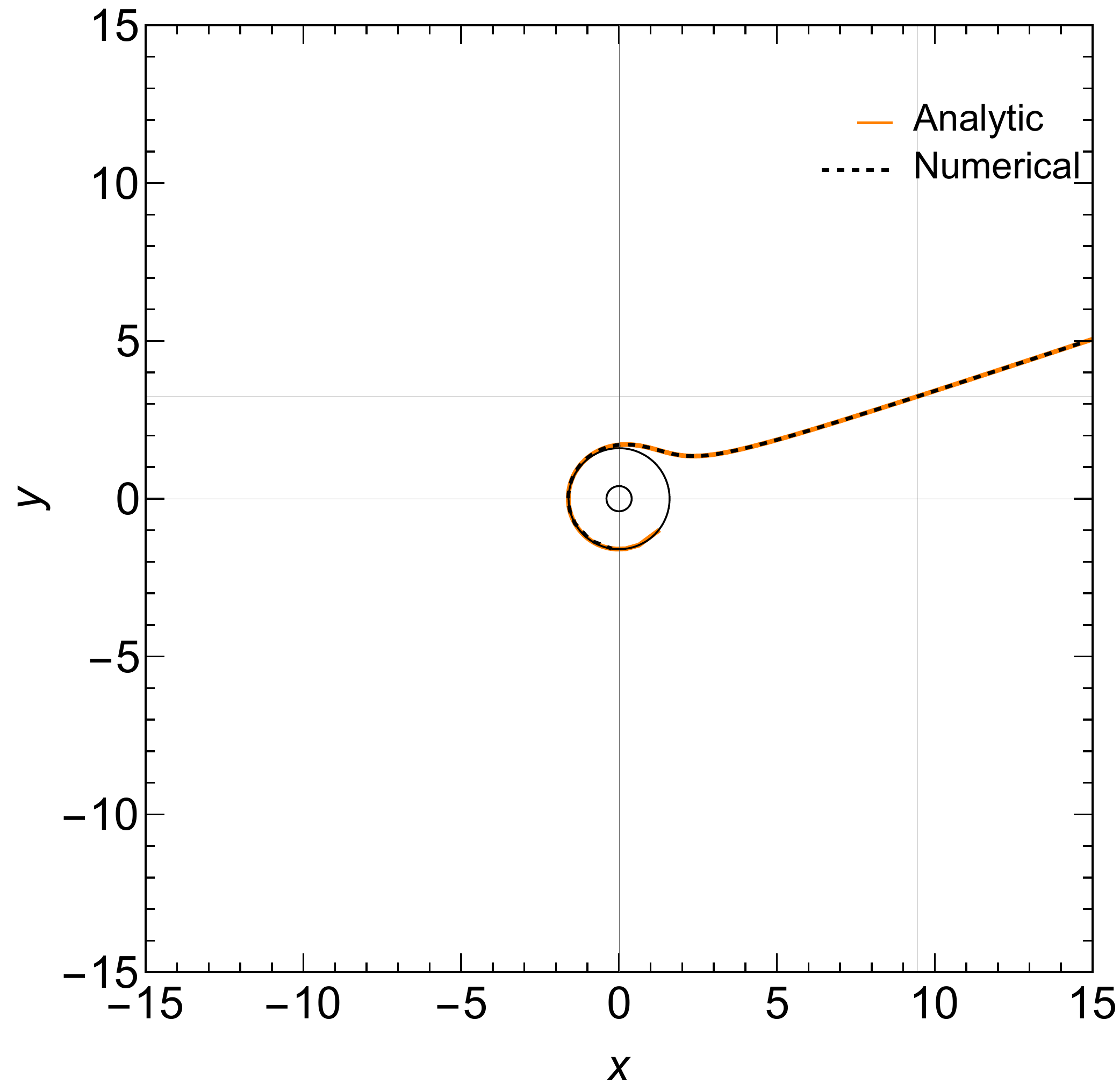}
\includegraphics[width=0.4\linewidth]{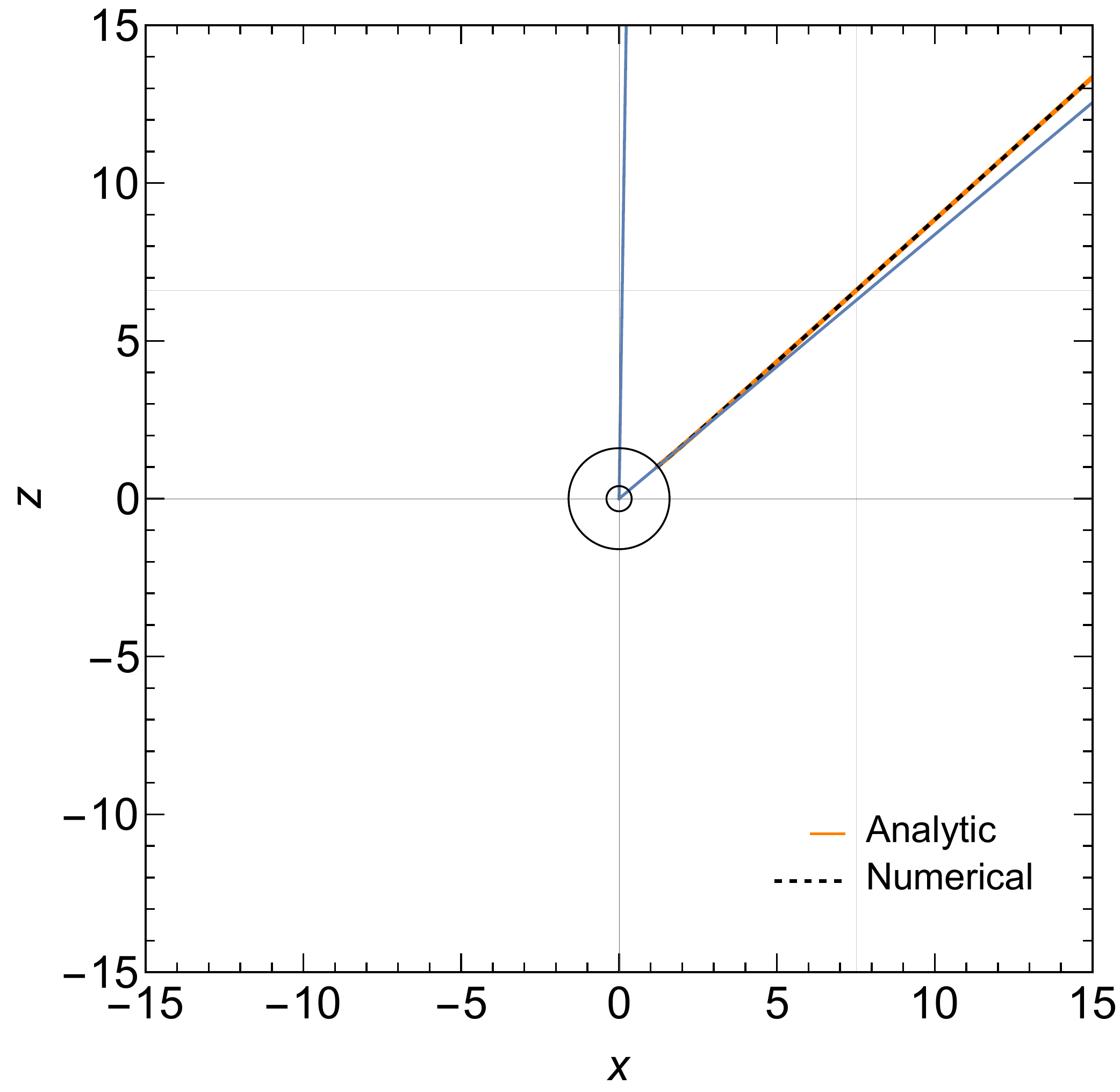}
\includegraphics[width=0.4\linewidth]{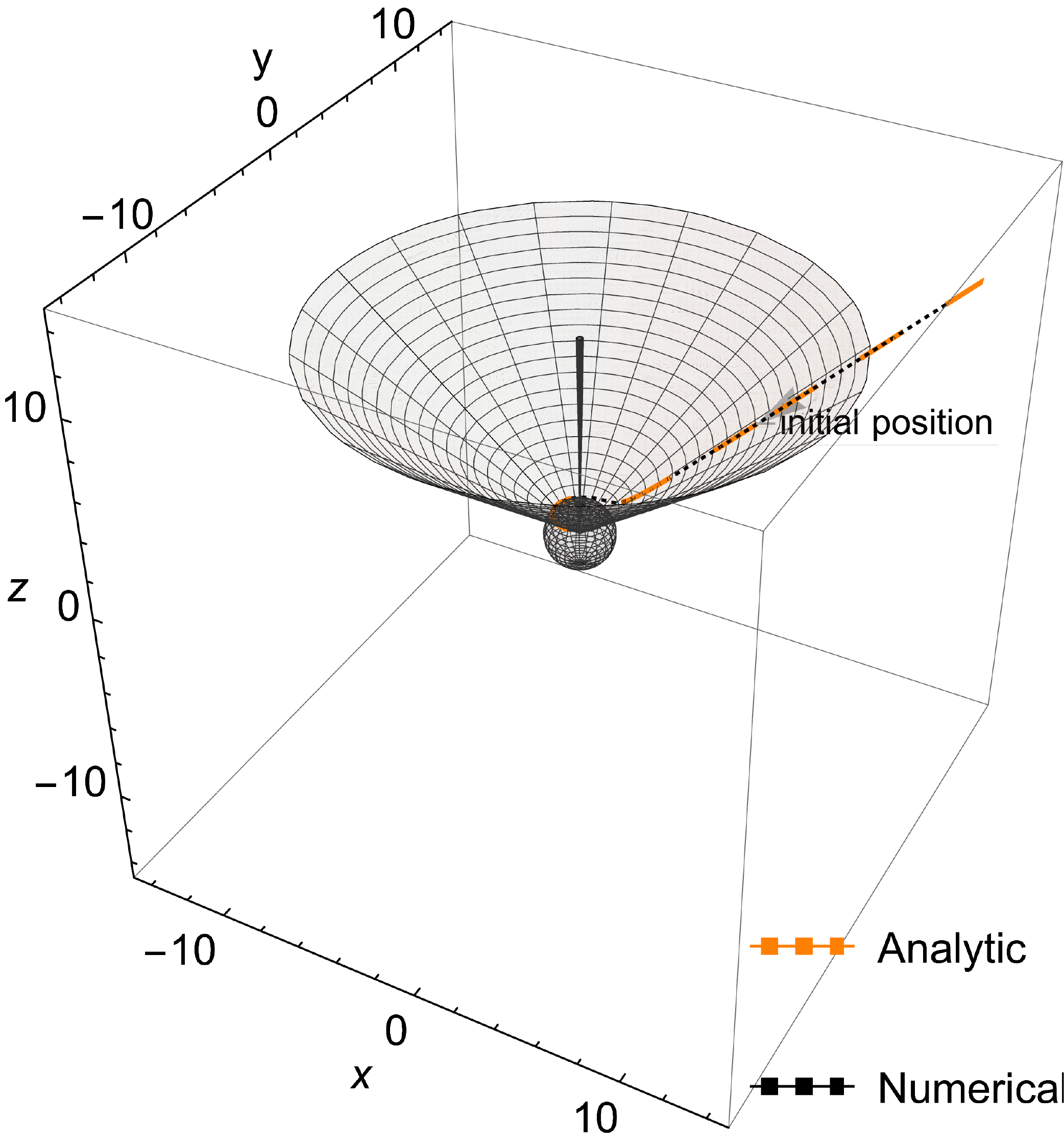}
\includegraphics[width=0.4\linewidth]{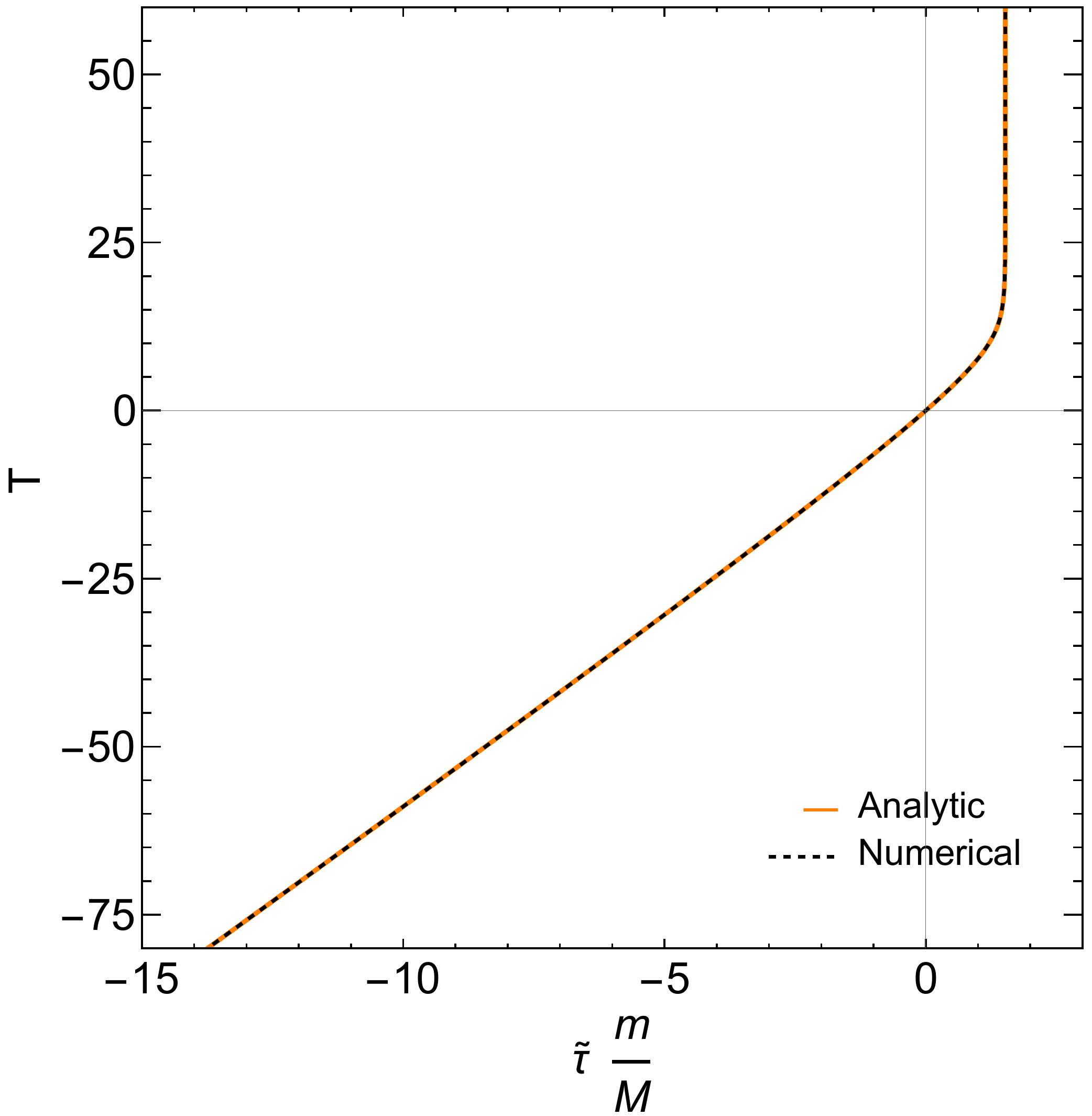}
\caption{\label{figF} Same as in Fig.\ \ref{figA}, but for a timelike unbound absorbed orbit with $\varepsilon^2 = 30$, $\lambda_z = -0.05$, $\alpha=0.8$, $\kappa=12$, and an initial position at $\xi_0 = 10$, $\theta_0 = 0.85$, $\varphi_0 = 0.33$, $\epsilon_r = -1$, $\epsilon_{\theta} = 1$.}
\end{figure}

\begin{figure}[t]
\centering
\includegraphics[width=0.4\linewidth]{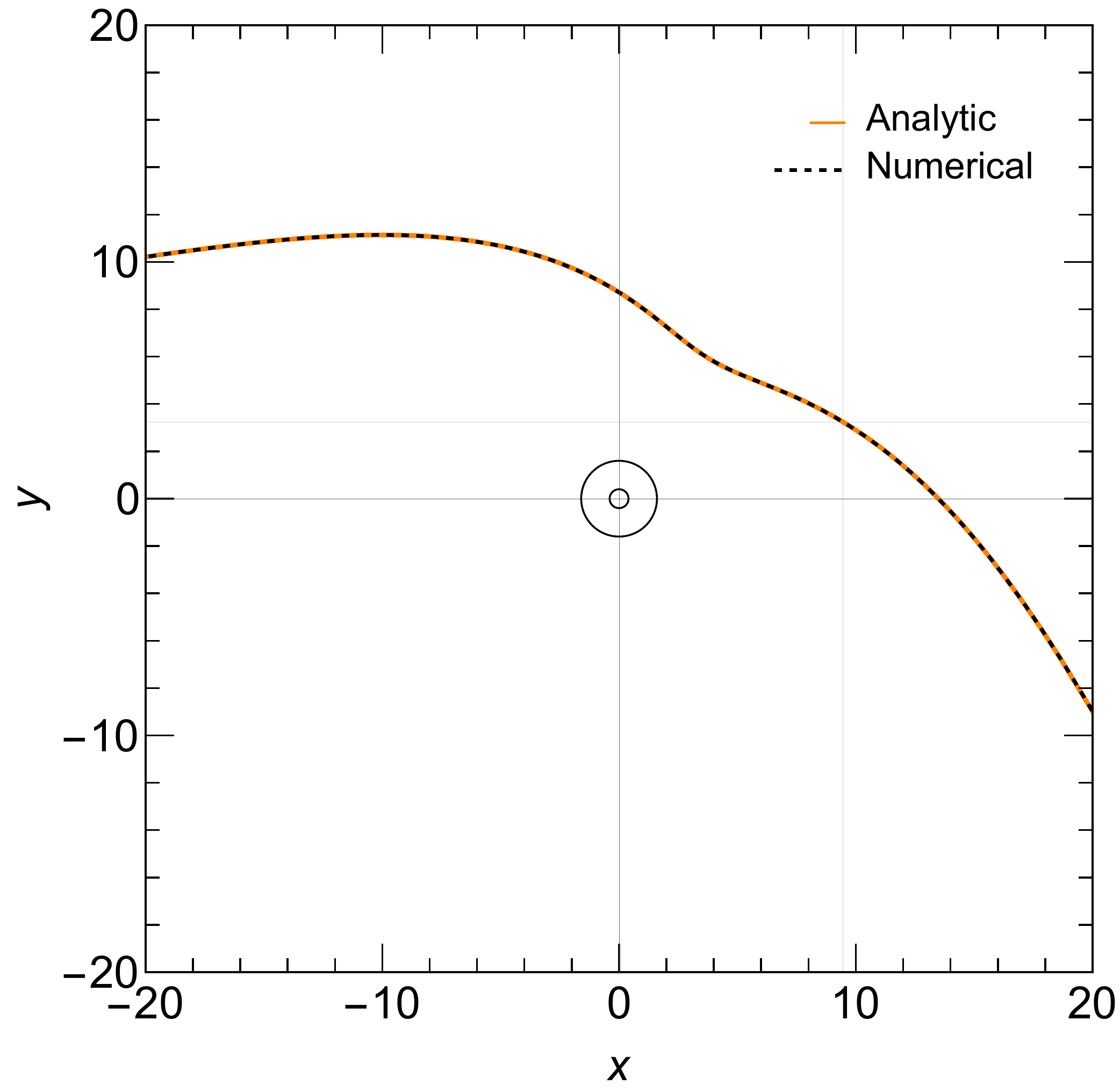}
\includegraphics[width=0.4\linewidth]{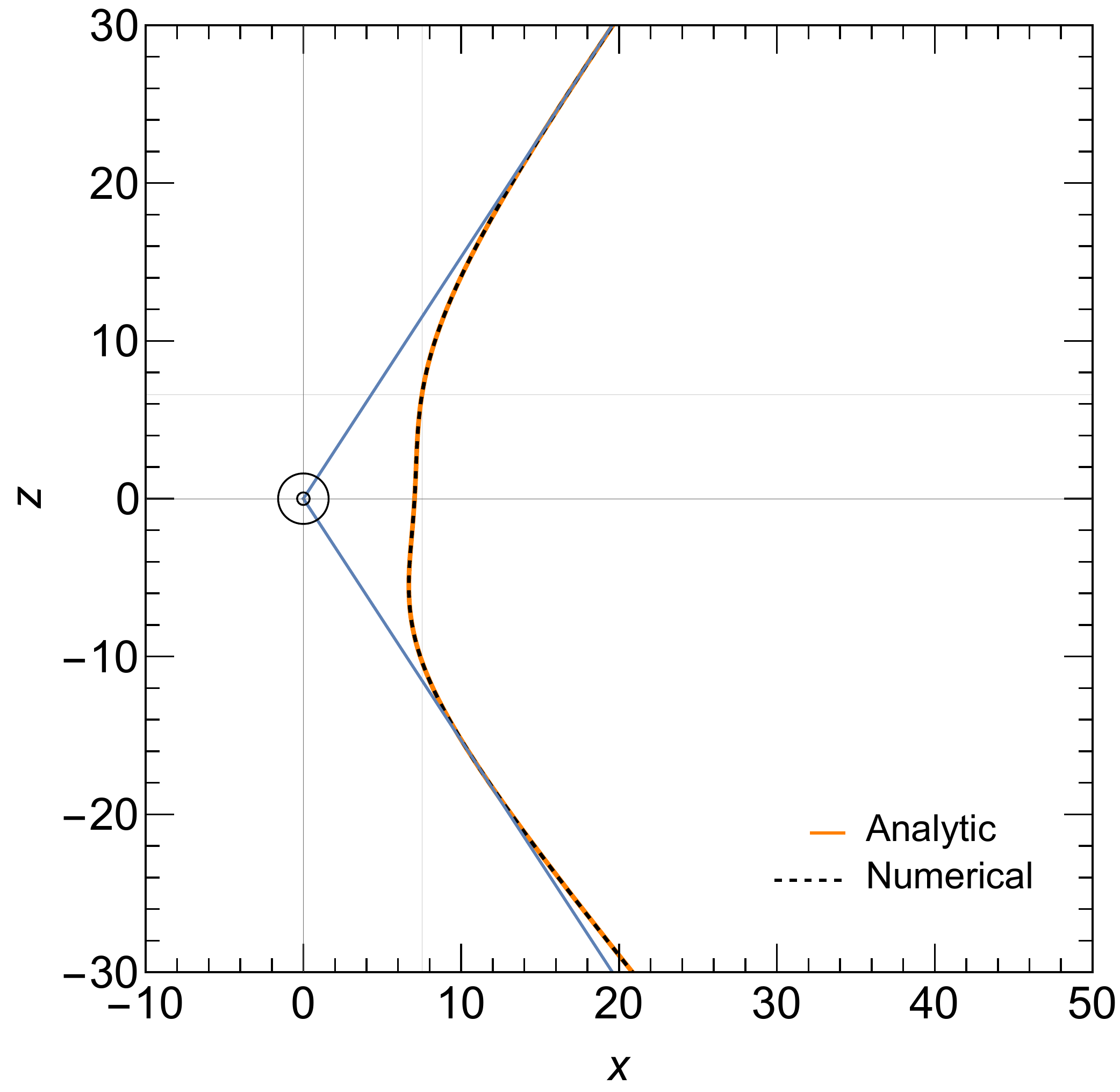}
\includegraphics[width=0.4\linewidth]{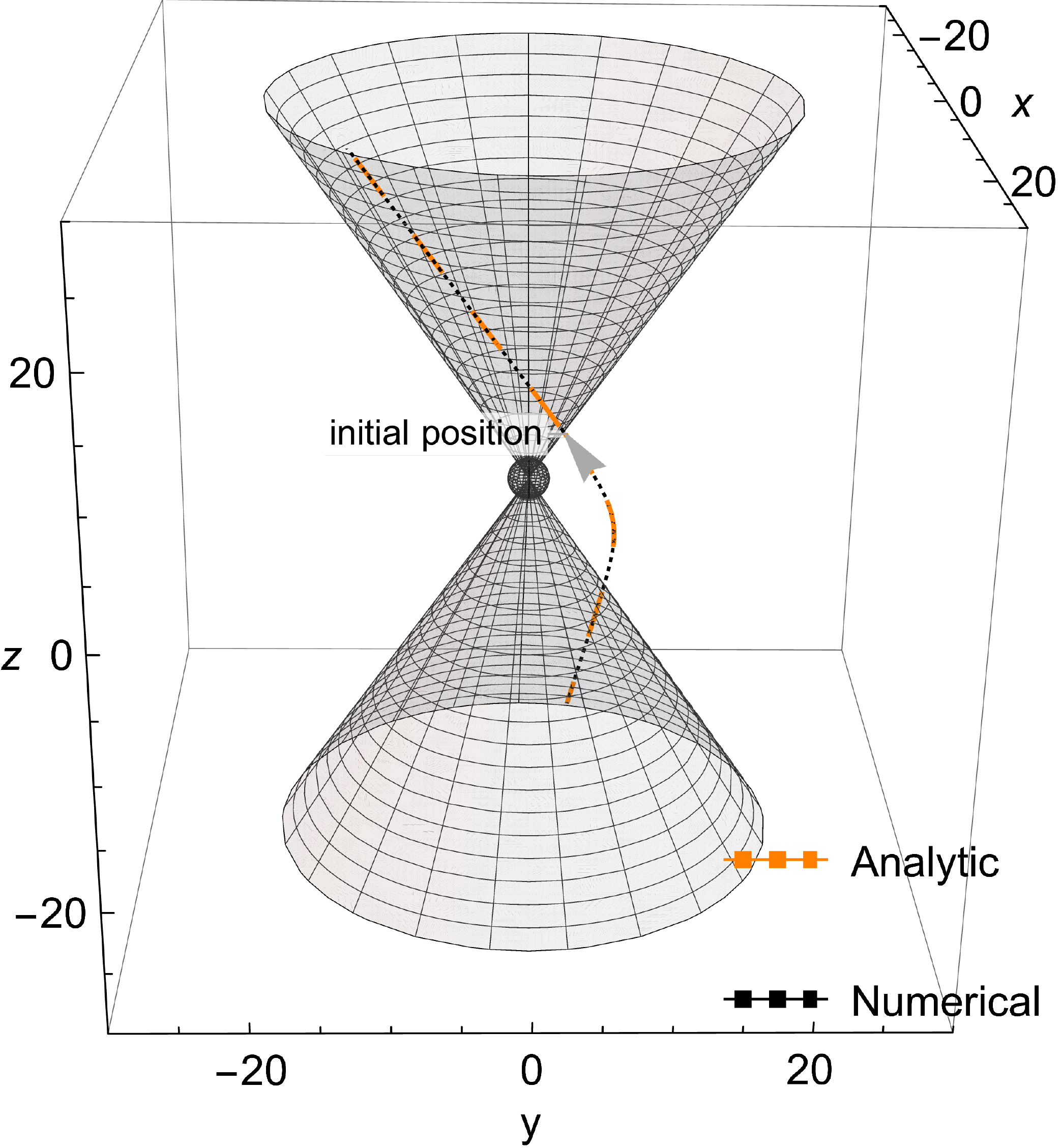}
\includegraphics[width=0.4\linewidth]{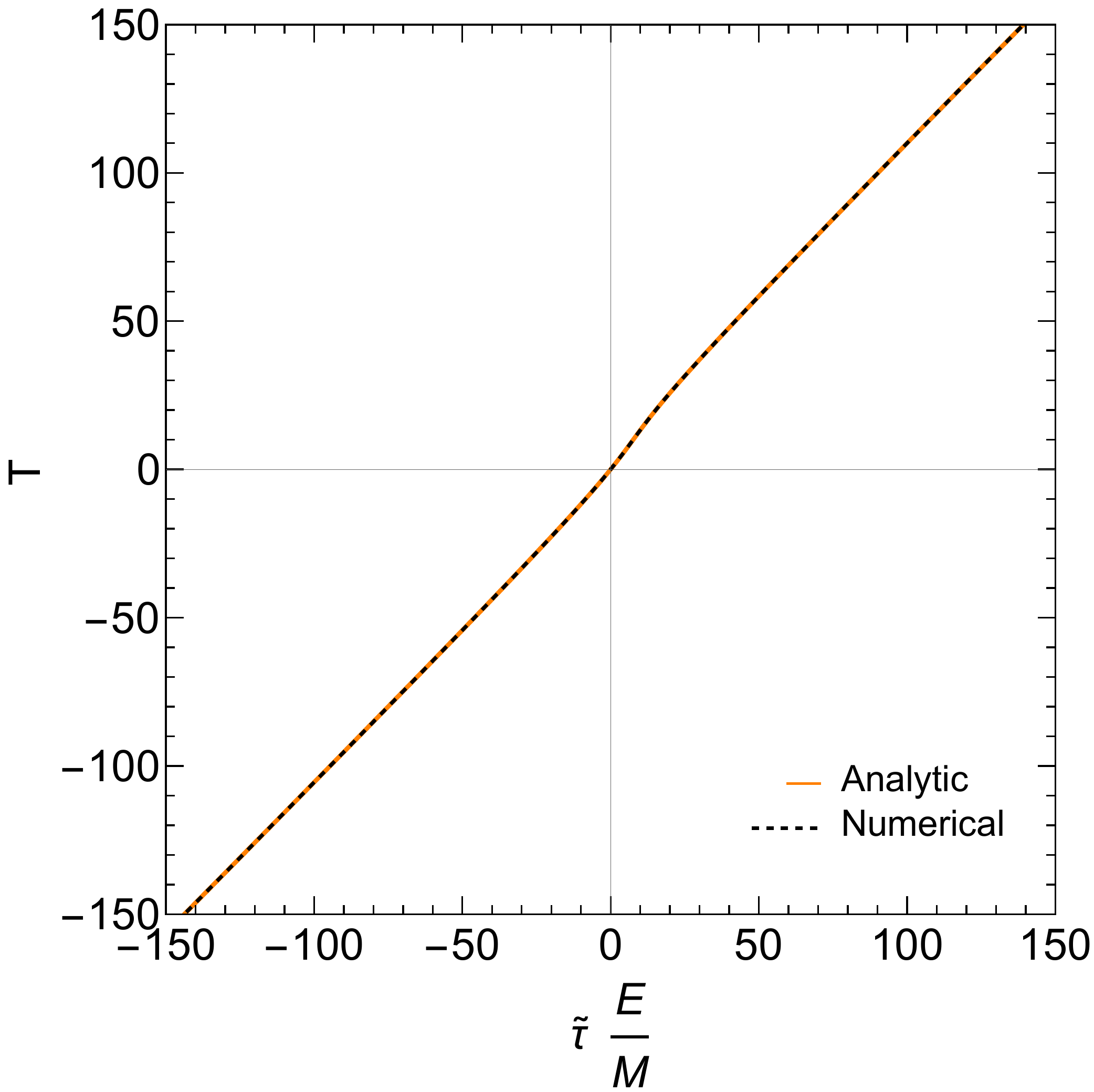}
\caption{\label{figNullA} Same as in Fig.\ \ref{figA}, but for a null unbound orbit with $\varepsilon^2 = 1$, $\lambda_z \approx 4.47214$, $\alpha=0.8$, $\kappa=60$, and an initial position at $\xi_0 = 10$, $\theta_0 = 0.85$, $\varphi_0 = 0.33$, $\epsilon_r = -1$, $\epsilon_{\theta} = 1$.}
\end{figure}

\begin{figure}[t]
\centering
\includegraphics[width=0.4\linewidth]{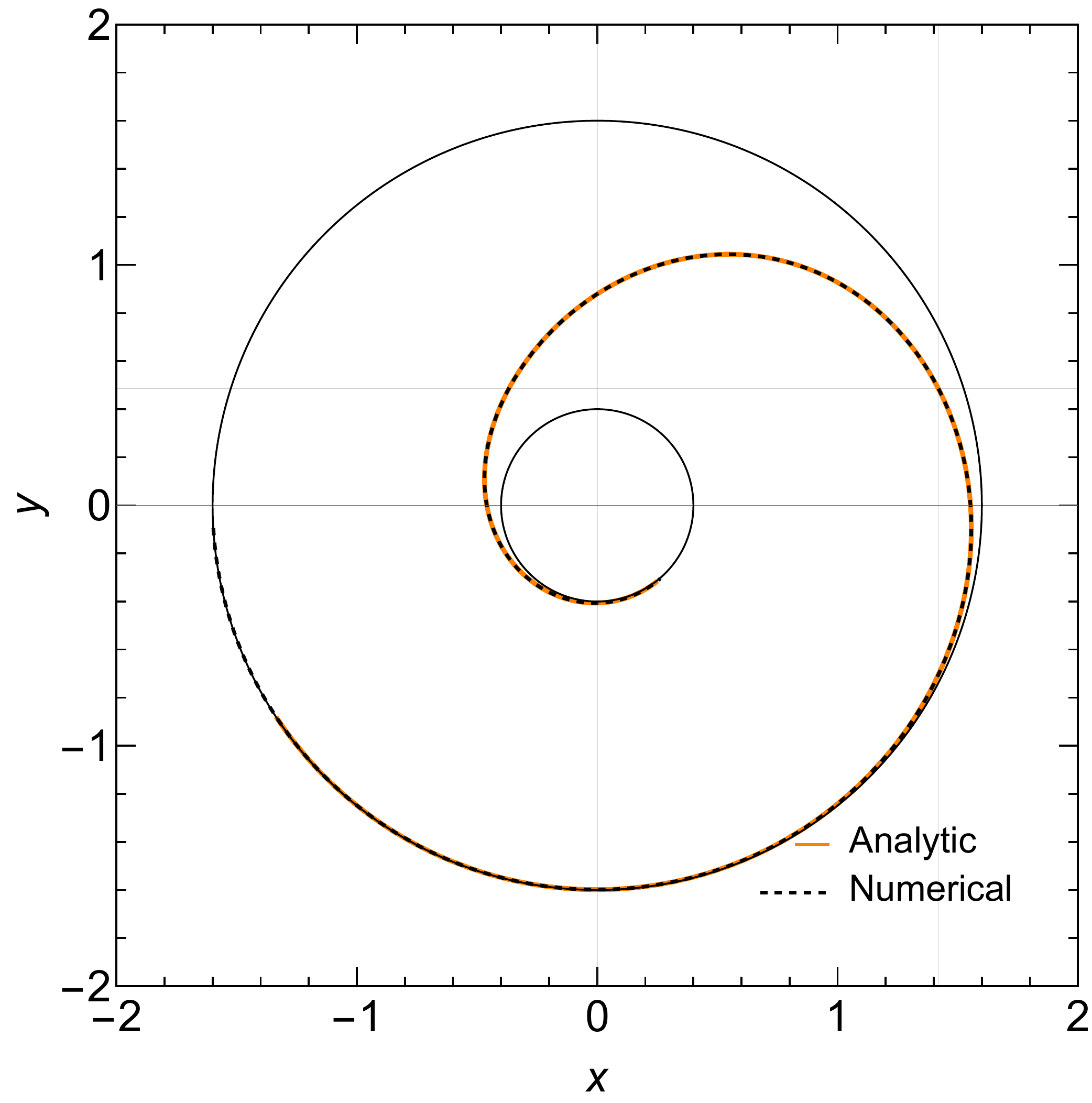}
\includegraphics[width=0.4\linewidth]{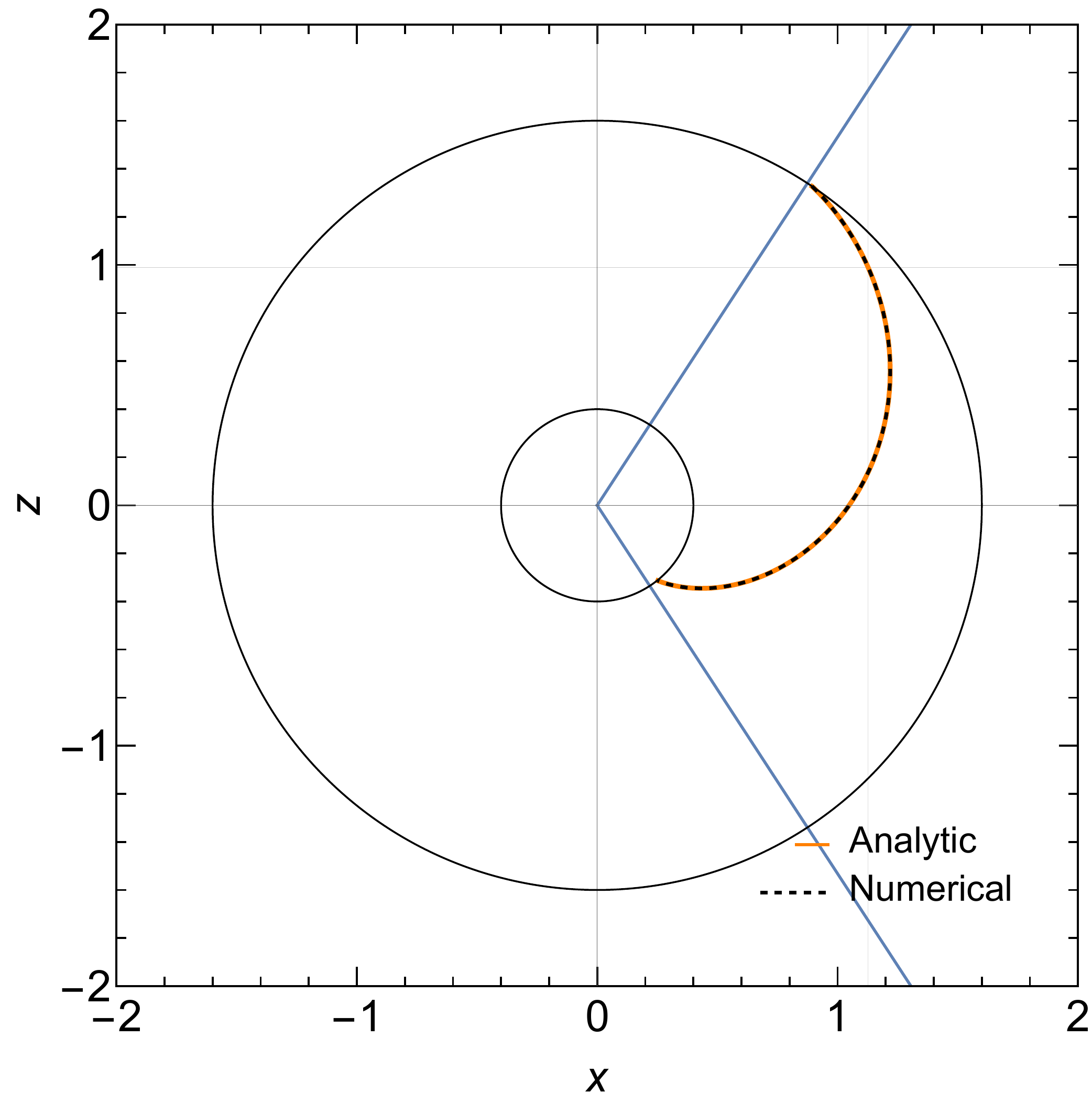}
\includegraphics[width=0.4\linewidth]{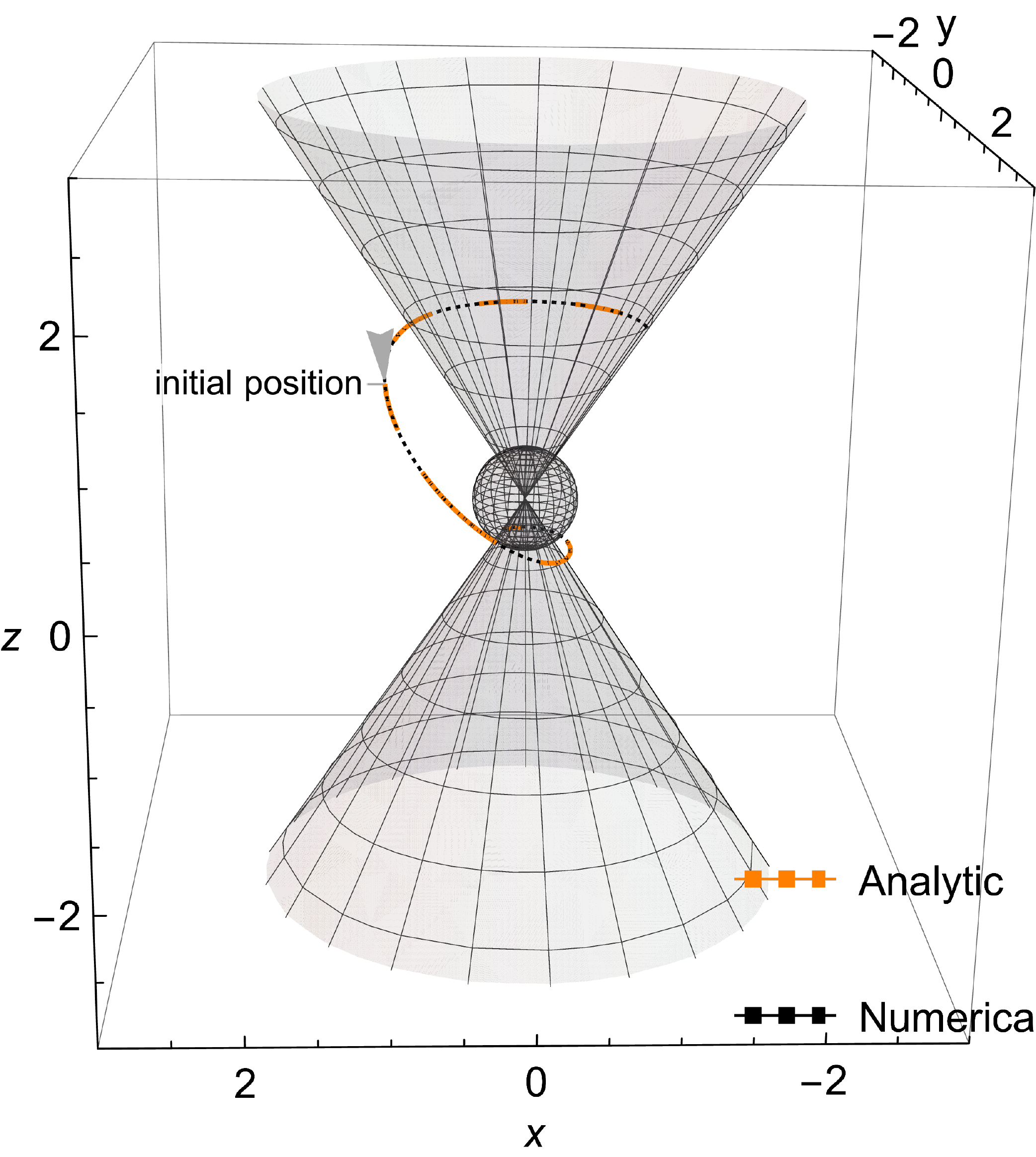}
\includegraphics[width=0.4\linewidth]{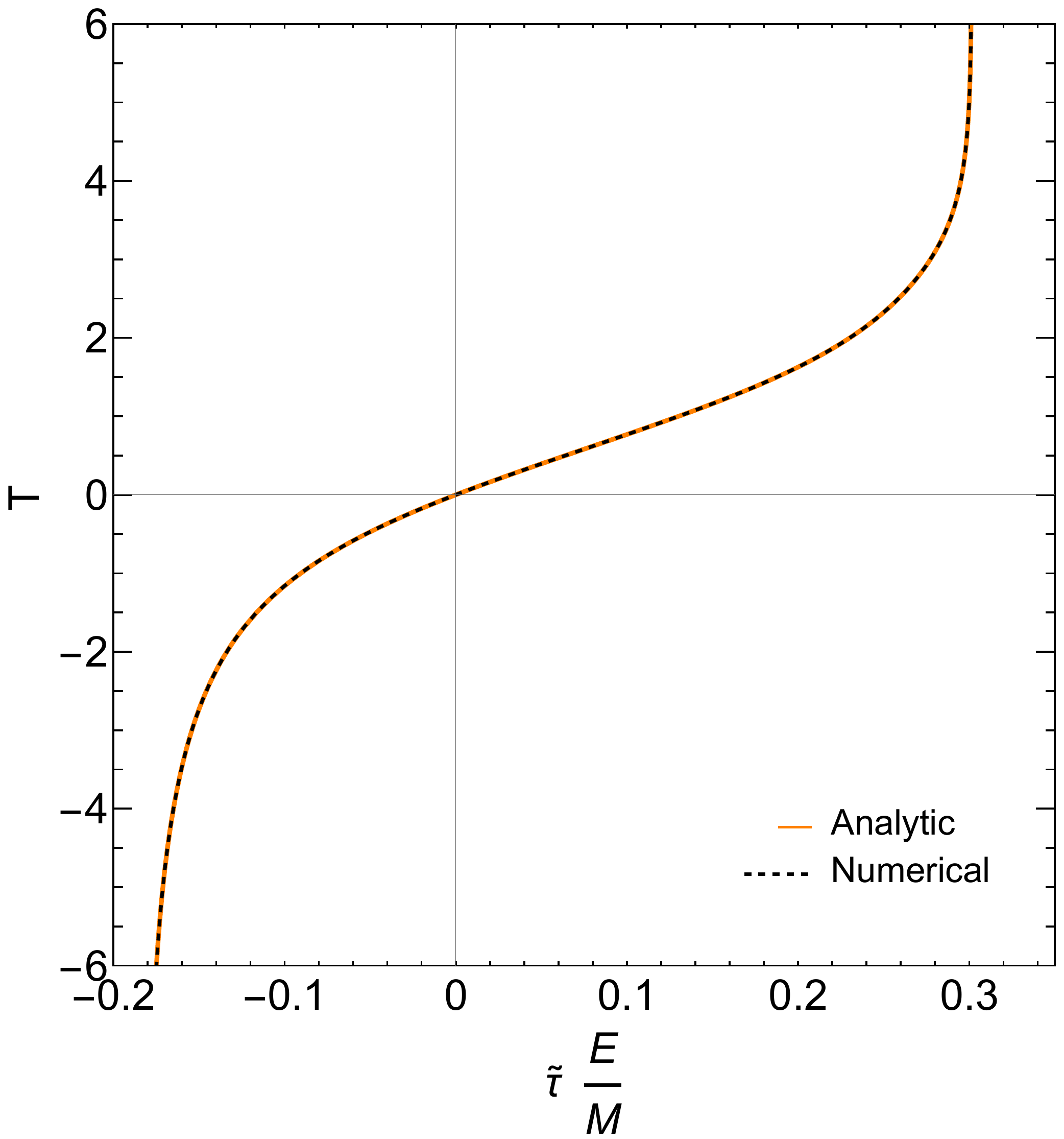}
\caption{\label{figNullB} Same as in Fig.\ \ref{figA}, but for a null bound orbit with $\varepsilon^2 = 1$, $\lambda_z \approx 4.47214$, $\alpha=0.6$, $\kappa=60$, and an initial position at $\xi_0 = 1.5$, $\theta_0 = 0.85$, $\varphi_0 = 0.33$, $\epsilon_r = -1$, $\epsilon_{\theta} = 1$.}
\end{figure}

\begin{figure}[t]
\centering
\includegraphics[width=0.4\linewidth]{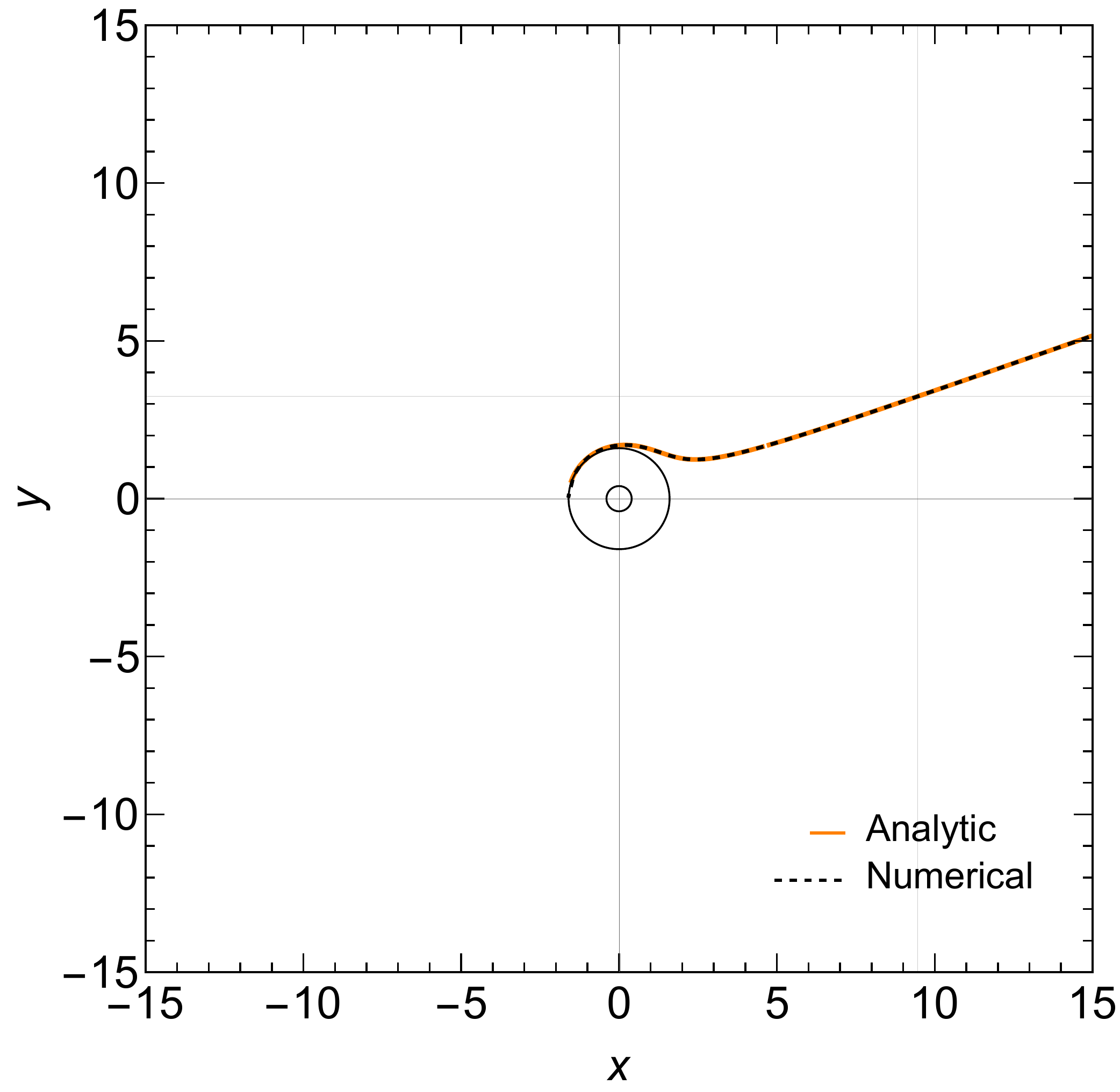}
\includegraphics[width=0.4\linewidth]{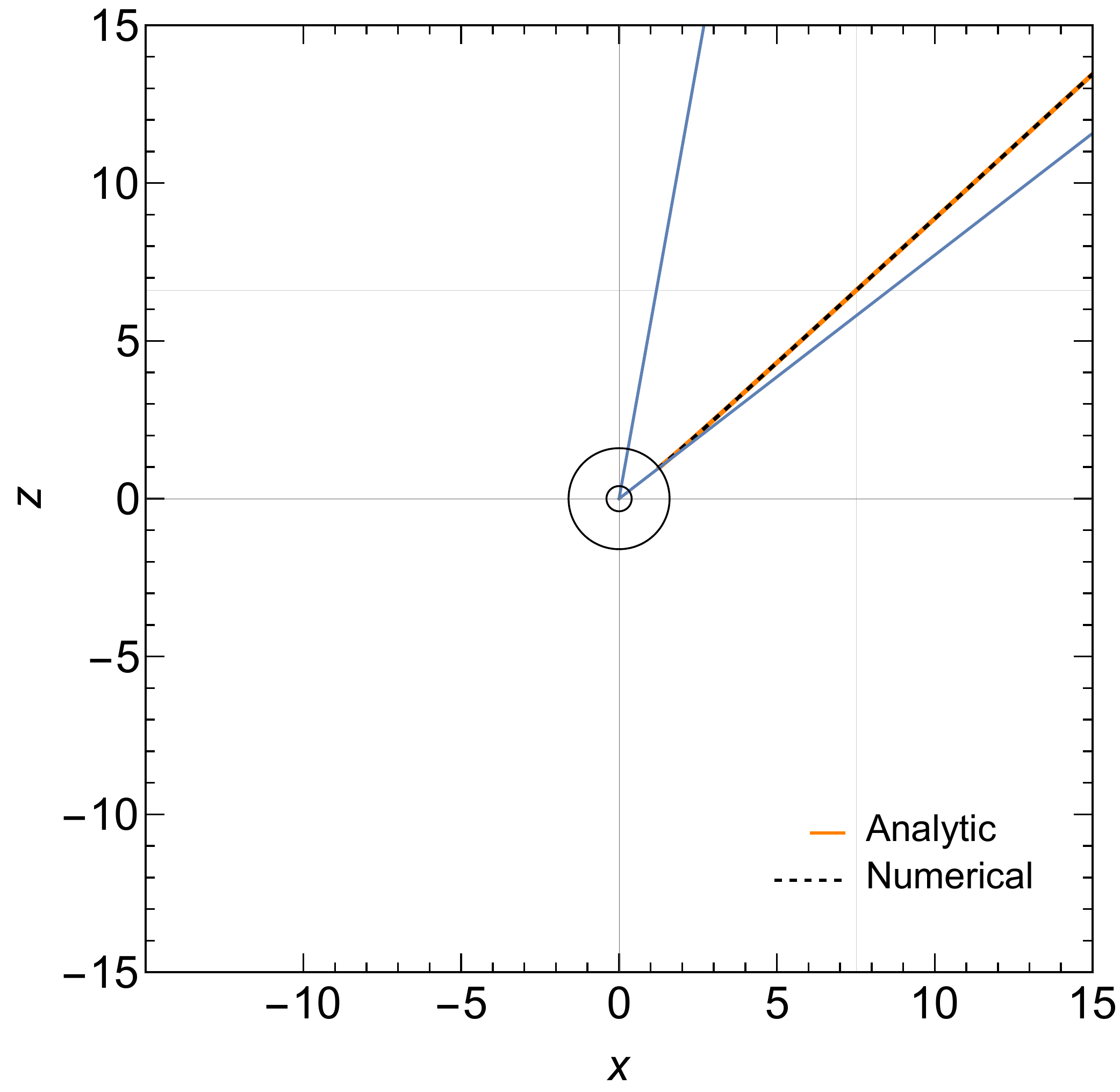}
\includegraphics[width=0.4\linewidth]{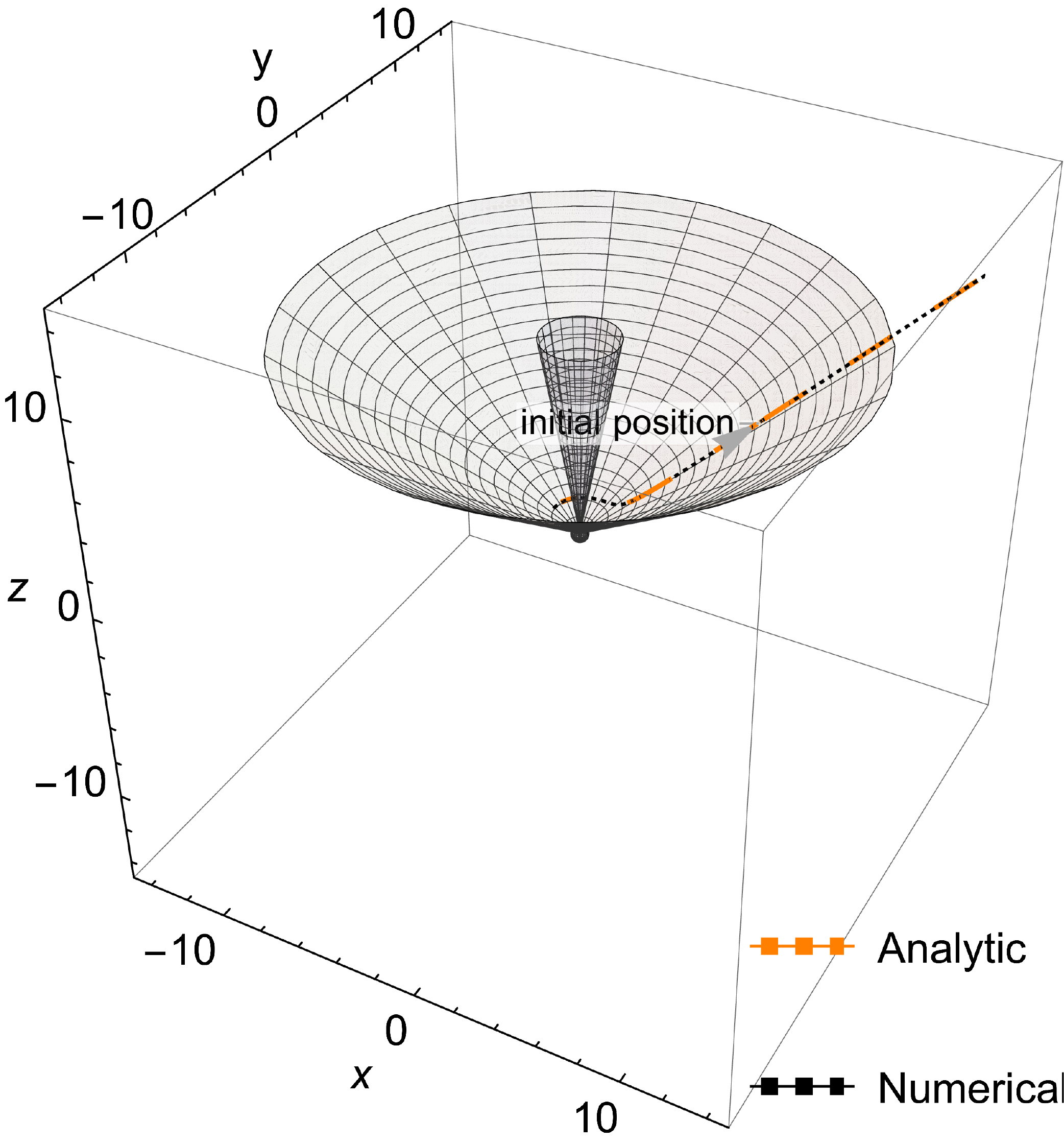}
\includegraphics[width=0.4\linewidth]{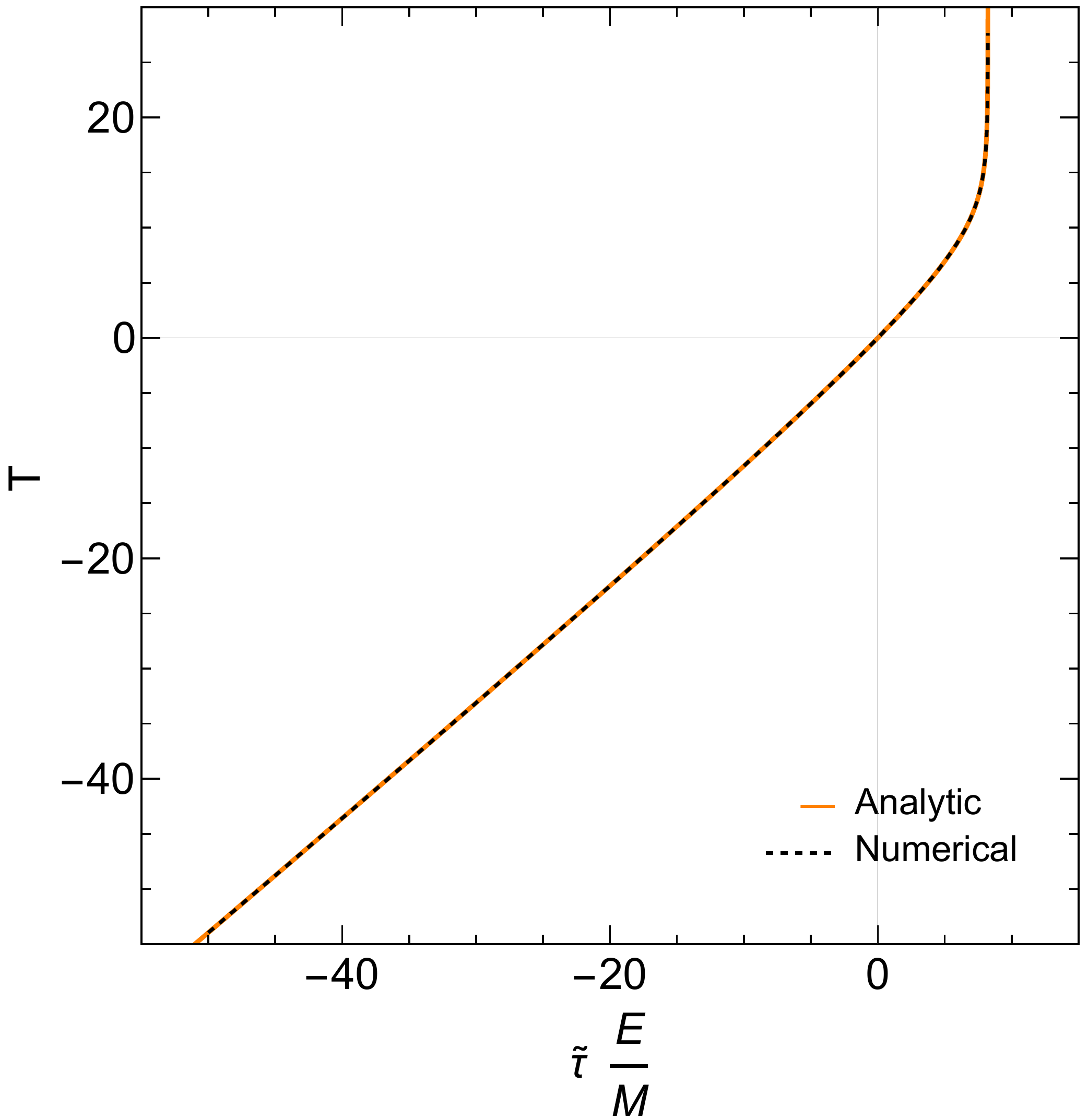}
\caption{\label{figNullC} Same as in Fig. \ref{figA}, but for a null unbound orbit with $\varepsilon^2 = 1$, $\lambda_z \approx -0.111803$, $\alpha=0.8$, $\kappa=0.6$, and an initial position at $\xi_0 = 10$, $\theta_0 = 0.85$, $\varphi_0 = 0.33$, $\epsilon_r = -1$, $\epsilon_{\theta} = 1$.}
\end{figure}

\begin{figure}[t]
\centering
\includegraphics[width=0.4\linewidth]{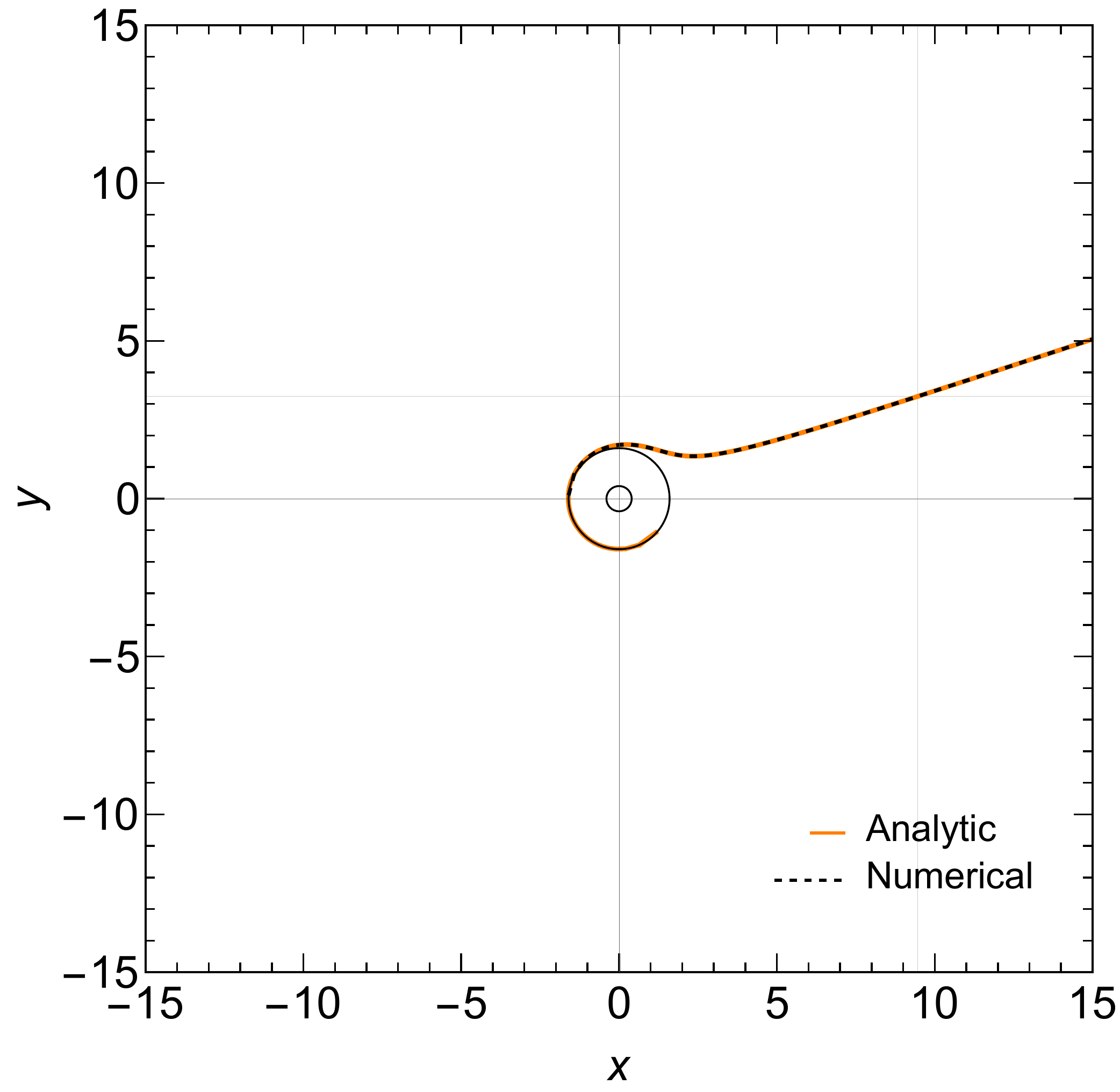}
\includegraphics[width=0.4\linewidth]{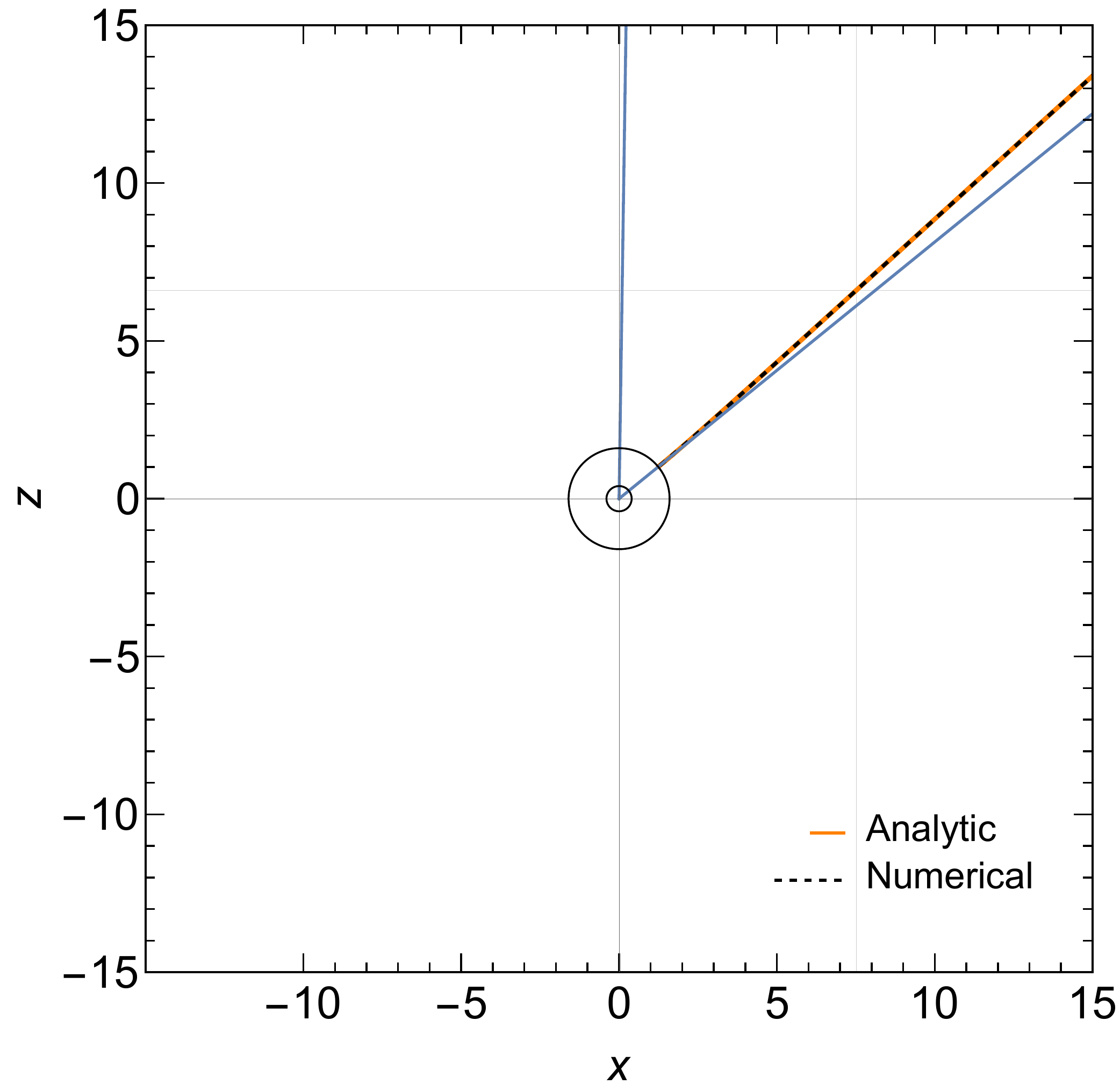}
\includegraphics[width=0.4\linewidth]{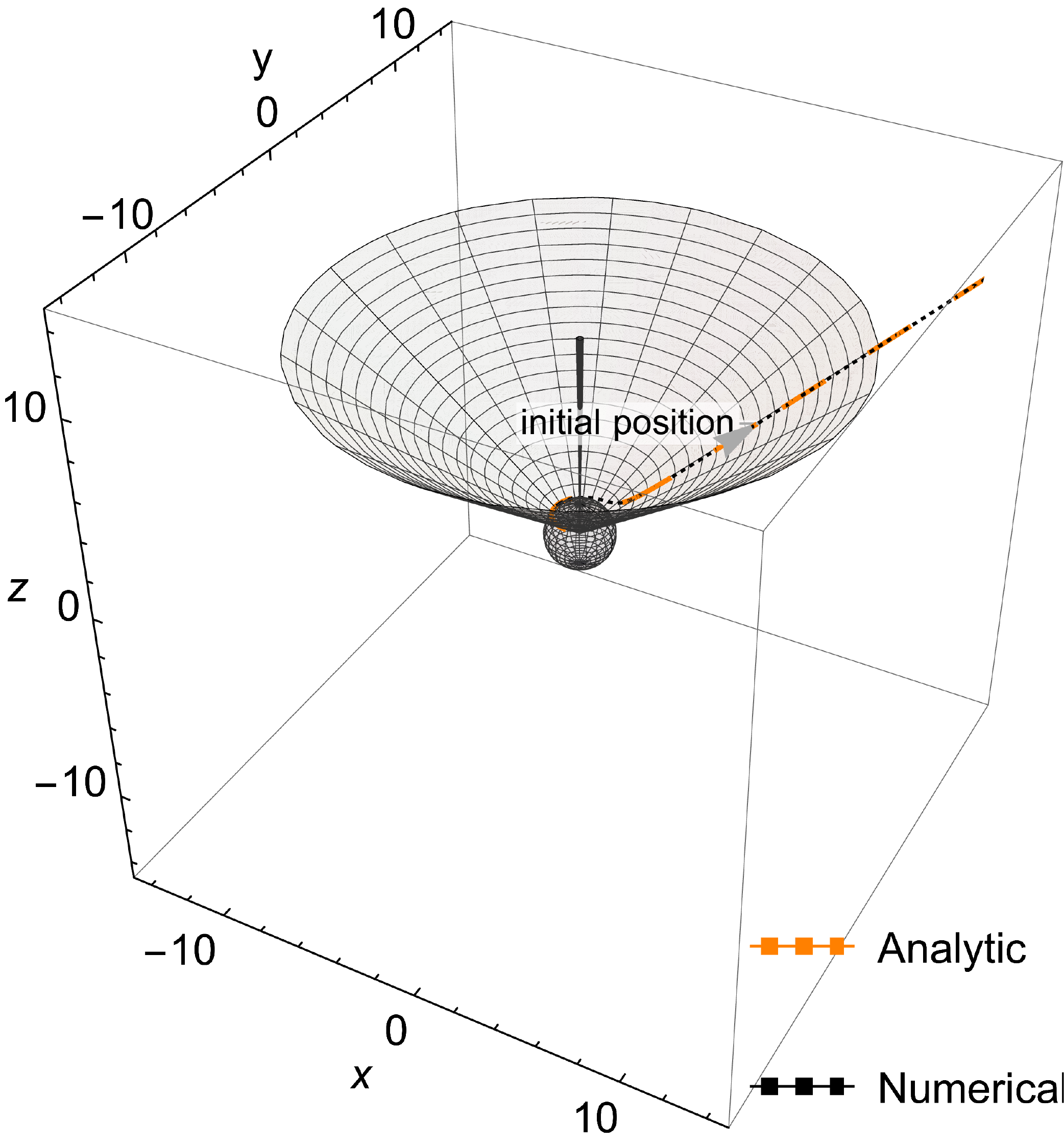}
\includegraphics[width=0.4\linewidth]{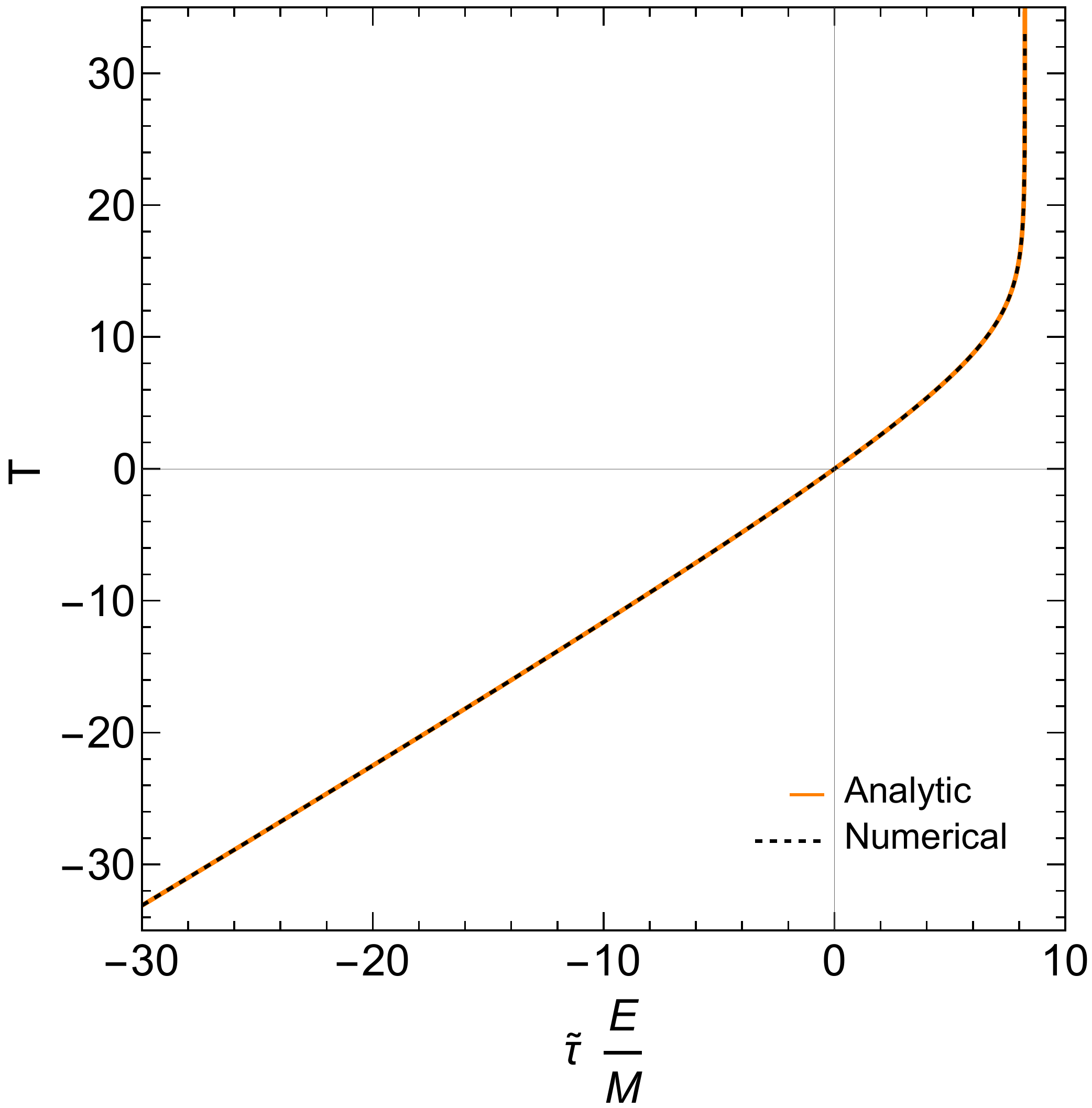}
\caption{\label{figNullD} Same as in Fig.\ \ref{figA}, but for a null unbound orbit with $\varepsilon^2 = 1$, $\lambda_z \approx -0.00912871$, $\alpha=0.8$, $\kappa=0.4$, and an initial position at $\xi_0 = 10$, $\theta_0 = 0.85$, $\varphi_0 = 0.33$, $\epsilon_r = -1$, $\epsilon_{\theta} = 1$.}
\end{figure}

\begin{figure}[t]
\centering
\includegraphics[width=0.4\linewidth]{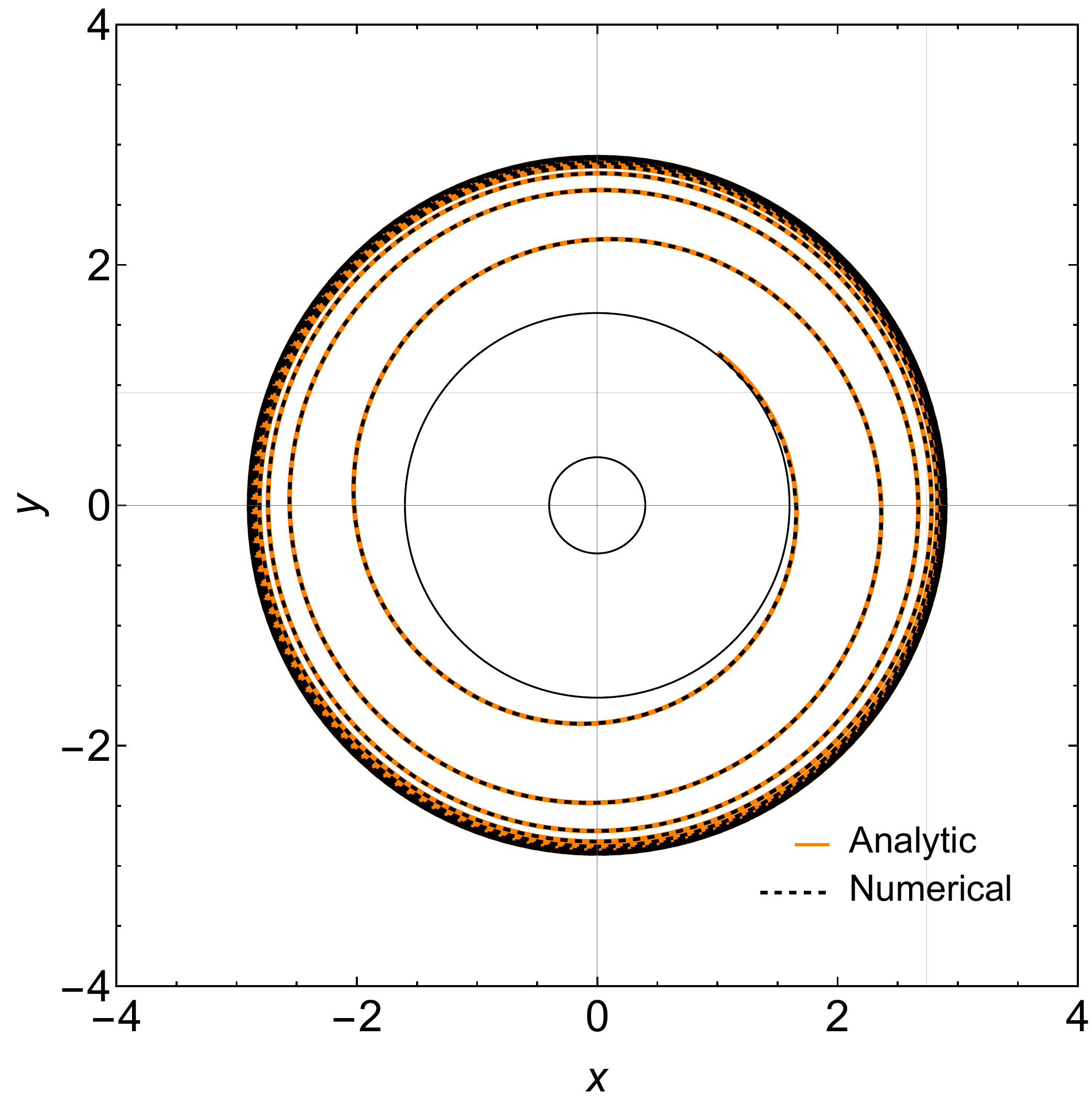}
\includegraphics[width=0.4\linewidth]{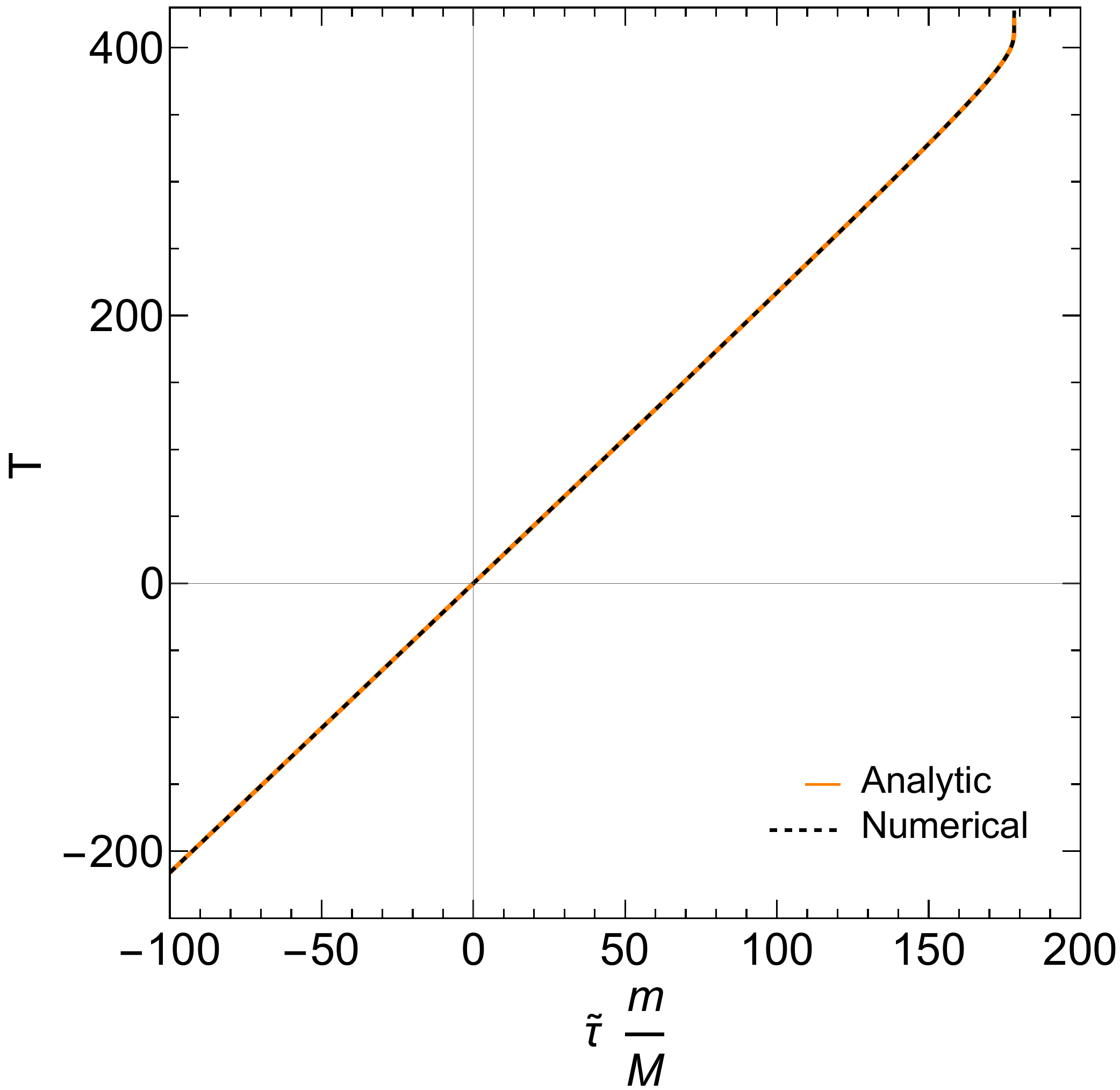}
\includegraphics[width=0.4\linewidth]{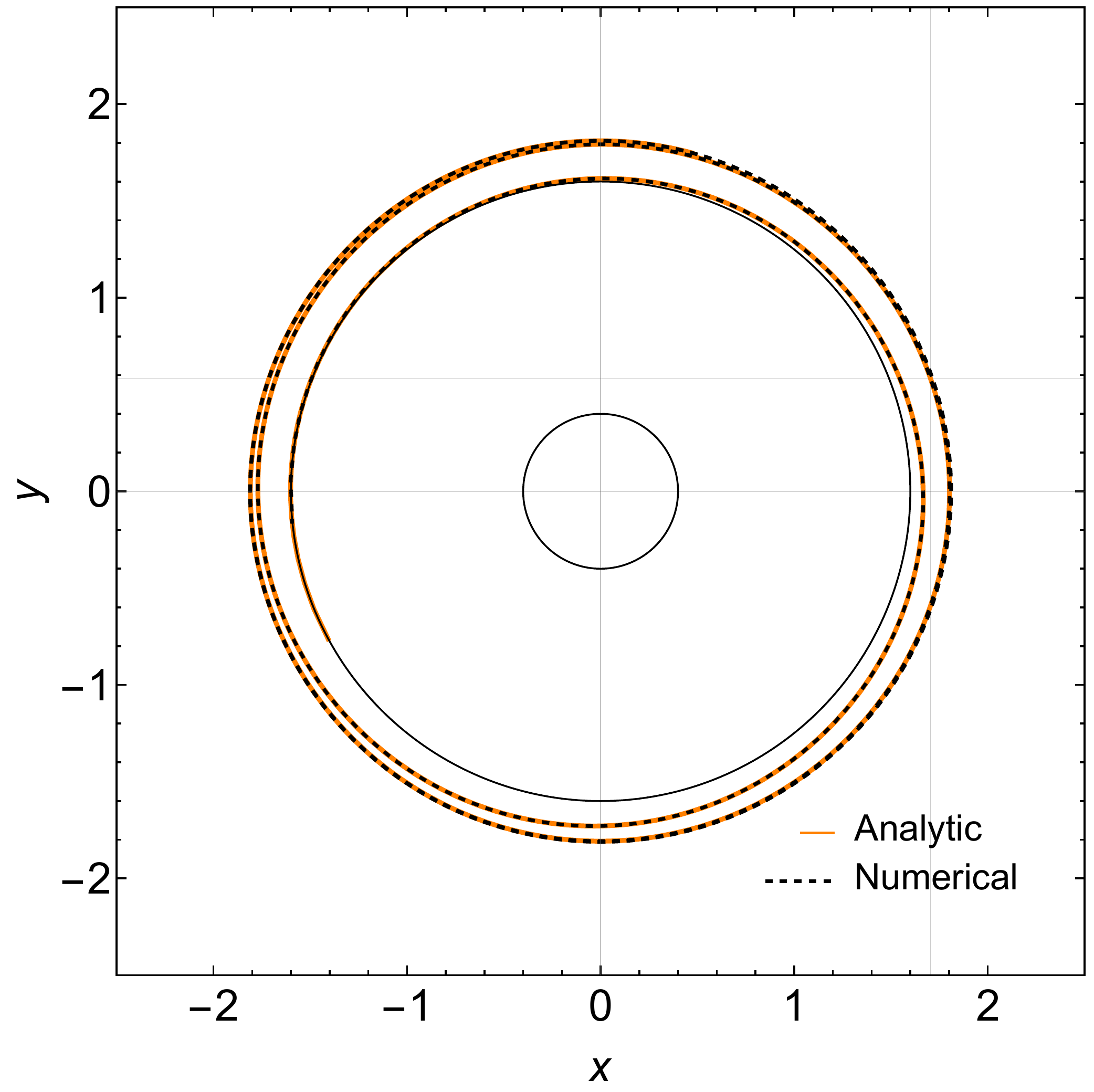}
\includegraphics[width=0.4\linewidth]{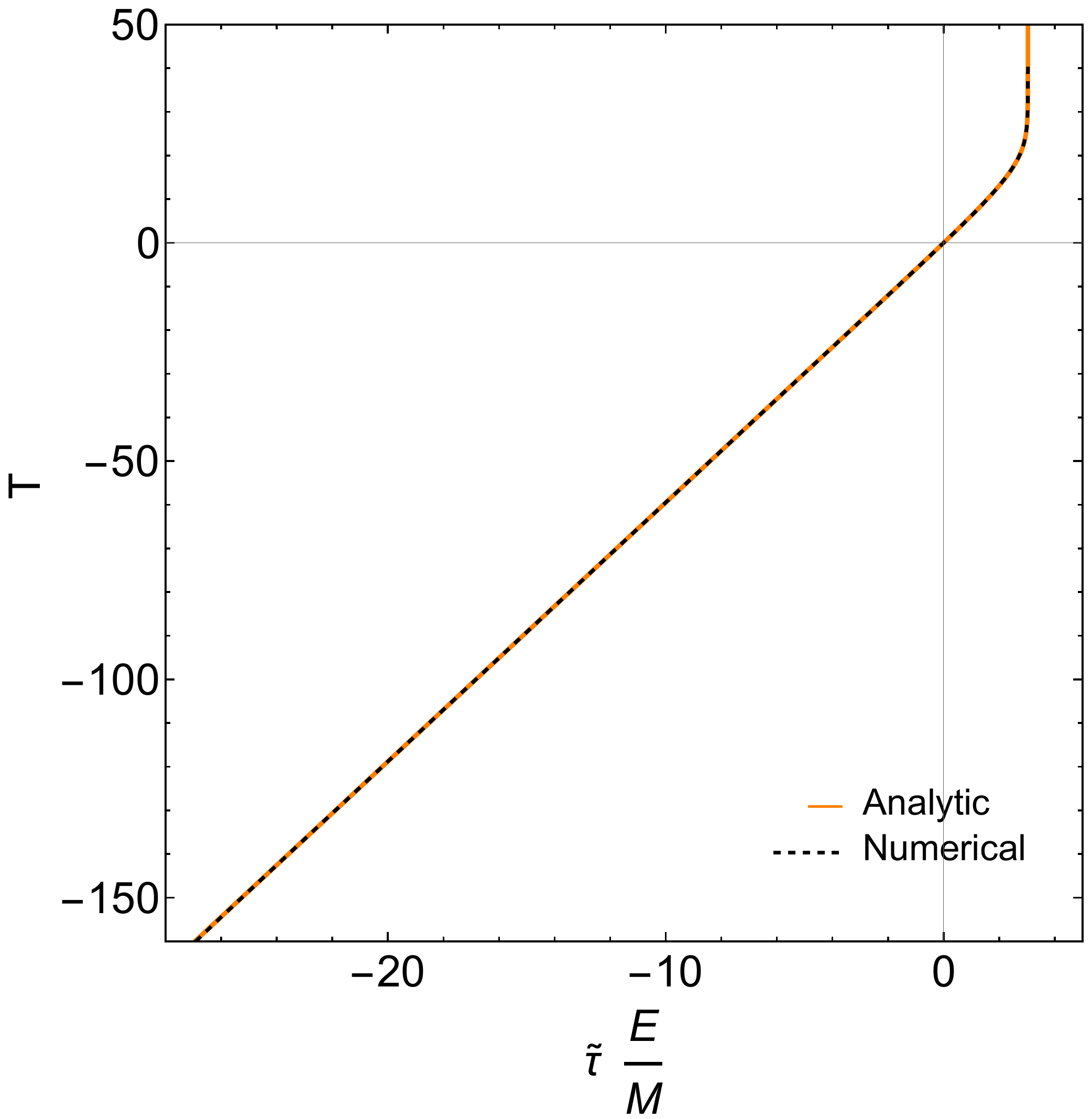}
\caption{\label{figZW} Examples of spiraling orbits in the equatorial plane. Upper figures correspond to a timelike orbit with $\varepsilon^2 \approx 0.77064$, $\lambda_z \approx 2.38044$, $\alpha=0.8$, $\kappa= (\lambda_z - \alpha \varepsilon)^2 \approx 2.81619$, and an initial position at $\xi_0 =\xi_\mathrm{ms} - 0.01\approx 2.89664$, $\theta_0 = \pi/2$, $\varphi_0 = 0.33$, $\epsilon_r = -1$. Lower figures correspond to a null orbit with $\varepsilon^2 =1$, $\lambda_z \approx 3.2373$, $\alpha=0.8$, $\kappa= (\lambda_z - \alpha \varepsilon)^2 \approx 5.94042 $ and an initial position at $\xi_0 =\xi_\mathrm{ph} - 0.01\approx 1.80109$, $\theta_0 = \pi/2$, $\varphi_0 = 0.33$, $\epsilon_r = -1$.} 
\end{figure}

Examples of solutions obtained in Sec.\ \ref{sec:solutions} are plotted in Figs.\ \ref{figA} to \ref{figZW}. In Figs.\  \ref{figA} to \ref{figNullD} we plot projections of the orbits on the $xy$ and $xz$ planes together with projections on the three-dimensional spaces of constant time. Here the Cartesian coordinates $(x,y,z)$ are defined as
\begin{subequations}
\begin{eqnarray}
x & = & \xi \cos \varphi \sin \theta, \\
y & = & \xi \sin \varphi \sin \theta, \\
z & = & \xi \cos \theta.
\end{eqnarray}
\end{subequations}
The last plot (lower right panel) in Figs.\ \ref{figA} to \ref{figNullD} shows the coordinate time $T$ versus the proper time $\tilde \tau$ for timelike geodesics and the affine parameter for null ones. Our analytic solutions are depicted with an orange line. For comparison, we also plot numerical solutions corresponding to the same initial data. They are depicted with black dotted lines. In all plots we mark initial positions $(\xi_0,\theta_0,\varphi_0)$, but in many cases the solutions are evolved both forward and backward in time. Blue lines or gray cones in the plots show the limiting values of the angle $\theta$, marking the region available for the motion. Black hole horizons (inner and outer) are drawn as black circles or gray spheres.

The coordinate singularities present in Boyer-Lindquist coordinates prevent us from continuing the solutions through horizons. The equations for $\xi(s)$ and $\theta(s)$ are unaffected, but the ones for $\varphi(s)$ and $T(s)$ are. Consequently, we only plot our solutions up to horizons.

For convenience, main parameters of solutions shown in Figs.\ \ref{figA} to \ref{figNullD} have been collected in Table I, together with information about their radial types and real zeros of the radial potential $\tilde R$. In all cases $\alpha = 0.8$, and consequently the horizons are located at $\xi_H^- = 1 - \sqrt{1 - \alpha^2} = 0.4$ and $\xi_H^+ = 1 + \sqrt{1 - \alpha^2} = 1.6$. Figures \ref{figA} to \ref{figF} show examples of timelike orbits ($\delta_1 = 1$). Null orbits ($\delta_1 = 0$) are shown in Figs.\ \ref{figNullA} to \ref{figNullD}.

Figure \ref{figA} shows a generic timelike flyby orbit of radial type II, plunging into the black hole. The orbit crosses the equatorial plane. Figure \ref{figB} depicts a generic timelike type III interval-bound orbit, restricted to a region outside the black hole horizon, oscillating around the equatorial plane. In Fig.\ \ref{figC} we assume the same constants of motion as in Fig.\ \ref{figB}, but the choice of the initial radius $\xi_0 = 1.55$ yields an inner bound orbit. We follow the motion both forward and backward in time, up to horizons at $\xi = \xi_H^\pm$. Figure \ref{figD} gives a generic example of a type IV flyby orbit. In this case the turning point is located at $\xi \approx 4.58 > \xi_H^+$. Consequently, the particle incoming from $\xi = + \infty$ is scattered by the black hole, and moves to $\xi = + \infty$. Figure \ref{figE} shows an example of an interval-bound orbit of radial type V, with $\xi_0 = 2.3 > \xi_H^+$. The motion is continued forward and backward in time up to $\xi = \xi_H^+$. Perhaps the most interesting case is shown in Fig.\ \ref{figF}. The trajectory is of type I and represents a transit orbit. The motion is restricted to the ``northern'' hemisphere. We emphasize once more that the radial motion is described by a manifestly real expression (\ref{xi_s}), even though the radial potential $\tilde R$ has no real zeros in this case.

In Fig.\ \ref{figNullA} we plot a generic null flyby trajectory of radial type IV. This trajecotry crosses the equatorial plane. For the null trajectory depicted in Fig.\ \ref{figNullB} we assume the same constants of motion as in Fig.\ \ref{figNullB}, but we take $\xi_0 = 1.5$. Consequently, the trajectory is of interval-bound type, and the part of the orbit shown in Fig.\ \ref{figNullB} is restricted to the region enclosed by the two horizons $\xi = \xi_H^-$ and $\xi = \xi_H^+$. Figures \ref{figNullC} and \ref{figNullD} show two examples of null orbits plunging into the black hole. Both trajectories originate at $\xi_0 = 10$. The trajectory shown in Fig.\ \ref{figNullC} is of radial type II (flyby). The one depicted in Fig.\ \ref{figNullD} belongs to radial type I (transit). Both trajectories are restricted to the ``northern'' hemisphere.

Figure \ref{figZW} shows examples of two special equatorial trajectories with constants of motion corresponding to circular orbits \cite{mummery_balbus_2022,mummery_balbus_2023}. Parameters of these solutions are collected in Table \ref{tab2}. Upper plots in Fig.\ \ref{figZW} depict a timelike spiraling orbit with constants of motion $\varepsilon$, $\lambda_z$, and $\kappa$ characteristic for a marginally stable co-rotating circular orbit with the radius $\xi_\mathrm{ms}$. In this case
\begin{equation}
    \tilde R = (\varepsilon^2 - 1) \xi (\xi - \xi_\mathrm{ms})^3,
\end{equation}
i.e., $\xi_\mathrm{ms}$ is a triple root of $\tilde R$. The radius $\xi_\mathrm{ms}$ can be computed as
\begin{equation}
\label{isco_formula}
\xi_\text{ms} = 3 + Z_2 - \epsilon_\lambda \alpha \sqrt{\frac{(3 - Z_1)(3 + Z_1 + 2 Z_2)}{\alpha^2}},
\end{equation}
where $Z_1 = 1+ \sqrt[3]{1-\alpha^2}(\sqrt[3]{1+\epsilon_\lambda\alpha} + \sqrt[3]{1-\epsilon_\lambda\alpha})$, $Z_2 = \sqrt{3 \alpha^2+Z_1^2}$ \cite{isco}. Here the sign $\epsilon_\lambda = \mathrm{sgn} \, \alpha$ corresponds to a co-rotating orbit, and $\epsilon_\lambda = -\mathrm{sgn} \, \alpha$ to a counter-rotating one. For $\alpha = 0.8$ and $\epsilon_\lambda = +1$, we get $\xi_\mathrm{ms} \approx 2.90664$. In this case a co-rotating marginally stable orbit is characterized by $\varepsilon^2 = 0.77064$, $\kappa = 2.81619$, and $\lambda_z = 2.38044$. For the timelike orbit plotted in Fig.\ \ref{figZW} we assume the same values of $\varepsilon$, $\lambda_z$, and $\kappa$, but the initial radius is set to be $\xi_0 = \xi_\mathrm{ms} - 0.01$.

Lower panels in Fig.\ \ref{figZW} show a similar picture for an equatorial null orbit. In this case we assume the constants $\varepsilon$, $\lambda_z$, and $\kappa$ corresponding to a circular null orbit of radius
\begin{equation}
\label{xiphoton}
    \xi_\mathrm{ph} = 2 + 2 \cos \left[ \frac{2}{3} \mathrm{arccos} \left( - \epsilon_\lambda  \alpha \right) \right].
\end{equation}
This gives, for $\alpha = 0.8$ and $\epsilon_\lambda = +1$, $\xi_\mathrm{ph} = 1.81109$, and the constants of motion $\lambda_z/\varepsilon = 3.2373$, and $\kappa/\varepsilon^2 = 5.94042$. The potential $\tilde R$ has a double zero at $\xi = \xi_\mathrm{ph}$. For the null orbit plotted in Fig.\ \ref{figZW} we assume the initial radius at $\xi_0 = \xi_\mathrm{ph} - 0.01$.

\section{Discussion}
\label{sec:discussion}

We have derived a single set of general, closed form analytic solutions describing all types of timelike and null Kerr geodesics in terms of Weierstrass elliptic functions. Our derivation follows largely the footsteps of Ref.\ \cite{hackmann_2010}, but a new ingredient is an application of the Biermann-Weierstrass theorem, which allows us to parametrize solutions with the constants of motion and arbitrary admissible initial coordinates. In particular, there is no need in our formalism to calculate any turning points of the motion, which is generally necessary in other methods.

We tried to be coherent in our approach, and to express our solutions in terms of Weierstrass elliptic functions only, but one is free to mix our formulas with the existing solutions, perhaps more convenient in some cases. Our method provides very general and clear expressions for radial and polar motions. As usual, expressions for $\varphi(s)$ and the time coordinate $T(s)$ are much more involved, as several elliptic integrals appear that are solved using logarithms of the Weierstrass sigma function. However, progress has been made in the required careful selection of appropriate branches of complex logarithms, as described in Sec.\ \ref{sec:implementation}. This has been achieved by extracting the linear parts and reducing the problem to the determination of a single constant for the complete trajectory. On the other hand, the big advantage of our derivation is that it does not differentiate between many possible algebraic types of solutions, perhaps except for (zero-measure) special cases with $\varepsilon = \delta_1$, described in Appendix \ref{appendix:Caseeeqd}, and a necessity to handle some aspects of the derivation in the extreme-Kerr case separately (Appendix \ref{appendix:extremeKerr}).

\begin{acknowledgments}
The authors would like to thank Andrzej Odrzywo\l{}ek for discussions. This work was partially supported by the Polish National Science Centre Grant No.\ 2017/26/A/ST2/00530. E.H. acknowledges support from the Deutsche Forschungsgemeinschaft (DFG, German Research Foundation) under Germany’s Excellence Strategy – EXC-2123 QuantumFrontiers (Project-ID 390837967), and SFB 1464 TerraQ (Project-ID 43461778) within project C03.
\end{acknowledgments}

%=================================================================================================================================
%                                                        APPENDIX
%=================================================================================================================================

\appendix

\section{Expressions for $J_\theta$, $N_1$, and $N_2$}
\label{mainappendix}

In this appendix we collect expressions for $J_\theta(s)$, $N_1(s)$, and $N_2(s)$, appearing in our formulas for $T(s)$. The derivations are similar to the ones for $N_H^\pm$ and $I_\theta(s)$, given in Sec.\ \ref{sec:varphi}.

\subsection{Expressions for $N_1$ and $N_2$}

Integrals $N_1$ and $N_2$ are defined as
\begin{equation}
    N_1(s) = \int_0^s \xi(\bar s) d \bar s
\end{equation}
and
\begin{equation}
    N_2(s) = \int_0^s \xi(\bar s)^2 d \bar s.
\end{equation}
As before, we define
\begin{subequations}
    \begin{eqnarray}
        \mathcal{C}_{1} &=& \frac{1}{48}\tilde R(\xi_0) \tilde R'''(\xi_0) - \frac{1}{96} \tilde R'(\xi_0)\tilde R''(\xi_0), \\    
        \mathcal{C}_{2} &=& \frac{1}{4}\tilde R'(\xi_0), \\
        \mathcal{C}_{3} &=& -\frac{1}{2}\epsilon_r \sqrt{\tilde R(\xi_0)},\\
        \mathcal{C}_{4} &=& \frac{1}{24}\tilde R''(\xi_0), \\
        \mathcal{C}_{5} &=&  \frac{1}{48}\tilde R(\xi_0)\tilde R^{(4)}(\xi_0), \label{C_5} \\
        \wp_{\tilde R,1} &=&   \mathcal{C}_{4} + \sqrt{\frac{ \mathcal{C}_{5}}{2}},\\
        \wp_{\tilde R,2} &=&   \mathcal{C}_{4} - \sqrt{\frac{ \mathcal{C}_{5}}{2}}.
    \end{eqnarray}
\end{subequations}
Hence Eq.\ \eqref{xi_s} takes the form
\begin{equation}
\label{xi_s_2}
    \xi(s) = \xi_0 + \frac{\mathcal{C}_{1} + \mathcal{C}_{2}\wp_{\tilde R}(s) + \mathcal{C}_{3} \wp_{\tilde R}'(s) }{ \left( \wp_{\tilde R}(s) - \wp_{\tilde R,1} \right)\left( \wp_{\tilde R}(s) - \wp_{\tilde R,2} \right)},
\end{equation}
and therefore
\begin{eqnarray}\label{Integral_N_1}
    N_1(s) & = & \int_0^s \xi(\bar s) d \bar s = \int_0^s \left\{ \xi_0 + \frac{\mathcal{C}_{1} + \mathcal{C}_{2}\wp_{\tilde R}(s) + \mathcal{C}_{3} \wp_{\tilde R}'(s) }{\left[ \wp_{\tilde R}(\bar s) - \wp_{\tilde R,1} \right] \left[ \wp_{\tilde R}(\bar s) - \wp_{\tilde R,2} \right]} \right\} d \bar s \nonumber \\
    & = &  \xi_{0} s +  \mathcal{C}_{1} K_{1}(s; p_{\tilde R,1}, p_{\tilde R,2}) + \mathcal{C}_{2}  K_{2}(s; p_{\tilde R,1}, p_{\tilde R,2})  + \mathcal{C}_{3}  K_{3}(s;p_{\tilde R,1}, p_{\tilde R,2}) ,
\end{eqnarray}
where $p_{\tilde R,1}$ and $p_{\tilde R,2}$ satisfy $\wp\left( p_{\tilde R,1}; g_{\tilde R,2}, g_{\tilde R,3} \right) = \mathcal{C}_{4} + \sqrt{\frac{ \mathcal{C}_{5}}{2}}$ and $\wp \left( p_{\tilde R,2}; g_{\tilde R,2}, g_{\tilde R,3} \right) = \mathcal{C}_{4} - \sqrt{\frac{ \mathcal{C}_{5}}{2}}$. Weierstrass functions appearing in expressions for $K_1$, $K_2$, $K_3$ in the above formula should be computed with the invariants $g_{\tilde R,2}$ and $g_{\tilde R,3}$.

The integral $N_2(s)$ can be calculated as follows. We have
\begin{eqnarray}\label{Integral_N_2}
    N_2(s) & = & \int_0^s \xi(\bar s)^2 d \bar s = \int_0^s \left\{ \xi_0 + \frac{\mathcal{C}_{1} + \mathcal{C}_{2}\wp_{\tilde R}(s) + \mathcal{C}_{3} \wp_{\tilde R}'(s) }{\left[ \wp_{\tilde R}(\bar s) - \wp_{\tilde R,1} \right] \left[ \wp_{\tilde R}(\bar s) - \wp_{\tilde R,2} \right]} \right\}^2 d \bar s  \nonumber \\
    & = & \int_0^s  \xi^2_0 + 2\xi_0 \frac{\mathcal{C}_{1} + \mathcal{C}_{2}\wp_{\tilde R}(s) + \mathcal{C}_{3} \wp_{\tilde R}'(s) }{\left[ \wp_{\tilde R}(\bar s) - \wp_{\tilde R,1} \right] \left[ \wp_{\tilde R}(\bar s) - \wp_{\tilde R,2} \right]} + \left\{\frac{\mathcal{C}_{1} + \mathcal{C}_{2}\wp_{\tilde R}(s) + \mathcal{C}_{3} \wp_{\tilde R}'(s) }{\left[ \wp_{\tilde R}(\bar s) - \wp_{\tilde R,1} \right] \left[ \wp_{\tilde R}(\bar s) - \wp_{\tilde R,2} \right]} \right\}^2 d \bar s \nonumber \\
    & = & \xi^2_0 s + 2\xi_0 \left[ N_1(s) - \xi_0 s \right] +  N_3(s) = 2\xi_0  N_1(s) - \xi^2_0 s +  N_3(s),
\end{eqnarray}
where
\begin{equation}\label{Integral_N_3}
    N_3(s) = \int_0^s \left\{\frac{\mathcal{C}_{1} + \mathcal{C}_{2}\wp_{\tilde R}(s) + \mathcal{C}_{3} \wp_{\tilde R}'(s) }{\left[ \wp_{\tilde R}(\bar s) - \wp_{\tilde R,1} \right] \left[ \wp_{\tilde R}(\bar s) - \wp_{\tilde R,2} \right]} \right\}^2 d \bar s.
\end{equation}
The integral $N_3$ can be written in the form
\begin{eqnarray}
     N_3(s) & = & \mathcal{C}^2_{1} L_{1}(s; p_{\tilde R,1}, p_{\tilde R,2}) + 2 \mathcal{C}_{1} \mathcal{C}_{2} L_{2}(s; p_{\tilde R,1}, p_{\tilde R,2}) + 2 \mathcal{C}_{1} \mathcal{C}_{3} L_{3}(s; p_{\tilde R,1}, p_{\tilde R,2}) \nonumber \\
     & & + 2 \mathcal{C}_{2} \mathcal{C}_{3} L_{4}(s; p_{\tilde R,1}, p_{\tilde R,2}) + \mathcal{C}^2_{2} L_{5}(s; p_{\tilde R,1}, p_{\tilde R,2}) + \mathcal{C}^2_{3} L_{6}(s; p_{\tilde R,1}, p_{\tilde R,2}),
\end{eqnarray}
where 

\begin{subequations}\label{L_int}
    \begin{eqnarray}
        L_{1} (s; p_{\tilde R,1}, p_{\tilde R,2}) &:=& \int_0^s \frac{d \bar s}{ \left[ \left(\wp(\bar s) -  \wp_{\tilde R,1}\right)\left(\wp(\bar s) - \wp_{\tilde R,2}\right)\right]^2},\\
        L_{2} (s; p_{\tilde R,1}, p_{\tilde R,2}) &:=& \int_0^s \frac{\wp(\bar s) d \bar s}{ \left[ \left(\wp(\bar s) -  \wp_{\tilde R,1}\right)\left(\wp(\bar s) - \wp_{\tilde R,2}\right)\right]^2},\\
        L_{3} (s; p_{\tilde R,1}, p_{\tilde R,2}) &:=& \int_0^s \frac{\wp'(\bar s) d \bar s}{ \left[ \left(\wp(\bar s) -  \wp_{\tilde R,1}\right)\left(\wp(\bar s) - \wp_{\tilde R,2}\right)\right]^2},\\
        L_{4} (s; p_{\tilde R,1}, p_{\tilde R,2}) &:=& \int_0^s \frac{\wp'(\bar s) \wp(\bar s) d \bar s}{ \left[ \left(\wp(\bar s) -  \wp_{\tilde R,1}\right)\left(\wp(\bar s) - \wp_{\tilde R,2}\right)\right]^2},\\
        L_{5} (s; p_{\tilde R,1}, p_{\tilde R,2}) &:=&   \int_0^s \frac{\wp(\bar s)^2 d \bar s}{ \left[ \left(\wp(\bar s) -  \wp_{\tilde R,1}\right)\left(\wp(\bar s) - \wp_{\tilde R,2}\right)\right]^2},\\
        L_{6} (s; p_{\tilde R,1}, p_{\tilde R,2}) &:=& \int_0^s \frac{\wp'(\bar s)^2 d \bar s}{ \left[ \left(\wp(\bar s) -  \wp_{\tilde R,1}\right)\left(\wp(\bar s) - \wp_{\tilde R,2}\right)\right]^2}.
    \end{eqnarray}
\end{subequations}
Integrals \eqref{L_int} are calculated in Appendix \ref{appendix:Integrals}.  For $\varepsilon =\delta_1$ integrals \eqref{Integral_N_1} and \eqref{Integral_N_3} have a different form. They are listed in Appendix  \ref{appendix:Caseeeqd}.

\subsection{Expression for $J_\theta$}

The integral
\begin{equation}
    J_\theta(s) = \int_0^s \sin^2 \theta(\bar s) d \bar s
\end{equation}
can be computed in a manner similar to $I_\theta$. Let
\begin{equation}
    z = \sin^2 \theta = 1 - \mu^2,
\end{equation}
where $\mu = \cos \theta$. This gives
\begin{equation}
    \frac{dz}{ds} = - 2 \mu \frac{d \mu}{ds} = \epsilon_\theta \sqrt{4 \mu^2 g(\mu) } = \epsilon_\theta  \sqrt{4 (1-z) \left[ b_0 (1 - z)^2 + 6 b_2 (1 - z) + b_4 \right]} \equiv \epsilon_\theta \epsilon_\mu \sqrt{v(z)}.
\end{equation}
Here
\begin{equation}
    v(z) \equiv 4 e_1 z^3 +6 e_2 z^2 + 4 e_3 z + e_4,
\end{equation}
and 
\begin{subequations}
    \begin{eqnarray}
        e_1 & = & \alpha ^2 \left( \varepsilon ^2 - \delta_1 \right) = d_3,\\
        e_2 & = & \frac{2}{3} \left[\alpha ^2 \left( 2 \delta_1 -  \varepsilon ^2 \right)  - \kappa - 2 \alpha \lambda_z \varepsilon  \right] = d_2,\\
        e_3 & = &  \kappa - \alpha ^2 \delta_1 + 2 \alpha  \lambda_z \varepsilon + \lambda_z^2 = d_1, \\
        e_4 & = & -4 \lambda_z^2 = d_0,
    \end{eqnarray}
\end{subequations}
where coefficients $d_0, \dots, d_3$ are given by Eqs.\ (\ref{coeffsd}). Weierstrass invariants of the polynomial $v(z)$ read
\begin{subequations}
\begin{eqnarray}
    g_{v,2} & = & - 4e_1 e_3 + 3e_2^2, \\
    g_{v,3} & = & 2e_1 e_2 e_3 - e_2^3 - e_1^2 e_4,
\end{eqnarray}
\end{subequations}
and not surprisingly, $g_{v,2} = g_{w,2}$ and $g_{v,3} = g_{w,3}$.
Consequently,
\begin{equation}
    z(s) = z_0 + \frac{ - \epsilon_\theta \epsilon_\mu \, \sqrt{v(z_0)} \wp_w'(s) + \frac{1}{2} v'(z_0 ) \left[ \wp_w(s) - \frac{1}{24}v''(z_0 )\right] + \frac{1}{24} v(z_0 ) v'''(z_0 )  }{2 \left[ \wp_w(s) - \frac{1}{24} v''(z_0 ) \right]^2 },
\end{equation}
where $z_0 = \sin^2 \theta_0$, $\wp_w(s) = \wp(s;g_{w,2},g_{w,3})$. As before, we define
\begin{subequations}
    \begin{eqnarray}
        \mathcal{D}_{1} &=& \frac{1}{48}v(z_{0}) v'''(z_{0}) - \frac{1}{96} v'(z_{0})v''(z_{0}), \\
        \mathcal{D}_{2} &=& \frac{1}{4}v'(z_{0}), \\
        \mathcal{D}_{3} &=& -\frac{1}{2}\epsilon_\theta \epsilon_\mu \sqrt{v(z_{0})},\\
        \mathcal{D}_{4} &=& \frac{1}{24}v''(z_{0}), \\
        \wp_{v,1} &=&  \mathcal{D}_{4} = \frac{1}{24}v''(z_{0}).
    \end{eqnarray}
\end{subequations}
Hence
\begin{eqnarray}
    J_\theta(s) & = & \int_0^s z(\bar s) d \bar s = \int_0^s \left\{ z_0 + \frac{ \mathcal{D}_1 + \mathcal{D}_2 \wp_w(\bar s) +\mathcal{D}_3 \wp_w'(\bar s) }{ \left[ \wp_w(\bar s) - \wp_{v,1} \right]^2} \right\} d \bar s \nonumber \\
    & = &  z_{0} s +  \mathcal{D}_{1} I_{3}(s; p_{v,1}) + \mathcal{D}_{2}  I_{8}(s; p_{v,1})  + \mathcal{D}_{3}  I_{4}(s;p_{v,1}),
\end{eqnarray}
where 
\begin{subequations}\label{I_int}
    \begin{eqnarray}
        I_{3}(s; p_{v,1}) &:=& \int_0^s \frac{ 1 }{ \left[ \wp(\bar s) - \wp_{v,1} \right]^2} d \bar s,\\
        I_{8}(s; p_{v,1}) &:=& \int_0^s \frac{\wp(\bar s) }{ \left[ \wp(\bar s) - \wp_{v,1} \right]^2} d \bar s,\\
        I_{4}(s;  p_{v,1}) &:=&  \int_0^s \frac{\wp'(\bar s) }{\left[ \wp(\bar s) - \wp_{v,1} \right]^2} d \bar s,
    \end{eqnarray}
\end{subequations}    
and $p_{v,1}$ satisfies $\wp\left( p_{v,1}; g_{w,2}, g_{w,3} \right) =\mathcal{D}_{4}$. Integrals \eqref{I_int} are calculated in Appendix \ref{appendix:Integrals}. 

\section{Limiting cases}
\label{appendix:limit}

In this appendix we discuss two special limiting cases: the Schwarzschild limit with $\alpha \to 0$, and the extreme Kerr limit with $\alpha \to \pm 1$.

\subsection{Schwarzschild limit}

The Biermann-Weierstrass formula has been used to provide an explicit description of Schwarzschild geodesics in \cite{CM2022} and \cite{cies_mach_acta_2023}. The description given in \cite{CM2022} exploits the fact that the motion occurs in a single plane and uses the true anomaly as a geodesic parameter. The solution discussed briefly in \cite{cies_mach_acta_2023} is adjusted to the setup of this paper---geodesics are parametrized by an equivalent of the Mino time. In this subsection we discuss $\alpha = 0$ limits of our solutions for $\xi(s)$, $\theta(s)$, $\varphi(s)$, and $T(s)$ and show that they coincide with the expressions given in \cite{cies_mach_acta_2023}.

\subsubsection{Solution for $\xi(s)$}

The form of the solution for $\xi(s)$---Eq.\ \eqref{xi_s}---does not change for $\alpha = 0$, but the coefficients \eqref{fcoeffs} and Weierstrass invariants \eqref{invariants_phys} of the polynomial $\tilde R$ acquire the following simpler form:
\begin{subequations}
    \begin{eqnarray}
        a_0 &=& \varepsilon^2 - \delta_1,\\
        a_1 &=& \frac{1}{2} \delta_1,\\
        a_2 &=& - \frac{1}{6} \kappa ,\\
        a_3 &=&  \frac{1}{2}\kappa,\\
        a_4 &=& 0,
    \end{eqnarray}
\end{subequations}
\begin{subequations}
\begin{eqnarray}   
        g_{\tilde R,2} & = & -4a_1 a_3 + 3a_2^2 = \frac{1}{12} \kappa^2 - \delta_1 \kappa, \\
        g_{\tilde R,3} & = &  2a_1 a_2 a_3 - a_2 ^3 - a_0a_3^2 = \frac{1}{6^3} \kappa^3 - \frac{\delta_1}{12} \kappa^2 - \frac{\varepsilon^2 - \delta_1}{4} \kappa^2 .
\end{eqnarray}
\end{subequations}

\subsubsection{Solution for $\theta(s)$}

For $\alpha = 0$ the solution for $\theta(s)$---Eq.\ (\ref{mu_s})---can be expressed in a simple form, involving elementary functions. This can be seen as follows. The coefficients of the polynomial $g(\mu)$ read, for $\alpha = 0$,
\begin{subequations}
\begin{eqnarray}
    b_0 & = & 0, \\
    b_2 & = &  - \frac{1}{6} \kappa, \\
    b_4 & = &  \kappa  - \lambda_z^2.
\end{eqnarray}
\end{subequations}
The appropriate Weierstrass invariants can be written as
\begin{eqnarray}
    g_{g,2} & = &  3 b_2^2 = \frac{1}{12} \kappa^2, \\
    g_{g,3} & = &  - b_2^3 = \frac{1}{6^3} \kappa^3.
\end{eqnarray}
The value of $\wp(s; g_{g,2}, g_{g,3})$ can be computed by noting the following two identities (\cite{abramowitz_handbook_1964}, p.\ 652, Eq.\ 18.12.27 and  \cite{DLMF}, Eq. 23.10.17):
\begin{equation}
      \wp \left(z;\frac{1}{12}, \frac{1}{6^3} \right) = \frac{1}{ 2\left( 1-\cos{z} \right)}  -\frac{1}{12},
\end{equation}
\begin{equation}\label{wp_rel_1}
      \wp \left( z;c^4 g_2, c^6 g_3 \right) =  c^2 \wp(c z; g_2, g_3 ),
\end{equation}
where $c$ is any nonzero real or complex constant.
This gives 
\begin{equation}\label{wp_simp_1}
     \wp \left(s; g_{g,2}, g_{g,3} \right) = \wp \left( s;\frac{1}{12} \kappa^2, \frac{1}{6^3} \kappa^3 \right) = \kappa \wp \left( \sqrt{\kappa} s;\frac{1}{12}, \frac{1}{6^3} \right) =\frac{\kappa}{ 2 \left[ 1-\cos(\sqrt{\kappa} s) \right]} - \frac{\kappa}{12}.
\end{equation}
Differentiating the above expression with respect to $s$, we get
\begin{equation}\label{wpp_simp_1}
    \wp'(s; g_{g,2}, g_{g,3}) = -\frac{\kappa^{3/2}}{2} \frac{1+\cos(\sqrt{\kappa} s)}{\left[ 1-\cos(\sqrt{\kappa} s) \right] \sin(\sqrt{\kappa} s)}.
\end{equation}
Using Eqs.\ \eqref{wp_simp_1} and \eqref{wpp_simp_1} in Eq.\ (\ref{mu_s}), it is easy to obtain the following expression for $\mu(s)$:
\begin{equation}
\label{muschw}
    \mu(s) = \mu_0 \cos\left(\sqrt{\kappa}s\right) - \epsilon_{\theta} \sqrt{1-\mu_0^2 - \frac{\lambda_z^2}{\kappa}} \sin\left(\sqrt{\kappa}s\right).
\end{equation}
Denoting $\beta_0 = \frac{\epsilon_\theta}{\sqrt{\kappa}} \arcsin{\frac{\mu_0}{\sqrt{1-\frac{\lambda_z^2}{\kappa}}}}$, we finally get
\begin{equation}
\label{muschwfinal}
     \mu(s)= - \epsilon_\theta \sqrt{1-\frac{\lambda_z^2}{\kappa}} \sin \left[\sqrt{\kappa}(s-\beta_0) \right].
\end{equation}
This form has already been obtained for the Schwarzschild metric in \cite{cies_mach_acta_2023}.

\subsubsection{Solution for $\varphi(s)$}

The only term remaining in the expression for $\varphi(s) - \varphi(0)$ in the Schwarzschild limit is the integral $I_\theta$. We have
\begin{equation}
    \varphi(s) - \varphi(0) = I_\theta = \lambda_z \int_0^s q(\bar s) d \bar s.
\end{equation}
The expression for $q(s) = 1/\sin^2 \theta(s)$---Eq.\ \ref{qs})---can also be simplified for $\alpha = 0$. The coefficients and Weierstrass invariantes of the polynomial $w(q)$ have the form
\begin{subequations}
\begin{eqnarray}
    d_0 & = & - 4 \lambda_z^2, \\
    d_1 & = & \kappa + \lambda_z^2, \\
    d_2 & = & - \frac{2}{3} \kappa, \\
    d_3 & = & 0,
\end{eqnarray}
\end{subequations}
and
\begin{subequations}
\begin{eqnarray}
    g_{w,2} & = & \frac{4}{3} \kappa^2, \\
    g_{w,3} & = & \left( \frac{2}{3} \right)^3 \kappa^3.
\end{eqnarray}
\end{subequations}
We find that
\begin{equation}
    \wp(s; g_{w,2}, g_{w,3}) = \wp \left(s; \frac{4}{3} \kappa^2, \frac{2^3}{3^3} \kappa^3 \right) = \kappa \left[ \frac{1}{\sin^2 ( \sqrt{\kappa} s)} - \frac{1}{3} \right]
\end{equation}
and
\begin{equation}
    \wp^\prime(s; g_{w,2}, g_{w,3}) = - 2 \kappa^\frac{3}{2} \frac{\cos (\sqrt{\kappa} s)}{\sin^3 (\sqrt{\kappa} s)}.
\end{equation}
An explicit expression for $q(s)$ is lenghty, but it can be shown that it simplifies to $q(s) = 1/[1 - \mu(s)^2]$, where $\mu(s)$ is given by Eq.\ (\ref{muschw}) or, equivalently, Eq.\ (\ref{muschwfinal}). This gives
\begin{equation}
    \varphi(s) - \varphi(0) = I_\theta = \lambda_z \int_0^s \frac{d \bar s}{1 - \mu^2(\bar s)} = \lambda_z \int_0^s \frac{d \bar s}{1 - \left( 1 - \frac{\lambda_z^2}{\kappa} \right) \sin^2 \left[ \sqrt{\kappa} (s - \beta_0) \right]}.
\end{equation}
Evaluating the last integral, we obtain
\begin{equation}
    \varphi(s) - \varphi(0) =  \arctan\left[ \frac{\lambda_z}{\sqrt{\kappa}} \tan[\sqrt{\kappa} (s-\beta_0)] \right] + \arctan\left[ \frac{\lambda_z}{\sqrt{\kappa}} \tan(\sqrt{\kappa} \beta_0) \right] + n \pi,
\end{equation}
where $n \in \mathbb Z$, in agreement with a formula given in \cite{cies_mach_acta_2023}.

\subsubsection{Solution for $T(s)$}

Assuming $\alpha = 0$, we get, form Eq.\ (\ref{tau_int}),
\begin{equation}
    T(s) - T(0) = \varepsilon \int_0^s \frac{\xi^3 (\bar s)}{\xi(\bar s) - 2} d \bar s.
\end{equation}
Noting that
\begin{equation}
    \frac{\xi^3}{\xi - 2} = \xi^2 + 2 \xi + 4 + \frac{8}{\xi - 2},
\end{equation}
one can write
\begin{eqnarray}
    T(s) - T(0) & = & \varepsilon \int_0^s \xi^2(\bar s) d \bar s + 2 \varepsilon \int_0^2 \xi(\bar s) d \bar s + 4 \varepsilon s + 8 \varepsilon \int_0^s \frac{d \bar s}{\xi(\bar s) - 2} \nonumber \\
    & = & \varepsilon N_2(s) + 2 \varepsilon N_1(s) + 4 \varepsilon s + 8 \varepsilon N_H^+(s).
\end{eqnarray}

Integrals $N_1$ and $N_2$ are given by Eqs.\ (\ref{N_1}) and (\ref{N_2}). The integral $N_H^+$ can still be computed as in Sec.\ \ref{sec:solutions}, i.e.,
\begin{equation}
    N_H^+(s) = \int_0^s \frac{d \bar s}{\xi(\bar s) - 2} = \int_0^s u_+(\bar s) d \bar s.
\end{equation}
Here $u_+$ satisfies Eq.\ (\ref{u_mot}) with the polynomial $h(u_+)$ given by Eq.\ (\ref{g_general}). On the other hand, coefficients $c_0, \dots, c_4$ have, for $\alpha = 0$, the following simple form
\begin{eqnarray}
    c_0 & = & 16 \varepsilon^2, \\
    c_1 & = & 8 \varepsilon^2 - \frac{1}{2} \kappa - 2 \delta_1, \\
    c_2 & = & 4 \varepsilon^2 - \frac{1}{6} \kappa - 2 \delta_1, \\
    c_3 & = & 2 \varepsilon^2 - \frac{3}{2} \delta_1, \\
    c_4 & = & \varepsilon^2 - \delta_1.
\end{eqnarray}
As in the general case, the Weierstrass invariants $g_{h,2}$ and $g_{h,3}$ of the polynomial $h(u_+)$ coincide with $g_{\tilde R,2}$ and $g_{\tilde R,3}$, respectively.

\subsection{Extreme Kerr limit}
\label{appendix:extremeKerr}

A few equations have to be modified in the limit of $\alpha \to \pm 1$ (extreme Kerr spacetime). In this case $\Delta = r^2 - 2 M r + a^2 = (r - M)^2$ and $\xi^2 - 2 \xi + \alpha^2 = (\xi - 1)^2$.

Equation (\ref{Ixi1}) reads
\begin{equation}
I_{\xi} (s) = \alpha \int_0^s \frac{ 2 \xi(\bar s) \varepsilon - \alpha \lambda_z}{[\xi(\bar s) - 1]^2} d \bar s. 
\end{equation}
A partial fraction expansion now gives
\begin{equation}
    I_{\xi} (s) = 2 \alpha \varepsilon N_4(s) + (2 \alpha \varepsilon - \lambda_z) N_5(s),
\end{equation}
where
\begin{equation}
    N_4(s) = \int_0^s \frac{d \bar s}{\xi(\bar s) - 1}, \quad N_5(s) = \int_0^s \frac{d \bar s}{[\xi(\bar s) - 1]^2}.
\end{equation}
Integral $N_4$ can be computed as in Eqs.\ (\ref{N_H})--(\ref{K_int}). Defining $u(s) = 1/[\xi(s) - 1]$, we get
\begin{equation}
    \frac{du}{ds} = - \epsilon_r \sqrt{h(u)},
\end{equation}
where
\begin{equation}
    h(u) = c_0 u^4 + 4 c_1 u^3 + 6 c_2 u^2 + 4 c_3 u + c_4,
\end{equation}
and
\begin{subequations}
\begin{eqnarray}
    c_0 & = & 4 \varepsilon^2 - 4 \alpha \varepsilon \lambda_z + \lambda_z^2,\\
    c_1 & = & 2 \varepsilon^2 - \alpha \varepsilon \lambda_z, \\
    c_2 & = & \frac{1}{6} \left( - \delta_1 + 8 \varepsilon^2 - \kappa - 2 \alpha \varepsilon \lambda_z \right), \\
    c_3 & = & \varepsilon^2 - \frac{1}{2} \delta_1, \\
    c_4 & = & \varepsilon^2 - \delta_1 .
\end{eqnarray}
\end{subequations}
Solution for $u$ is then given, as in Eq.\ (\ref{u_s}), and integrals $N_4$ and $N_5$ can be computed as
\begin{equation}
    N_4(s) = \int_0^s u(\bar s) d \bar s, \quad N_5(s) = \int_0^s u^2(\bar s) d \bar s.
\end{equation}

Similarly, the integral $J_\xi(s)$ appearing in Eq.\ (\ref{tau_int}) reads, for $\alpha = \pm 1$,
\begin{eqnarray}
     J_\xi(s) = \int_0^s \frac{\left[\xi^2(\bar s) + 1 \right]^2 \varepsilon - 2\alpha \lambda_z \xi(\bar s)}{[ \xi(\bar s) - 1 ]^2} d \bar s  
     & = & 5 \varepsilon s + 2 \varepsilon \int_0^s \xi(\bar s) d \bar s + \varepsilon \int_0^s \xi^2(\bar s) d \bar s \nonumber \\
    & & + 2 (4 \varepsilon - \alpha \lambda_z) N_4(s) + 2 (2 \varepsilon - \alpha \lambda_z) N_5(s) .
\end{eqnarray}

\subsection{The case with $\varepsilon = \delta_1$}
\label{appendix:Caseeeqd}

In this subsection we compute integrals \eqref{Integral_N_1} and \eqref{Integral_N_3} in the special case with $\varepsilon = \delta_1$. In this case $\mathcal{C}_{5}=0$ [Eq.\ \ref{C_5}], and Eq.\ \eqref{xi_s_2} has the following form
\begin{equation}
    \xi(s) = \xi_0 + \frac{\mathcal{C}_{1} + \mathcal{C}_{2}\wp_{\tilde R}(s) + \mathcal{C}_{3} \wp_{\tilde R}'(s)   }{ \left[ \wp_{\tilde R}(s) - \wp_{\tilde R,3} \right]^2},
\end{equation}
where $\wp_{\tilde R,3} = \wp\left( p_{\tilde R,3}; g_{\tilde R,2}, g_{\tilde R,3} \right) = \mathcal{C}_{4}$.
Thus
\begin{eqnarray}
    N_1(s) & = & \int_0^s \xi(\bar s) d \bar s = \int_0^s \left\{ \xi_0 + \frac{\mathcal{C}_{1} + \mathcal{C}_{2}\wp_{\tilde R}(s) + \mathcal{C}_{3} \wp_{\tilde R}'(s) }{ \left[ \wp_{\tilde R}(s) - \wp_{\tilde R,3} \right]^2} \right\} d \bar s \nonumber \\
    & = &  \xi_{0} s +  \mathcal{C}_{1} I_{3}(s; p_{\tilde R,3}) + \mathcal{C}_{2}  I_{8}(s; p_{\tilde R,3})  + \mathcal{C}_{3}  I_{4}(s;p_{\tilde R,3}),
\end{eqnarray}
and 

\begin{eqnarray}
     N_3(s) & = & \int_0^s \left\{\frac{\mathcal{C}_{1} + \mathcal{C}_{2}\wp_{\tilde R}(s) + \mathcal{C}_{3} \wp_{\tilde R}'(s) }{\left[ \wp_{\tilde R}(\bar s) - \wp_{\tilde R,3} \right]^2 } \right\}^2 d \bar s = \mathcal{C}^2_{1} I_{19}(s; p_{\tilde R,3}) + 2 \mathcal{C}_{1} \mathcal{C}_{2} I_{20}(s; p_{\tilde R,3}) \nonumber \\
     & & + 2 \mathcal{C}_{1} \mathcal{C}_{3} I_{21}(s; p_{\tilde R,3}) + 2 \mathcal{C}_{2} \mathcal{C}_{3} I_{22}(s; p_{\tilde R,3}) + \mathcal{C}^2_{2} I_{23}(s; p_{\tilde R,3}) + \mathcal{C}^2_{3} I_{25}(s; p_{\tilde R,3}),
\end{eqnarray}
where
\begin{subequations}
    \begin{eqnarray}
        I_{3}(s; p_{\tilde R,3}) &:=& \int_0^s \frac{ 1 }{ \left[ \wp(\bar s) -  \wp_{\tilde R,3} \right]^2 } d \bar s,\\
        I_{4}(s; p_{\tilde R,3}) &:=& \int_0^s \frac{ \wp'(\bar s) }{ \left[ \wp(\bar s) -  \wp_{\tilde R,3} \right]^2 } d \bar s,\\
        I_{8}(s; p_{\tilde R,3}) &:=& \int_0^s \frac{ \wp(\bar s) }{ \left[ \wp(\bar s) -  \wp_{\tilde R,3} \right]^2 } d \bar s,\\
        I_{19}(s; p_{\tilde R,3}) &:=& \int_0^s \frac{ 1}{ \left[ \wp(\bar s) -  \wp_{\tilde R,3} \right]^4 } d \bar s,\\
        I_{20}(s; p_{\tilde R,3}) &:=& \int_0^s \frac{  \wp(\bar s)}{ \left[ \wp(\bar s) -  \wp_{\tilde R,3} \right]^4 } d \bar s,\\
        I_{21}(s; p_{\tilde R,3}) &:=& \int_0^s \frac{  \wp'(\bar s)}{ \left[ \wp(\bar s) -  \wp_{\tilde R,3} \right]^4 } d \bar s,\\
        I_{22}(s; p_{\tilde R,3}) &:=& \int_0^s \frac{ \wp'(\bar s)  \wp(\bar s)}{ \left[ \wp(\bar s) -  \wp_{\tilde R,3} \right]^4 } d \bar s,\\
        I_{23}(s; p_{\tilde R,3}) &:=& \int_0^s \frac{ \wp(\bar s)^2}{ \left[ \wp(\bar s) -  \wp_{\tilde R,3} \right]^4 } d \bar s,\\
        I_{25}(s; p_{\tilde R,3}) &:=& \int_0^s \frac{ \wp'(\bar s)^2}{ \left[ \wp(\bar s) -  \wp_{\tilde R,3} \right]^4 } d \bar s.
    \end{eqnarray}
\end{subequations}

A analogous analysis should be carried out for initial parameters yielding $\mathcal{A}_{5\pm} = 0$ or  $\mathcal{B}_{5} =0$.

\section{Integrals}
\label{appendix:Integrals}

In this appendix, we collect the formulas for all elliptic integrals appearing in this paper and used to solve Kerr geodesic equations. We adopt a convention according to which in all following formulas $x$ denotes the variable, while $y$ and $z$ are treated as parameters. This is emphasized by separating the variable $x$ from the parameters $y$ and $z$ with the semicolon. The integrals are based on the results presented in \cite{gradshtein_table_2007} (pp. 626) \cite{byrd_handbook_1971} (pp. 311), \cite{Tannery_1893} (vol.4, pp. 109--110)

\begin{alignat}{1}
\label{intC1}
I_1 (x;y) =& \int \frac{dx}{\wp(x) - \wp(y)} = \frac{1}{\wp'(y)} \left( 2\zeta(y) x + \ln{\frac{\sigma( x - y)}{\sigma(x + y)}} \right),\\
\label{intC2}
I_2 (x;y) =& \int \frac{\wp'(x) dx}{\wp(x) - \wp(y)} = \ln{\left(\wp(x) - \wp(y) \right) },\\
\label{intC3}
I_3 (x;y) =& \int \frac{dx}{\left(\wp(x) - \wp(y)\right)^2} = - \frac{\wp''(y)}{\wp'(y)^3}\ln{\frac{\sigma\left(x - y\right)}{\sigma\left(x + y\right)}} - \nonumber \\ &  -\frac{1}{\wp'\left( y\right)^2} \Bigg( \zeta\left(x +y\right) + \zeta\left(x - y\right) + \left( 2\wp\left( y\right) + \frac{2 \wp''\left( y\right)\zeta\left(y\right)}{\wp'\left(y\right)}    \right) x \Bigg),\\
I_4 (x;y) =& \int \frac{\wp'(x) dx}{\left(\wp(x) - \wp(y)\right)^2} = - \frac{1}{\wp(x) - \wp(y) },\\
I_5 (x;y) =& \int \frac{\wp'(x) \wp(x) dx}{\wp(x) - \wp(y)} = \wp(y)  \ln{\left(\wp(x) - \wp(y) \right) } + \wp(x),\\
I_6 (x;y) =& \int \frac{\wp'(x) \wp(x) dx}{\left(\wp(x) - \wp(y)\right)^2} =  \ln{\left(\wp(x) - \wp(y) \right)} - \frac{\wp(y)}{\wp(x) - \wp(y)},\\
\label{intC7}
I_7 (x;y) =& \int \frac{\wp(x) dx}{\wp(x) - \wp(y)} = x + \frac{\wp(y)}{\wp'(y)}\left( 2\zeta(y)x + \ln{\frac{\sigma\left(x - y\right)}{\sigma\left(x + y\right)}} \right),\\
I_8 (x;y) =& \int \frac{\wp(x) dx}{\left(\wp(x) - \wp(y)\right)^2} = I_1(x;y) + \wp(y) I_3 (x;y),\\
I_9 (x;y) =& \int \frac{\wp^2(x) dx}{\wp(x) - \wp(y)} = \wp(y) I_7 (x;y) - \zeta(x),\\
I_{10} (x;y) =& \int \frac{\wp^3(x) dx}{\wp(x) - \wp(y)} = \wp(y)I_9(x;y) + \frac{1}{6}\wp'(x) + \frac{g_2}{12}x,\\
I_{11} (x;y) =&  \int \frac{\wp'^2(x) dx}{\wp(x) - \wp(y)} = 4I_{10}(x;y) -g_2 I_7(x;y) -g_3 I_1(x;y),\\
I_{12} (x;y) =& \int \frac{\wp^2(x) dx}{\left(\wp(x) - \wp(y)\right)^2} = I_7(x;y) + \wp(y)I_8(x;y),\\
I_{13} (x;y) =& \int \frac{\wp^3(x) dx}{\left(\wp(x) - \wp(y)\right)^2} = I_{9}(x;y) + \wp(y)I_{12}(x;y),\\
I_{14} (x;y) =& \int \frac{\wp'^2(x) dx}{\left(\wp(x) - \wp(y)\right)^2} = 4I_{13}(x;y) - g_2 I_8(x;y) - g_3 I_3(x;y),\\
I_{15} (x;y) =& \int \frac{ dx}{\left(\wp(x) - \wp(y)\right)^3} = \frac{1}{2\wp'(y)^3}\Big( \wp(x+y) - \wp(x-y) - 2\wp'(y) x - \nonumber \\ & - 12 \wp(y)\wp'(y) I_1(x;y) - 3\wp'(y) \wp''(y) I_3 (x;y)\Big) ,\\
I_{16} (x;y) =& \int \frac{ \wp'(x) dx}{\left(\wp(x) - \wp(y)\right)^3} = - \frac{1}{2\left(\wp(x) - \wp(y)\right)^2},\\
I_{17} (x;y) =& \int \frac{ \wp(x) dx}{\left(\wp(x) - \wp(y)\right)^3} = I_3(x;y) + \wp(y) I_{15}(x;y),
\end{alignat}
\begin{alignat}{1}
I_{18} (x;y) =& \int \frac{ \wp^2(x) dx}{\left(\wp(x) - \wp(y)\right)^3} = I_8(x;y) + \wp(y) I_{17}(x;y),\\
I_{19} (x;y) =& \int \frac{  dx}{\left(\wp(x) - \wp(y)\right)^4} =  \frac{1}{6\wp'(y)^4}\Big[ \wp'(x+y) + \wp'(x-y) - 2\wp''(y) x - \nonumber \\ & -  \wp^{(4)}(y) I_1(x;y) - \left( 3\wp''(y)^2 + 48\wp(y)\wp'(y) \right) I_3 (x;y) - 12 \wp''(y) \wp'(y)^2 I_{15}(x;y)\Big] ,\\
I_{20} (x;y) =& \int \frac{ \wp(x) dx}{\left(\wp(x) - \wp(y)\right)^4} =  I_{15}(x;y) + \wp(y) I_{19}(x;y),\\
I_{21} (x;y) =& \int \frac{ \wp'(x) dx}{\left(\wp(x) - \wp(y)\right)^4} = - \frac{1}{3\left(\wp(x) - \wp(y)\right)^3},\\
I_{22} (x;y) =& \int \frac{ \wp(x) \wp'(x) dx}{\left(\wp(x) - \wp(y)\right)^4} = - \frac{3\wp(x) - \wp(y)}{6\left(\wp(x) - \wp(y)\right)^3} ,\\
I_{23} (x;y) =& \int \frac{ \wp^2(x) dx}{\left(\wp(x) - \wp(y)\right)^4} =  I_{17}(x;y) + \wp(y) I_{20}(x;y),\\
I_{24} (x;y) =& \int \frac{ \wp^3(x) dx}{\left(\wp(x) - \wp(y)\right)^4} =  I_{18}(x;y) + \wp(y) I_{23}(x;y),\\
I_{25} (x;y) =& \int \frac{ \wp'^2(x) dx}{\left(\wp(x) - \wp(y)\right)^4} =  4 I_{24}(x;y) - g_2 I_{20}(x;y) - g_3 I_{19}(x;y),\\
\label{intK1}
K_{1} (x;y,z) =& \int \frac{dx}{\left(\wp(x) - \wp(y)\right)\left(\wp(x) - \wp(z)\right)} = \frac{I_1(x;y) - I_1(x;z)}{\wp(y) - \wp(z)},\\
\label{intK2}
K_{2} (x;y,z) =& \int \frac{\wp(x)dx}{\left(\wp(x) - \wp(y)\right)\left(\wp(x) - \wp(z)\right)} = \frac{I_7(x;y) - I_7(x;z)}{\wp(y) - \wp(z)},\\
\label{intK3}
K_{3} (x;y,z) =& \int \frac{\wp'(x)dx}{\left(\wp(x) - \wp(y)\right)\left(\wp(x) - \wp(z)\right)} = \frac{I_2(x;y) - I_2(x;z)}{\wp(y) - \wp(z)},\\
K_{4} (x;y,z) =& \int \frac{\wp(x) \wp'(x)dx}{\left(\wp(x) - \wp(y)\right)\left(\wp(x) - \wp(z)\right)} = \frac{I_5(x;y) - I_5(x;z)}{\wp(y) - \wp(z)},\\
K_{5} (x;y,z) =& \int \frac{\wp^2(x)dx}{\left(\wp(x) - \wp(y)\right)\left(\wp(x) - \wp(z)\right)} = \frac{I_9(x;y) - I_9(x;z)}{\wp(y) - \wp(z)},\\
K_{6} (x;y,z) =& \int \frac{\wp'^2(x)dx}{\left(\wp(x) - \wp(y)\right)\left(\wp(x) - \wp(z)\right)} = \frac{I_{11}(x;y) - I_{11}(x;z)}{\wp(y) - \wp(z)},\\
L_{1} (x;y,z) =& \int \frac{dx}{ \left[ \left(\wp(x) - \wp(y)\right)\left(\wp(x) - \wp(z)\right)\right]^2} = \frac{I_3(x;y) -2 K_1(x;y,z) + I_3(x;z)}{\left(\wp(y) - \wp(z)\right)^2},\\
L_{2} (x;y,z) =& \int \frac{\wp(x) dx}{ \left[ \left(\wp(x) - \wp(y)\right)\left(\wp(x) - \wp(z)\right)\right]^2} = \frac{I_8(x;y) -2 K_2(x;y,z) + I_8(x;z)}{\left(\wp(y) - \wp(z)\right)^2},\\
L_{3} (x;y,z) =& \int \frac{\wp'(x) dx}{ \left[ \left(\wp(x) - \wp(y)\right)\left(\wp(x) - \wp(z)\right)\right]^2} = \frac{I_4(x;y) -2 K_3(x;y,z) + I_4(x;z)}{\left(\wp(y) - \wp(z)\right)^2},\\
L_{4} (x;y,z) =& \int \frac{\wp'(x) \wp(x) dx}{ \left[ \left(\wp(x) - \wp(y)\right)\left(\wp(x) - \wp(z)\right)\right]^2} = \frac{I_6(x;y) -2 K_4(x;y,z) + I_6(x;z)}{\left(\wp(y) - \wp(z)\right)^2},\\
L_{5} (x;y,z) =&   \int \frac{\wp(x)^2 dx}{ \left[ \left(\wp(x) - \wp(y)\right)\left(\wp(x) - \wp(z)\right)\right]^2} = \frac{I_{12}(x;y) -2 K_5(x;y,z) + I_{12}(x;z)}{\left(\wp(y) - \wp(z)\right)^2},\\
L_{6} (x;y,z) =& \int \frac{\wp'(x)^2 dx}{ \left[ \left(\wp(x) - \wp(y)\right)\left(\wp(x) - \wp(z)\right)\right]^2} = \frac{I_{14}(x;y) -2 K_6(x;y,z) + I_{14}(x;z)}{\left(\wp(y) - \wp(z)\right)^2}.
\end{alignat}

%===========================================================================================
%===========================================================================================

\end{document}